\documentclass[reprint,floatfix,aps,prd,showpacs,nofootinbib,amsmath,amssymb,amsfonts,superscriptaddress]{revtex4-1}
\usepackage{bm}
\usepackage{graphicx,color}
\usepackage{physics}
\usepackage{dsfont}
\usepackage[normalem]{ulem}
\usepackage{hyperref,url}
\usepackage{mathrsfs}
\usepackage{calrsfs}
\usepackage{subfig}

\renewcommand{\vec}[1]{\mathbf{#1}}
\usepackage{varioref}
\usepackage[active]{srcltx}

\expandafter\ifx\csname package@font\endcsname\relax\else
\expandafter\expandafter
\expandafter\usepackage
\expandafter\expandafter
\expandafter{\csname package@font\endcsname}%
\fi
\hypersetup{
	colorlinks=true,       % false: boxed links; true: colored links
	linkcolor=blue,          % color of internal links
	citecolor=blue,        % color of links to bibliography
}

\usepackage[section]{placeins}

\usepackage{fancybox}
\labelformat{figure}{Fig.~#1}

\begin{document}
\title{One-to-one correspondence between entanglement mechanics and black hole thermodynamics}
\author{S. Mahesh Chandran} 
\email{maheshchandran@iitb.ac.in}
\affiliation{Department of Physics, Indian Institute of Technology Bombay, Mumbai 400076, India}
\author{S. Shankaranarayanan}
\email{shanki@phy.iitb.ac.in}
\affiliation{Department of Physics, Indian Institute of Technology Bombay, Mumbai 400076, India}
%%%
\begin{abstract}
We establish a one-to-one mapping between entanglement entropy, energy, and temperature (quantum entanglement mechanics) with black hole entropy, Komar energy, and Hawking temperature, respectively. We show this explicitly for 4-D spherically symmetric asymptotically flat and non-flat space-times with single and multiple horizons.  We exploit an inherent scaling symmetry of entanglement entropy and identify scaling transformations that generate an infinite number of systems with the same entanglement entropy, distinguished only by their respective energies and temperatures. We show that this scaling symmetry is present in most well-known systems starting from the two-coupled harmonic oscillator to quantum scalar fields in spherically symmetric space-time. The scaling symmetry allows us to identify the cause of divergence of entanglement entropy to the generation of (near) zero-modes in the systems. We systematically
isolate the zero-mode contributions using suitable boundary conditions.  We show that the entanglement entropy and energy of quantum scalar field scale differently in space-times with horizons and flat space-time.
The relation $E=2TS$, in analogy with the horizon's thermodynamic structure, is also found to be universally satisfied in the entanglement picture. We then show that there exists a one-to-one correspondence leading to the Smarr-formula of black hole thermodynamics for asymptotically flat and non-flat space-times.
\end{abstract}
\pacs{}

\maketitle

\section{Introduction}
Black hole entropy, or broadly black hole thermodynamics, is expected to provide
information about quantum gravity~\cite{2001-Wald-LivingReviewsinRelativity}. In the absence of
a workable theory of quantum gravity, it is necessary to have an
approach that incorporates key features of quantum gravity, 
yet does not depend on the details of any approach to
quantum gravity. Quantum entanglement is one such approach~\cite{1986-Bombelli.etal-Phys.Rev.D,1993-Srednicki-Phys.Rev.Lett.,Das2010,2011-Solodukhin-LivingRev.Rel.}.

While entanglement is a common characteristic of quantum mechanics, 
it is a core feature in quantum field theory~\cite{Bekenstein1994,2017-Unruh.Wald-ReportsonProgressinPhysics}. The two-point function 
of a quantum field $\varphi(x^{\mu})$ propagating in an arbitrary background
space-time, in  a quantum state $| \Psi \rangle$ is given by~\cite{1982-Birrell.Davies-QuantumFieldsCurved}:
\[
\left\langle\Psi\left|\phi\left(x_{1}\right) \phi\left(x_{2}\right)\right| \Psi\right\rangle \propto 
\frac{1}{\sigma\left(x_{1}, x_{2}\right)}
\]
where $\sigma(x_1,x_2)$ is the square of the geodesic distance between the two points 
$x_1$ and $x_2$ in 4-D space-time. In the limit of $x_1 \to x_2$, the two-point function diverges as power-law. However, the expectation of the scalar field is finite, i. e., 
\[
\left\langle\Psi\left|\phi\left(x_{1}\right) \right| \Psi\right\rangle 
\left\langle\Psi\left|\phi\left(x_{2}\right)\right| \Psi\right\rangle 
= \left( \left\langle\Psi\left|\phi\left(x\right)\right| \Psi\right\rangle \right)^2 \, .
\]
Thus, 
\[
\left\langle\Psi\left|\phi\left(x_{1}\right) \phi\left(x_{2}\right)\right| \Psi\right\rangle
\neq \left( \left\langle\Psi\left|\phi\left(x\right)\right| \Psi\right\rangle \right)^2 \, .
\]
In other words, there exists entanglement between the quantum field in two causally complementary regions~\cite{Bekenstein1994}. While quantum entanglement may not be of significance in Minkowski space-time, it seems to have important implications for space-times with horizons. More specifically, the entanglement entropy associated with a region of space in quantum field theory (QFT) may correspond with black-hole entropy~\cite{1986-Bombelli.etal-Phys.Rev.D,Das2010,2011-Solodukhin-LivingRev.Rel.}. There are three primary reasons for the relevance of entanglement for black-hole entropy.

First, entanglement, like black-hole entropy, is a quantum effect with no classical analog~\cite{2010-Eisert.etal-Rev.Mod.Phys.}. 
Second, entanglement entropy and black-hole entropy are associated
with the existence of the horizon. To elaborate this, let us consider a
scalar field on a background of a collapsing star. Before the
collapse, an outside observer, at least theoretically, has all the
information about the collapsing star. Hence, the entanglement entropy
is zero. During the collapse and once the horizon forms, black-hole entropy is
non-zero~\cite{Bekenstein1998}.  The outside observers at spatial infinity do not have the
information about the quantum degrees of freedom inside the horizon.
Thus, entanglement entropy is also non-zero. Therefore, both entropies are associated with the existence of the horizon. Note that it is
possible to obtain a non-vanishing entanglement entropy for a scalar field in flat space-time by artificially creating a horizon~\cite{1993-Srednicki-Phys.Rev.Lett.}. However, 
in a black-hole, the event horizon is a physical boundary beyond
which the observers do not access.
Third, entanglement is a generic feature of quantum theory and
hence should be present in any quantum theory of gravity. Although
the results presented below do not involve quantization of gravity,
they have implications for the full theory.

One of the key problems with entanglement entropy of quantum fields (in more than two space-time dimensions) is the unavailability of robust analytical tools~\cite{2010-Eisert.etal-Rev.Mod.Phys.,2012-Peschel-BrazilianJournalofPhysics}. 
Currently, there are two approaches to calculate entanglement
entropy of free quantum fields in the literature. One approach is the replica trick,
which rests on evaluating the partition function on an $n-$fold cover of the
background geometry where a cut is introduced throughout the exterior of the entangling surface~\cite{2004-Calabrese.Cardy-JournalofStatisticalMechanicsTheoryandExperiment,2009-Calabrese.Cardy-JournalofPhysicsAMathematicalandTheoretical,2011-Solodukhin-LivingRev.Rel.}. 
Second is a direct approach, where the Hamiltonian of the field is discretized, and the reduced density matrix is evaluated in the real space~\cite{1993-Srednicki-Phys.Rev.Lett.,Das2010,2011-Solodukhin-LivingRev.Rel.}.
The divergence of entanglement entropy is one thing that is common in both approaches.

The divergence of entanglement entropy is associated with the \emph{ultraviolet divergence}~\cite{2010-Padmanabhan-Phys.Rev.D}. 
However, there is no obvious way to renormalize the entanglement entropy. In this work, we take a different approach to the issue of divergence of entanglement entropy. Consider a classical model that possesses symmetries. Upon quantization, some of these classical symmetries may disappear when the quantum theory is properly defined in the presence of its infinities. These are referred to as anomalies or \emph{quantum mechanically broken symmetries}~\cite{1995-Jackiw-DiverseTopicsTheoretical,Jackiw1999}. In this work, we take a step in this direction and show that a massive scalar field in a fixed background possesses new scaling symmetries, which can be attributed to the divergence of the entanglement entropy. 

In Ref.~\cite{2014-Mallayya.etal-Phys.Rev.D}, the authors studied the origin of divergence of entanglement entropy in 
$(1 + 1)-$dimensional space-time. It was shown that the entanglement entropy is invariant under a scaling transformation even when the Hamiltonian is not. The divergence in entanglement entropy in $(1 + 1)-$ dimensions in the continuum limit is due to the presence of many near-zero modes 
(and is not of UV origin as commonly believed). This work shows that the entanglement entropy of a massive scalar field in space-times with the horizon is invariant under a set of scaling transformation while the entanglement energy (disturbed vacuum energy in the presence of a boundary)\cite{1998-Mukohyama.etal-Phys.Rev.D} is not. Exploiting the scaling symmetry, we attribute the entropy divergence in scalar field theory to the generation of zero modes in the system. More specifically, the entanglement entropy has a scaling symmetry that the entanglement energy of the system does not have. This implies that entanglement entropy has an \emph{infinite degeneracy} w.r.t these transformations, and \emph{only} the entanglement energy of the system can break this infinite degeneracy. 

The second key issue we address in this work is to obtain a consistent structure of horizon thermodynamics from quantum entanglement. While the association of black-hole entropy to entanglement entropy has been well-established on account of area-law, the subsystem analogs of horizon energy and temperature are much less studied. On fixing the definition of ``entanglement energy", we show that this energy scales as the Komar energy of the horizon. The relation $E=2TS$ is universally satisfied in the entanglement picture~\cite{1973-Bardeen.etal-CMP,2004-Padmanabhan-ClassicalandQuantumGravity,2005-Padmanabhan-PhysicsReports,2009-Kastor.etal-CQG,2010-Banerjee.Majhi-Phys.Rev.D}. 
Despite entanglement energy and entropy not being equal to the Komar energy and Bekenstein-Hawking entropy respectively, we show that there is exists a one-to-one correspondence that leads to the Smarr-formula of black hole thermodynamics from \emph{quantum entanglement mechanics} 
(for brevity, we refer this as entanglement mechanics).

For the hurried reader, we give below the key results in each section and point to relevant equations and figures:
\begin{itemize}
	\item Section \ref{seccho}: On exploiting the scaling symmetry of entanglement, we show that entropy divergence can be attributed to zero modes, even in the strong coupling limit. See Eq. \eqref{eq:SE-CHO} and \ref{cho1a}.
	\item Section \ref{secmath} : Tools for calculating entanglement energy \eqref{eq:def-EntE}, and temperature \eqref{eq:def-EntT} is developed. Non-uniqueness of entanglement energy definition is accounted for by pre-factor $\epsilon$, which may be fixed by comparing with already established thermodynamic results. \ref{cho2} shows that the definition of entanglement energy and temperature leads to physically consistent results for two-coupled harmonic oscillator.
\item Section \ref{sec1d}:  To isolate the zero-mode contributions, we study massive scalar fields in $(1 + 1)-$dimensions with two different boundary conditions. The results are as follows: \\
(i) Entropy and energy scale similarly for both massless and massive cases. See Eq. \eqref{eq:SE-1DScalar} and \ref{1dfin1}. \\ 
(ii) For both massless and massive cases, entanglement temperature is a constant fixed by the UV cut-off $a$. See  Eq. \eqref{eq:Temp-1DScalar}. \\
iii) Critical scaling is almost exactly captured for Dirichlet condition while there are spurious zero-mode effects in Nuemann condition.  The finite chain with Dirichlet condition exactly captures the conformal limit scaling and provides the correct value of the entanglement entropy~\cite{2004-Calabrese.Cardy-JournalofStatisticalMechanicsTheoryandExperiment}. See Eq. \eqref{scaling1d}. \\
 iv) Entropy divergence is always associated with the generation of zero-modes in the system, for both Dirichlet and Neumann conditions. See Eq. \eqref{scaling1d}.
 %%%
	\item Section \ref{sec3d} : We study massive scalar fields in $(3 + 1)-$dimension Minkowski space-time for Dirichlet and Neumann conditions as in Section \ref{sec1d}. The results are as follows: \\
(i) Entropy and energy always follow area-law. See Eq. (\ref{eq:SE-Scalar3dMink01}). \\
(ii) Temperature is a constant fixed by the UV-cutoff $a$. See Eq. \eqref{eq:SE-Scalar3dMink02}. \\
(iii) Entanglement entropy for both Dirichlet and Neumann conditions converge even when $\Lambda$ is very small ($\sim10^{-10}$). See \ref{3dse}.  \\
(iv) Entanglement energy is far more sensitive to zero modes, wherein the pre-factor of area law increases drastically. See Eq. \eqref{eq:SE-Scalar3dMink02}. \\
(v) Entropy divergence is always associated with the generation of zero-modes in the system, for both Dirichlet and Neumann conditions.
%%%%
	\item Section \ref{secgr}: We extend the scale-invariant treatment of entanglement entropy, energy, and temperature to static, spherically symmetric space-times with horizons. Key results are:\\
(i) Unlike flat space-time, entropy and energy scale differently. See Eqs. (\ref{eq:SE-Schw02}, \ref{eq:SE-dS02},  \ref{rnscale}, \ref{eq:SE-SAdS02}, \ref{eq:SE-SdS02}, \ref{eq:SE-SdS04}). \\
(ii) Unlike flat space-time, entanglement temperature is independent of the UV cutoff. See Eqs. (\ref{eq:SE-Schw02}, \ref{eq:SE-dS02},  \ref{rnscale}, \ref{eq:SE-SAdS02}, \ref{eq:SE-SdS02}, \ref{eq:SE-SdS04}). \\
(iii)  For all space-times studied, $T\sim1.26T_H$. See Eqs. (\ref{eq:EntTemp-Sch}, \ref{eq:EntTemp-dS}, \ref{eq:EntTemp-RN}, \ref{eq:EntTemp-SAdS}, \ref{eq:EntTemp-SdS01}, \ref{eq:EntTemp-SdS02}). \\
(iii)  There is a one-to-one mapping between entanglement energy, entropy, temperature and Komar energy, Bekenstein-Hawking entropy, Hawking temperature, respectively. See Table \ref{tab:mapping}.
%%%%%%%  
	\item Section \ref{sec:Ent-BHconnection} : We absorb the pre-factor $1.26$ into $\epsilon$ in the definition of entanglement energy \eqref{eq:EntTem-HawTemp}. As a result: \\
(i) $T=T_H$ and $E=2TS$ are universally obeyed.  See. Eq. \eqref{eq:SmarrFor01}. \\
(ii) Smarr formula of black hole thermodynamics for asymptotically flat and non-flat space-times
can be derived from entanglement mechanics. See Eqs. (\ref{eq:SmarrFor02}, \eqref{eq:SmarrFor03}, \eqref{eq:SmarrFor04}). See Table \eqref{tab:mapping02}.
%%%
\item Section \ref{secconc} contains the implications of the results. 
\end{itemize}

Throughout this work, the metric signature we adopt is $(-,+,+,+)$ and set $G=\hbar = c = k_B = 4 \pi \epsilon_0 = 1$.

\section{Warm up: Scaling symmetry and zero modes in CHO}\label{seccho}
The coupled harmonic oscillator system serves as a fundamental testing ground for various techniques that have important field theory applications, as we will see in later sections. Here, we concentrate on a class of transformations that leave entanglement entropy of a system invariant, not its energy (${\cal E}$). 
In the next section, we define the entanglement energy ($E$)~\cite{1997-Mukohyama.etal-Phys.Rev.D} and analyze the properties under the same class of transformations.
As mentioned earlier, this additional symmetry that the entropy possesses provides better insight into a certain aspect of quantum systems that is less explored. We show that the entropy divergence in the large coupling limit is due to {zero modes}.

In order to understand how this comes about, let us begin with the Hamiltonian for a coupled harmonic oscillator:
\begin{equation}
\label{eq:CHO-Hamil}
H=\frac{p_1^2}{2m}+\frac{p_2^2}{2m}+\frac{1}{2}m\omega^2\left(x_1^2+x_2^2\right)+\frac{\alpha^2}{2}\left(x_1-x_2\right)^2.
\end{equation}
Under the transformations $x_{\pm}=(x_1\pm x_2)/\sqrt{2}$, the above Hamiltonian reduces to:
\begin{equation}
\label{eq:CHO-Hamil02}
H=\frac{p_+^2}{2m}+\frac{p_-^2}{2m}+\frac{1}{2}m\omega_+^2x_+^2+\frac{1}{2}m\omega_-^2x_-^2,
\end{equation}
where the normal modes are:
\begin{equation}
\omega_-=\sqrt{\omega^2+\frac{2\alpha^2}{m}}; \qquad  \omega_+=\omega.
\end{equation}
For the ground state wave-function of the above Hamiltonian, the entanglement entropy is given by~\cite{1993-Srednicki-Phys.Rev.Lett.}:
\begin{align}
\label{eq:CHO-EntS}
%\label{ent}
	S(\lambda)&=-\sum_{n=1}^\infty p_n(\lambda)\ln{p_n(\lambda)}\nonumber\\
	&=-\ln{\left[1-\xi(\lambda)\right]}-\frac{\xi(\lambda)}{1-\xi(\lambda)}\ln{\xi(\lambda)} \, ,
\end{align}
where 
\begin{equation}\label{xi}
\xi(\lambda) = \left[\frac{\left(\lambda+2\right)^{1/4}-\lambda^{1/4}}{\left(\lambda+2\right)^{1/4}+\lambda^{1/4}}\right]^2;~\lambda=\frac{m\omega^2}{\alpha^2}.
\end{equation}
[For completeness, we have provided the key steps of the derivation of $S(\lambda)$ in Appendix \eqref{wkb}.]

The ground state energy corresponding to the Hamiltonian \eqref{eq:CHO-Hamil02} is given by:
\begin{equation}
\mathscr{E}_0=\frac{\hbar\alpha}{2\sqrt{m}}\left(\sqrt{\lambda+2}+\sqrt{\lambda}\right)=\frac{\hbar\omega}{2}\left(1+\sqrt{1+\frac{2}{\lambda}}\right)
\end{equation}
Note that entanglement entropy depends only on $\lambda$, which is the ratio of frequency ($\omega$), 
mass ($m$), and the coupling strength ($\alpha$). However, $\mathscr{E}_0$ depends on $m$, $\alpha$ 
and $\lambda$ {(or on $\omega$ and $\lambda$)}. As we will show, this feature is present for the scalar fields in $(1 + 1)-$ dimension 
and $(3 + 1)-$dimensions space-times. 

We now want to identify the symmetries associated with entanglement entropy of this system. There may be multiple scaling transformations that keep $\lambda$ invariant, here we focus on one particular scaling transformation that has implications in scalar field theory:
\begin{equation}\label{trans}
\omega\to\eta\omega;\quad \alpha\to\eta\alpha
\end{equation}
where $\lambda$ is a constant. Under this transformation, we see the entropy \eqref{eq:CHO-EntS} is invariant, however the energy scales as $\mathscr{E}_0\to\eta \mathscr{E}_0$. Before we discuss more about this scaling symmetry, we obtain the same result by factorizing the Hamiltonian \eqref{eq:CHO-Hamil} as:
\begin{equation}
H=\frac{\alpha}{\sqrt{m}}\tilde{H}
\end{equation}
The Hamiltonian $\tilde{H}$ is necessarily the original Hamiltonian $H$ rescaled by a constant\footnote{$\tilde{H}$ and $H$ do not have the same dimensions. We will still refer to $\tilde{H}$ as Hamiltonian}. Under the canonical transformations 
\[
p_i=(\alpha^2m)^{1/4}\tilde{p}_i,~x_i=(\alpha^2m)^{-1/4}\tilde{x}_i \,~\mbox{where}~i = 1, 2 \, ,
\]
the above rescaled Hamiltonian ($\tilde{H}$) is characterized by a single parameter $\lambda$:
\begin{equation}\label{cho1}
\tilde{H}=\frac{1}{2}\left\{\tilde{p}_1^2+\tilde{p}_2^2+\lambda\left(\tilde{x}_1^2+\tilde{x}_2^2\right)+\left(\tilde{x}_1-\tilde{x}_2\right)^2\right\}.
\end{equation}
The normal modes of this system are:
\begin{equation}
\label{eq:H2-Normalmodes}
\tilde{\omega}_-=\sqrt{\lambda+2};\quad\tilde{\omega}_+=\sqrt{\lambda}.
\end{equation}
Note that this Hamiltonian ($\tilde{H}$) and its normal modes are invariant under the scaling transformations in (\ref{trans}). Furthermore, the entanglement entropy for $\tilde{H}$ is given by Eq.~(\ref{eq:CHO-EntS}). 
Let us now compare the results for systems $H$ and $\tilde{H}$:
\begin{equation}
\label{eq:SE-CHO}
S=\tilde{S}(\lambda);\quad \mathscr{E}_0=\frac{\alpha}{\sqrt{m}}\tilde{\mathscr{E}_0}(\lambda).
\end{equation}
From this, we can conclude that on rescaling the Hamiltonian by a constant, the entropy remains unchanged. In other words, the entanglement entropy has a scaling symmetry that the energy of the system does not have. This implies that entanglement entropy has an \emph{infinite degeneracy} w.r.t the transformations (\ref{trans}), and \emph{only} the energy of the system can break this infinite degeneracy.  More specifically, entanglement entropy can not distinguish these infinite possible physical systems; all systems having the same $\lambda$ can be grouped. The scaling transformations (\ref{trans}) can be used to generate an infinite number of systems belonging to this group.
%In fact, this implies that we cannot say anything about the energy of such oscillator systems if we were only to measure its entanglement entropy.

To tune ourselves to the UV-divergence in field theory, we consider the strong coupling limit $\alpha\to\infty$ of Hamiltonian \eqref{eq:CHO-Hamil}. For the rescaled Hamiltonian \eqref{cho1}, this corresponds to $\lambda \to 0$. As mentioned above, $\lambda = 0$ does not correspond to one specific system but a group of systems with different energies; however, all of them have the same entanglement entropy \eqref{eq:CHO-EntS}. In this limit, the normal modes of $\tilde{H}$ \eqref{eq:H2-Normalmodes} are $\tilde{\omega}_+=0$ and $\tilde{\omega}_-=\sqrt{2}$. Thus, one of the normal modes vanishes in this strong coupling limit, and the presence of a \emph{zero-mode} leads to the divergence of the entanglement entropy. In the following sections, we will show that this feature is present for massive scalar field in $(1 + 1)-$ dimension and $(3 + 1)-$dimensional space-times.

\begin{figure}[!hbt]
	\centering
	\includegraphics[scale=0.58]{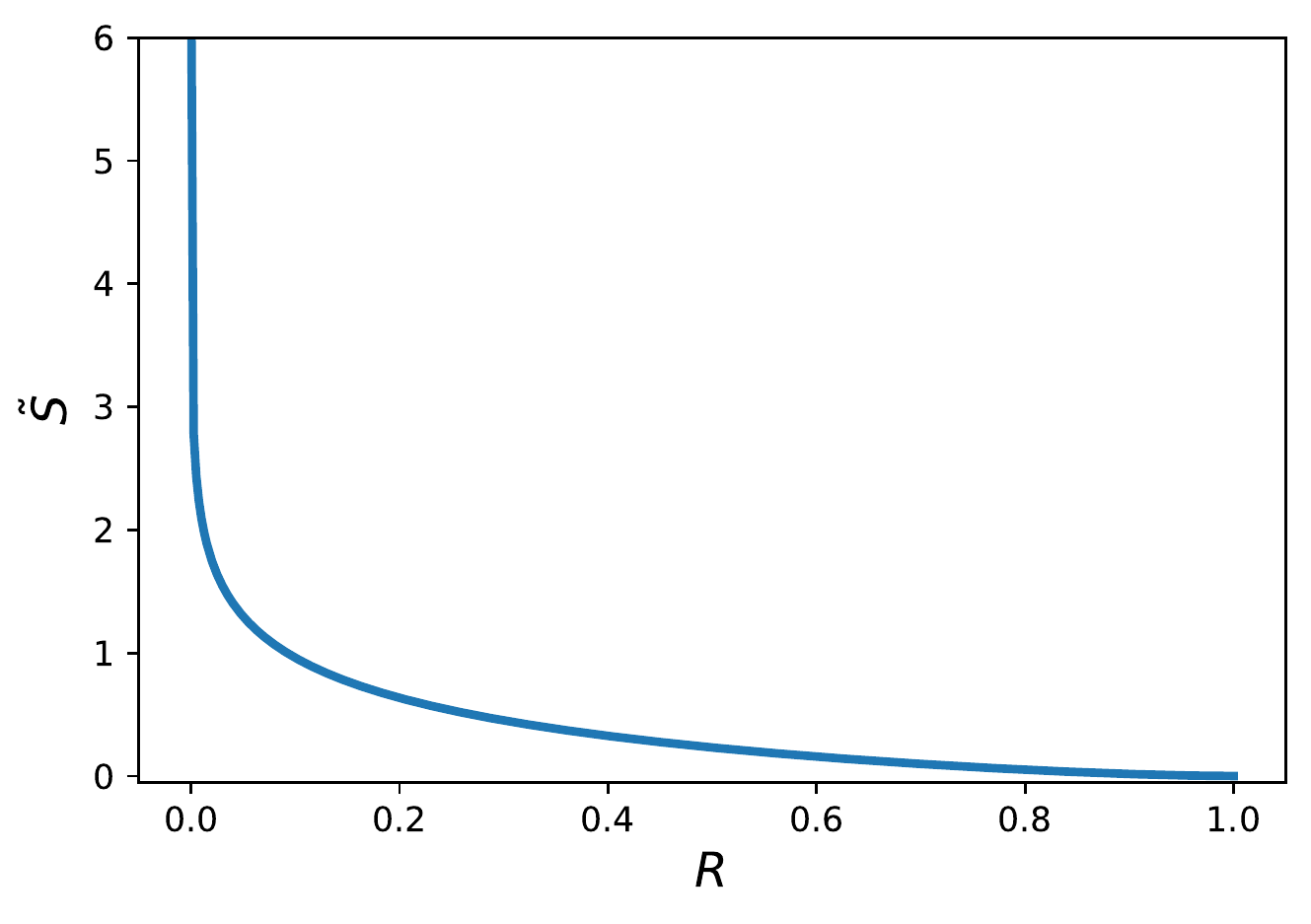}
	\caption{Entanglement entropy of the coupled harmonic oscillator. $R=0$ corresponds to zero mode limit $\lambda=0$ whereas $R=1$ is the decoupling limit $\lambda\to\infty$.}
	\label{cho1a}
\end{figure}

To capture the behavior of entanglement entropy and energy, it is better suited to define entropy in terms of $R=\omega_+/\omega_-$. In terms of $R$, we have:
\begin{equation}
\lambda=\frac{2R^2}{1-R^2};\qquad\xi=\left\{\frac{1-\sqrt{R}}{1+\sqrt{R}}\right\}^2
\end{equation}
\ref{cho1a} is the plot of entanglement entropy as a function of $R$ and shows that the divergence of the entanglement entropy which can be attributed to the presence of zero modes. 

Identifying the divergence of the entanglement entropy to zero-modes provides a way to understand the infinite degeneracy. To understand this, we now evaluate the ground state energy in the strong coupling limit and is given by:
\begin{equation}
\mathscr{E}_0=\mathscr{E}_++\frac{\hbar\alpha}{\sqrt{2m}},
\end{equation}
where $\mathscr{E}_+$ is the energy of the free-particle. Since the free-particle wave-function is non-normalizable, its energy is fixed by the initial condition. Hence, there is are infinite possible ways to fix the energy of the free particle. This corresponds to the infinite degenerate systems for a specific value of entanglement entropy. The current authors have shown the occurrence of zero modes for the coupled harmonic oscillator with negative coupling constant~\cite{2019-Chandran.Shankaranarayanan-Phys.Rev.D}. Here, we have shown the presence of zero modes in the strong coupling limit 
($\alpha \to \infty$). We like to stress that entropy divergence is a direct consequence of a free-particle being generated in the system.

The zero-mode analysis in Ref. \cite{2019-Chandran.Shankaranarayanan-Phys.Rev.D} can be extended to the strong coupling limit. Appendix  \eqref{app:zeromode} contains the WKB approximation of harmonic oscillator~\cite{2019-Chandran.Shankaranarayanan-Phys.Rev.D} and have shown that introducing IR cutoff in $\lambda$ will lead to finite entanglement entropy. 

This is the \emph{first key result} of this work and leads to interesting consequences that were overlooked in the literature. $\left\{\lambda=0\right\}$ correspond to two physically different limits:\footnote{It should be noted that the $m=0$ divergence is ignored. In subsequent sections, we use $m=1$ units.}
\begin{itemize}
	\item $\omega\to0$ : Entropy diverges, but the energy of the system is, in general, finite.
	\item $\alpha\to\infty$ : Both entropy and energy diverge. This corresponds to the UV-limit of the system.
\end{itemize}
As far as entanglement entropy is considered, both these limits are the same. The two sub-groups 
can \emph{only be} distinguished by their respective energies. As we explicitly show in the later sections,  such zero-mode divergences take center stage in the scalar field theory. 

As mentioned in the Introduction, the second aim of this work is to build a consistent structure of horizon thermodynamics from entanglement. In the next section,  we discuss the basic concepts of \emph{entanglement mechanics} applied to harmonic oscillator chains, which will then be extended to field theory.

\section{Entanglement mechanics in a Harmonic Chain}
\label{secmath}

The Hamiltonian corresponding to the harmonic chains is given by:
\begin{equation}
\label{eq:HC-Hamil}
H=\frac{1}{2}\left[\sum_i p_i^2+\sum_{ij}x_iK_{ij}x_j\right] \, .
\end{equation}
$K_{ij}$ is the coupling matrix, which contains all the relevant information about the interactions, and in which all information about entanglement entropy is encoded. The exact form of $K$ depends on the boundary conditions used. In the next section, we will discuss the effects of the boundary conditions on the entanglement entropy.   

\begin{figure*}[!ht]
	\begin{center}
		
			\subfloat[\label{cho2a}]{%
				\includegraphics[width=0.4\textwidth]{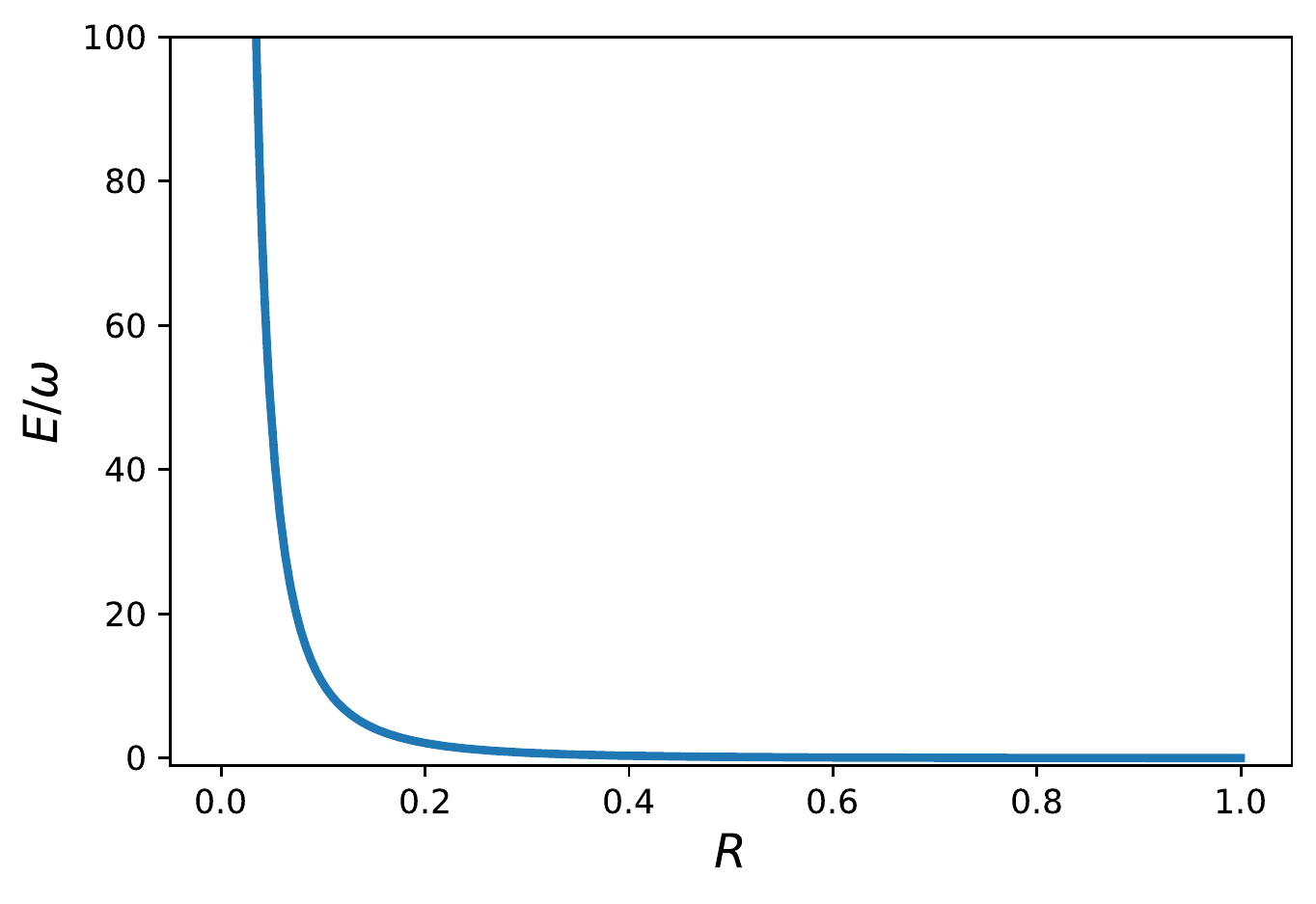}
			}
			\subfloat[\label{cho2b}]{%
				\includegraphics[width=0.4\textwidth]{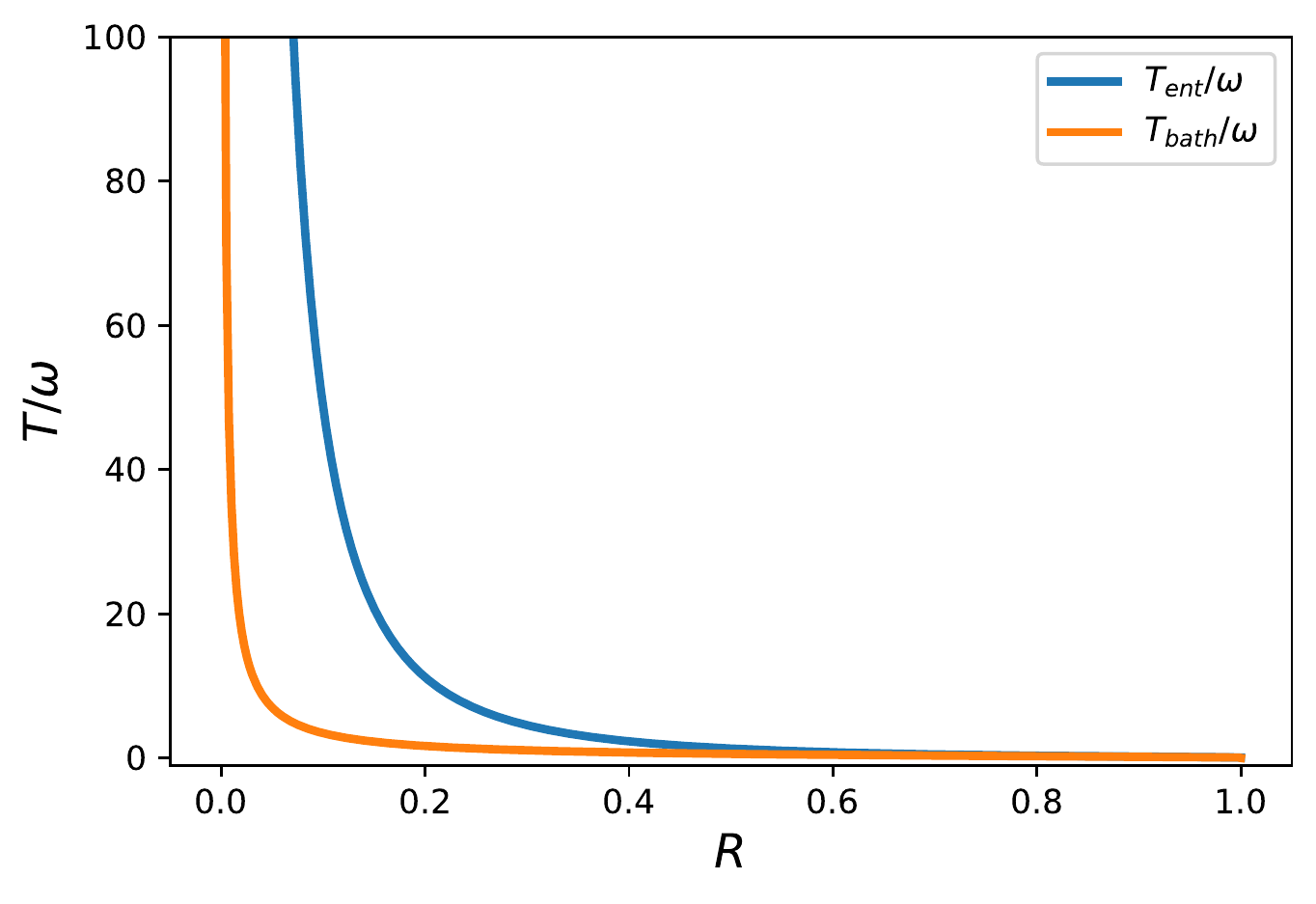}
			}\hfill
			
			\caption{Entanglement energy and entanglement temperature for a coupled harmonic oscillator.}
			\label{cho2}	
	\end{center}
\end{figure*}

While the von Neumann entropy is the most common entanglement measure, it is neither the most general nor unique. There are other measures, such as the R\'enyi and Tsallis entropies; however, 
all these are generalizations and, under certain limits, reduce to the von Neumann entropy. Thus, we take von Neumann entropy as a concrete definition for bipartite entanglement entropy in a quantum system. The algorithm to numerically calculate entanglement entropy from this coupling matrix is well known~\cite{1986-Bombelli.etal-Phys.Rev.D,1993-Srednicki-Phys.Rev.Lett.,Das2010}. For Completeness, we have provided the key steps of the derivation in Appendix \eqref{app:Ent-HC}. The entanglement entropy is given by:
\begin{equation}
\label{eq:Ent-HC}
S=\sum_iS(\xi_i);\quad \xi_i=\frac{\bar{\beta}_i}{1+\sqrt{1-\bar{\beta}_i^2}},
\end{equation}
where $\{\bar{\beta}_i\}$ are the eigenvalues of matrix $\bar{\beta}$ defined in \eqref{beta}, and $S(\xi_i)$ has the same form as in the coupled harmonic oscillator \eqref{eq:CHO-EntS}.

However, there is no concrete definition for \emph{entanglement energy}, as isolating subsystem energies may not be as trivial. Taking the CHO as an example, we see that the interaction energy $H_{int}$ is shared between the two oscillators; as we increase $N$, allocating the interaction energy among subsystems is highly non-trivial. Various definitions have been proposed to isolate subsystem energies in entangled systems \cite{2016-Alipour.etal-ScientificReports,2019-Najafi.Rajabpour-Phys.Rev.B,2017-Stokes.etal-JournalofModernOptics}. In this work, we follow the definition of \emph{entanglement energy}  proposed in Ref.~\cite{1998-Mukohyama.etal-Phys.Rev.D}:
\begin{equation}
\label{eq:def-EntE}
E_{\rm ent}=\epsilon\left\langle:H_{in}:\right\rangle.
\end{equation} 
Physically, the entanglement energy $E_{\rm ent}$ is the disturbed vacuum energy in the presence of a boundary. The above definition can be taken as a generalization of Casimir energy. Like in
Casimir effect, the entanglement energy of the field in vacuum, is altered by material (boundary) around it. However, unlike Casimir, we are interested in the energy change due to the lack of information from 
the complementary region. 

We have added a constant pre-factor $\epsilon$ since the above definition of entanglement energy is not unique. In Sec. \eqref{sec:Ent-BHconnection}, we will show that $\epsilon$ is fixed by matching the entanglement mechanics results with the well-known results from black-hole thermodynamics. 

To calculate $E_{\rm ent}$, we first write down the normal ordered Hamiltonian $: {H}_{in}:$ as follows:
\begin{equation}
:H_{in}:=-\frac{1}{2}\delta^{ab}\left(\frac{\partial }{\partial x^a}-\omega_{ac}^{in}x^c\right)\left(\frac{\partial }{\partial x^b}+\omega_{bd}^{in}x^d\right),
\end{equation}
where $\omega^{in}=K_{in}^{1/2}$. Transforming to a new basis $\{\bar{x}^A\}$,
\begin{align}
	\bar{x}^A&=\delta^{AB}(\Omega^{1/2})_{BC}x^C\nonumber\\
	U^{AB}&=\delta^{AC}(\Omega^{1/2})_{Ca}\delta^{ab}(\Omega^{1/2})_{bD}\delta^{DB}\nonumber\\
	\bar{\omega}_{AB}^{in}&=\delta_{AC}(\Omega^{-1/2})^{Ca}\omega_{ab}^{in}(\Omega^{-1/2})^{bD}\delta_{DB} \, ,
\end{align}
the normal ordered Hamiltonian becomes:
\begin{equation}
:H_{in}:=-\frac{1}{2}U^{ab}\left(\frac{\partial }{\partial \bar{x}^A}-\bar{\omega}_{AC}^{in}\bar{x}^C\right)\left(\frac{\partial }{\partial \bar{x}^B}+\bar{\omega}_{BD}^{in}\bar{x}^D\right)
\end{equation}
The entanglement energy reduces to:
\begin{align}\label{entenergy}
E_{\rm ent}&=\epsilon\int \prod_{A=1}^Nd\bar{x}^A\left\langle\{\bar{x}^B\} \middle|:H_{in}:\rho\middle|\{\bar{x}^C\}\right\rangle\nonumber\\
&=\frac{\epsilon}{4}\Tr\left[K_{in}\tilde{A}+A-2\omega^{in}\right]
\end{align}
To have a physical understanding of the entanglement energy and entanglement temperature, we 
consider CHO described by the rescaled Hamiltonian \eqref{cho1}. Thus, the entanglement energy is given by:
\begin{equation}
\tilde{E}=\frac{1}{8}\left[\left(\sqrt{\lambda}+\sqrt{\lambda+2}\right)\left\{1+\frac{\lambda+1}{\sqrt{\lambda(\lambda+2)}}\right\}-4\sqrt{\lambda+1}\right]
\end{equation} 
where we have set $\epsilon=1$. Hereafter, for easy reading, we drop the subscript ${\rm ent}$ in $E_{\rm ent}$. The entanglement energy for the original Hamiltonian $H$ \eqref{eq:CHO-Hamil} is:
\begin{equation}
E=\frac{\omega}{\sqrt{\lambda}}\tilde{E}.
\end{equation}
The left panel of \ref{cho2} is the plot of $E/\omega$ as a function of $R$. Like entropy $S$, the entanglement energy  diverges in the zero-mode limit ($R\to0$ for a fixed finite $\omega$) 
and vanishes in the decoupling limit ($R\to1$). 

In analogy with microcanonical ensemble picture of equilibrium statistical
mechanics, evaluation of the entanglement energy $E$, corresponds to different coupling constant $R$ (or $\lambda$) . In analogy we define \emph{entanglement temperature}~\cite{1989-Sakaguchi-PTP,2016-Kumar.Shankaranarayanan-EPJC}:
\begin{equation}
\label{eq:def-EntT}
T=\frac{dE}{dS}=\frac{\left(dE/d\lambda\right)}{\left(dS/d\lambda\right)}.
\end{equation}
In the case of CHO, we can map the reduced density matrix to a thermal density matrix~\cite{1993-Srednicki-Phys.Rev.Lett.}. The temperature corresponding to the thermal density matrix 
is given by:
\begin{equation}
 T_{\rm Bath}=\frac{\sqrt{\omega_+\omega_-}}{\ln{1/\xi}}=\frac{\omega}{2\sqrt{R}\ln{\frac{1+\sqrt{R}}{1-\sqrt{R}}}} \, ,
\end{equation}
where $\xi$ and $R$ are defined in Eq. \eqref{xi}. 
The right panel of \ref{cho2} is the plot of $T/\omega$ and $T_{\rm Bath}/\omega$ as a function of 
$R$. Like the entropy and energy, $T/\omega$ and $T_{\rm Bath}/\omega$ diverge in the 
zero-mode limit ($R\to0$ for a fixed finite $\omega$) and vanishes in the decoupling limit ($R\to1$). 
While both temperatures exhibit similar monotonic behavior, they do not coincide in the strong-coupling limit. This is due to arbitrariness in fixing the value of $\epsilon$ in Eq. \eqref{eq:def-EntE}. This will persist in field theory calculations, and, as mentioned earlier, 
we will fix the value of $\epsilon$ by matching the results with the results from the black-hole thermodynamics. 

From the above, we make the following important conclusions: First,  the entanglement temperature can be associated with the system's thermodynamic temperature. Second, the entanglement energy is a physically relevant quantity. In the subsequent sections, 
we will put the definitions of entanglement energy and entanglement temperature to test for the scalar fields propagating in space-times with the horizon.

\section{Massive Scalar Field in (1+1)-dimensions}
\label{sec1d}

The Hamiltonian of a massive scalar field in $(1+1)-$dimensions is given by:
\begin{eqnarray}
H=\frac{1}{2}\int dx \left[\pi^2+(\nabla\varphi)^2+m_f^2\varphi^2\right]
\end{eqnarray}
where $m_f$ is the mass of the scalar field. To evaluate the real-space entanglement entropy of the scalar field, we discretize the above Hamiltonian into a chain of harmonic oscillators by imposing a UV cut-off $a$ as well as an IR cutoff $L=(N+1)a$. On employing a mid-point discretization procedure, the resultant Hamiltonian takes the following form~\cite{Das2010}: 
\begin{equation}
\label{eq:1DHami-Mink}
H=\frac{1}{2a}\sum_j\left[\pi_j^2+\Lambda\varphi_j^2+(\varphi_j-\varphi_{j+1})^2\right]=\frac{1}{a}\tilde{H},
\end{equation}
where $\Lambda=a^2m_f^2$. From the definition of $\Lambda$, it is clear that $\Lambda$ is invariant under the scaling $(\eta)$ transformations:
\begin{equation}
a\to \eta a;\quad m_f\to\eta^{-1}m_f
\end{equation} 
Like in CHO, we have factorized the original Hamiltonian into a scale-dependent part ($1/a$) and a scale-independent part ($\tilde{H}$). Following the discussion in the previous sections, the entanglement entropy and entanglement energy corresponding to $H$ and $\tilde{H}$ are related by:
\begin{eqnarray}
S=\tilde{S}(\Lambda);\quad E=\frac{1}{a}\tilde{E}(\Lambda) \, .
\end{eqnarray}
As mentioned in the previous section, all information about the entanglement entropy is encoded in the coupling matrix $K$. The form of $K$ also depends on the boundary conditions. Periodic boundary conditions and the resulting zero mode divergence have already been studied in earlier works~\cite{2014-Mallayya.etal-Phys.Rev.D,2017-Yazdi-JournalofHighEnergyPhysics}. Keeping its generalization to spherical systems in mind, we will instead study the Dirichlet and Neumann boundary conditions:
\begin{itemize}
	\item \textbf{Dirichlet Condition} (DBC) : Here, we impose the condition $\varphi_0=\varphi_{N+1}=0$. The coupling matrix $K_{ij}$ becomes a symmetric Toeplitz matrix with the following non-zero elements:
	\begin{align}
	K_{jj}&=\Lambda+2\nonumber\\
	K_{j,j+1}=K_{j+1,j}&=-1
	\end{align}
	The normal modes are calculated to be~\cite{2008-Willms-SIAMJournalonMatrixAnalysisandApplications}:
	\begin{eqnarray}
	\tilde{\omega}_k^2=\Lambda+4\cos^2{\frac{k\pi}{2(N+1)}}\quad k=1,..N
	\end{eqnarray}
	We immediately see that the system does not develop any zero modes even when $\Lambda=0$ as long as $N$ is finite. In the thermodynamic limit ($N \to \infty$), the Dirichlet chain develops exactly one zero-mode ($\tilde{\omega}_N$) and a large number of near-zero modes.
	
	\item \textbf{Neumann Condition} (NBC): We impose the condition $\partial_x \varphi=0$ at the 
	two ends of the chain by setting $\varphi_0=\varphi_1$ and $\varphi_{N+1}=\varphi_N$. 
	The resultant coupling matrix is therefore a perturbed symmetric Toeplitz matrix whose non-zero elements are given below:
	\begin{align}
	 K_{jj\neq1,N}&=\Lambda+2\nonumber\\
	 K_{11}=K_{NN}&=\Lambda+1\nonumber\\
	 K_{j,j+1}=K_{j+1,j}&=-1
	\end{align}
	The normal modes for this system are found to be~\cite{2008-Willms-SIAMJournalonMatrixAnalysisandApplications}:
	\begin{eqnarray}
	\tilde{\omega}_k^2=\Lambda+4\cos^2{\frac{k\pi}{2N}};\quad k=1,...N
	\end{eqnarray}
	We see that the system develops exactly one zero mode ($\tilde{\omega}_N$) when $\Lambda=0$, even for a finite $N$. This is similar to periodic boundary conditions~\cite{2014-Mallayya.etal-Phys.Rev.D}. Since the entanglement entropy diverges, we must impose a small cutoff value for $\Lambda$ to extract the scaling behavior of entanglement.  Thus, Neumann  boundary condition can be used to study the effects of zero-modes on entanglement scaling. Note that the oscillator system described in Section \eqref{seccho} belongs to this category when $N=2$.
\end{itemize}

\begin{figure*}[!ht]
		\begin{center}
			\subfloat[\label{1dfin1a}][$\Lambda=10^{-6}$]{%
				\includegraphics[width=0.4\textwidth]{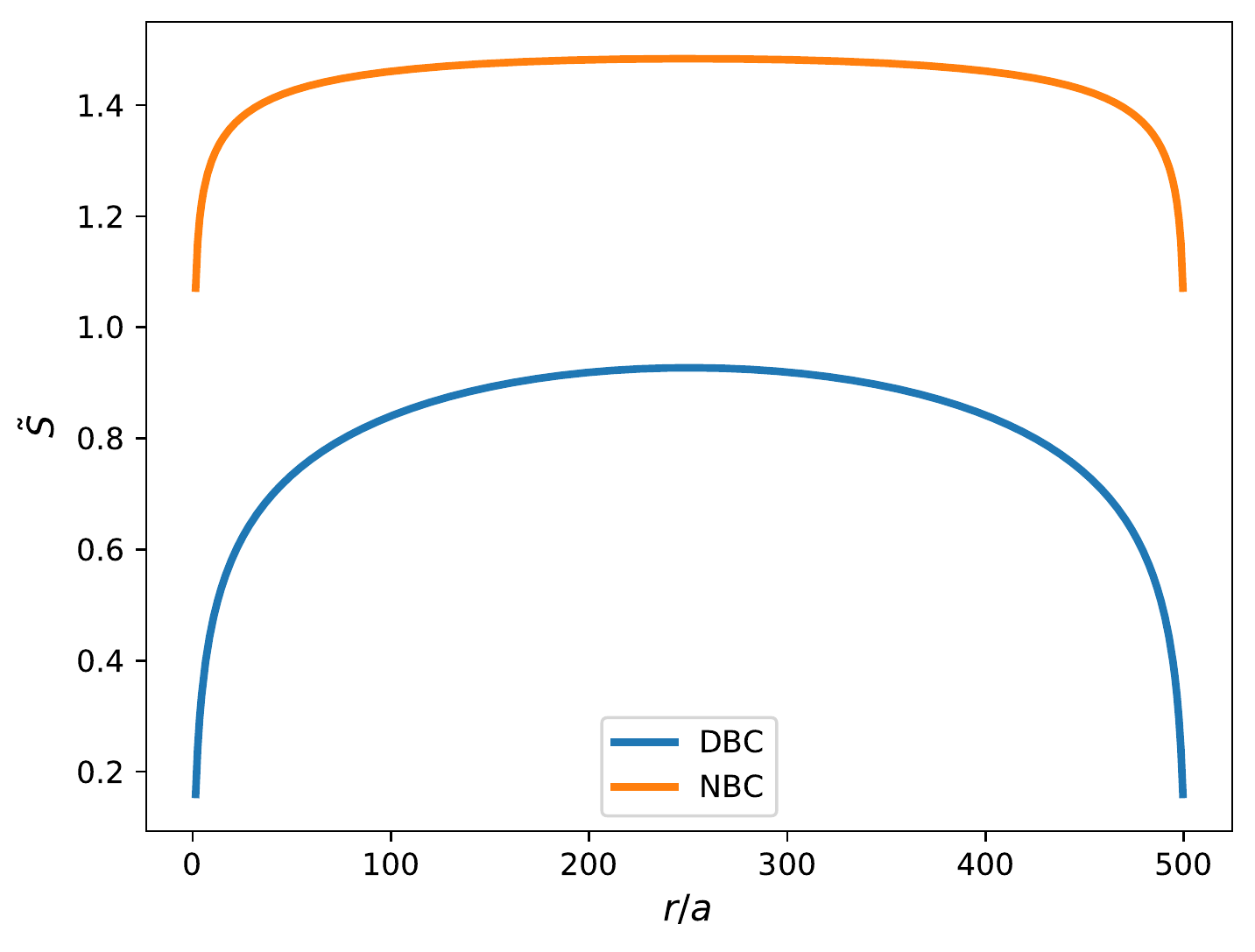}
			
			}
			\subfloat[\label{1dfin1b}][$\Lambda=10^{-6}$]{%
				\includegraphics[width=0.4\textwidth]{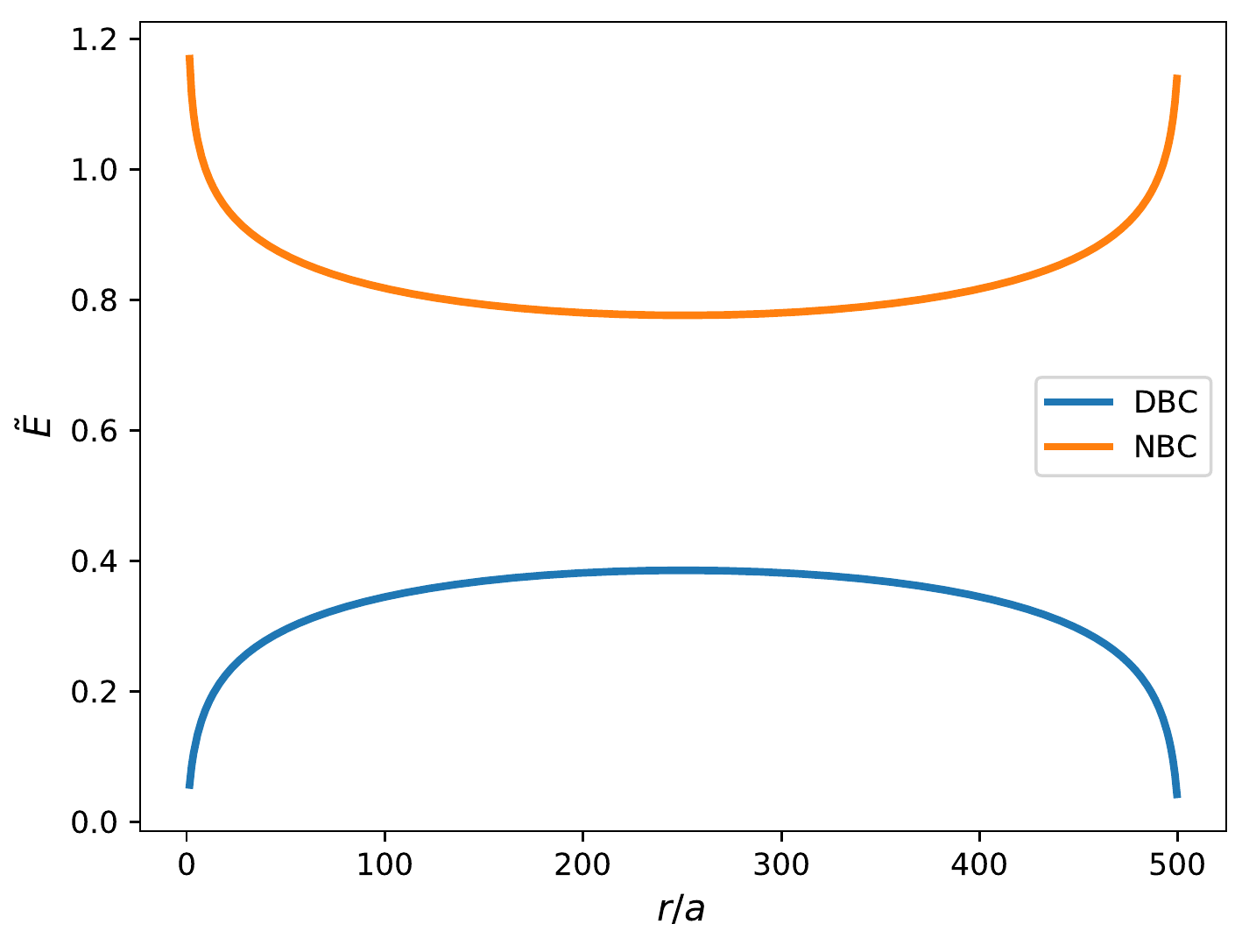}
			}\hfill		
			\subfloat[\label{1dfin1c}][$\Lambda=1$]{%
				\includegraphics[width=0.4\textwidth]{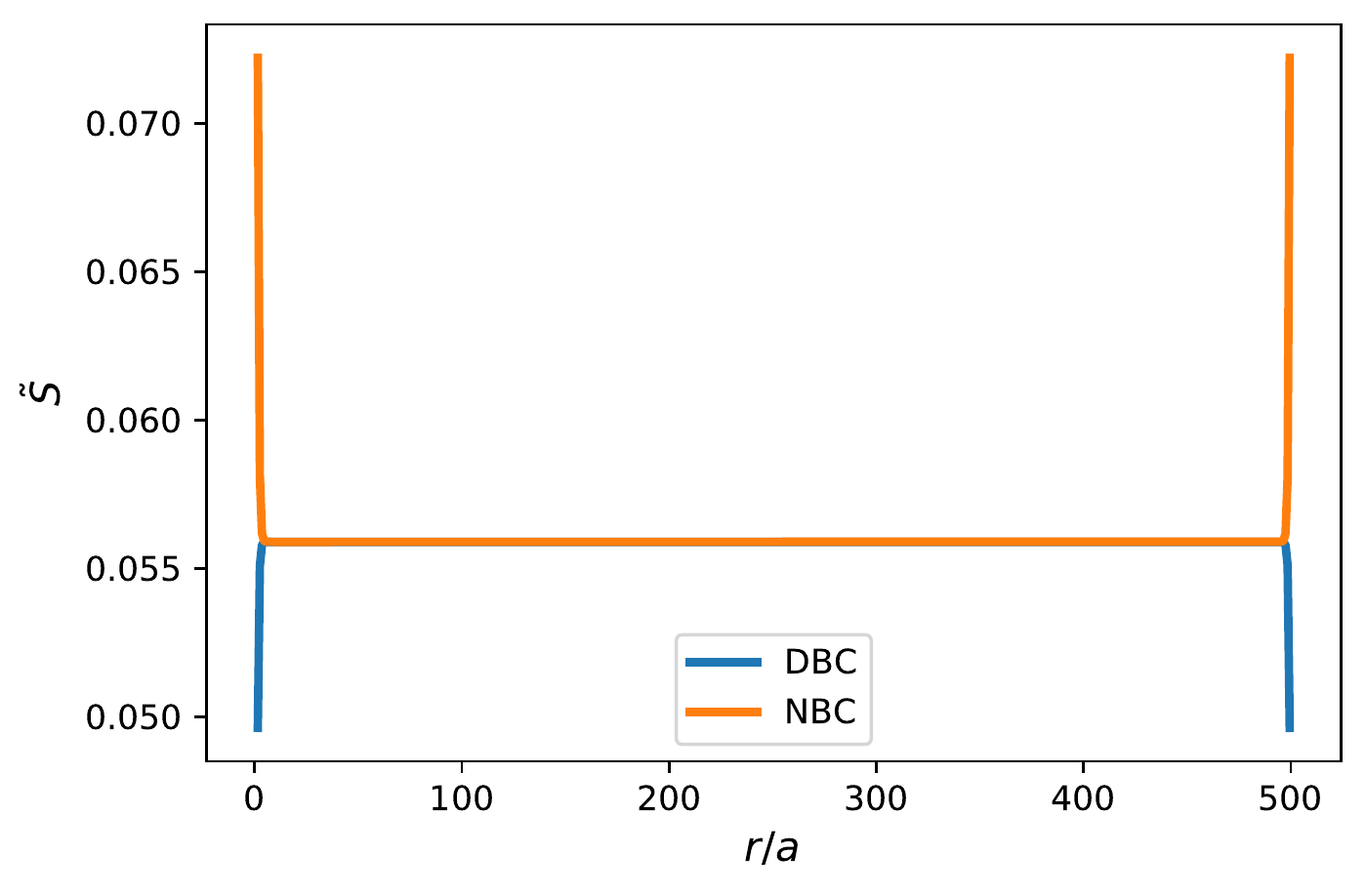}
			}			
			\subfloat[\label{1dfin1d}][$\Lambda=1$]{%
				\includegraphics[width=0.4\textwidth]{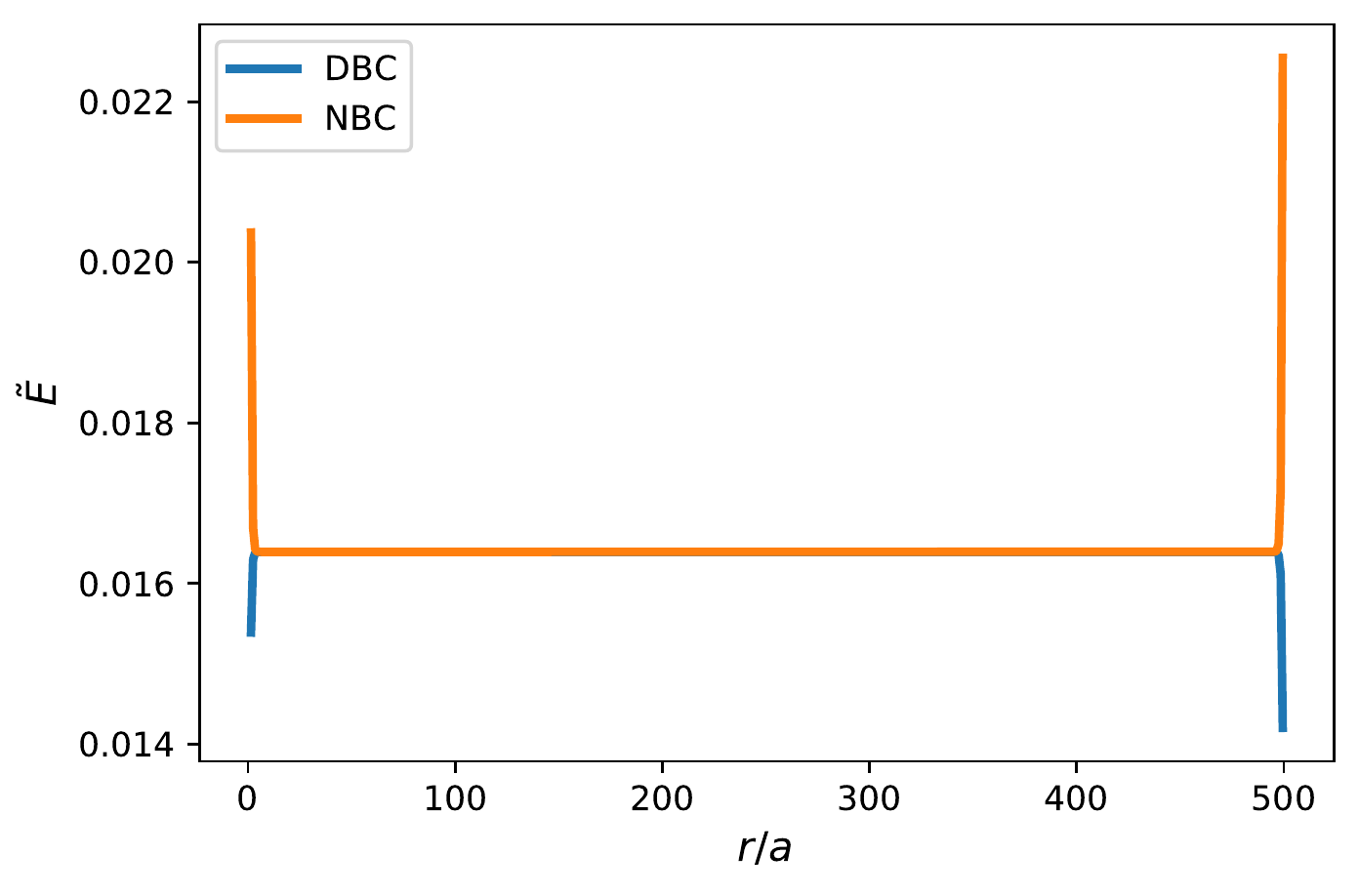}
			}\hfill
			
			\caption{Subsystem scaling of entanglement entropy and entanglement energy for $N=500$.}
			\label{1dfin1}
			\end{center}
		\end{figure*}

Choosing a boundary condition fixes the elements of the coupling matrix. To evaluate the entanglement entropy and entanglement energy, we follow the procedure in Section \eqref{secmath}.  All numerical computations have been carried out in Python 3.0, with the standard double precision ($10^{-16}$) of the NumPy package, thereby fixing the lower bound on the mass regulator as $\Lambda \sim 10^{-15}$ in NBC. 

\ref{1dfin1} contains the plot of entanglement entropy and entanglement energy for two different values of $\Lambda$. From \ref{1dfin1}, we deduce the following:
First, we observe a violation of ``area law" close to the critical point $\Lambda=0$, where near-zero modes dominate. [We will discuss more on this at the end of this section.]
In this regime, the results for the chains considered differ significantly --- the entropy for NBC diverges in this limit while that of DBC is finite.  Second, the entropy scaling for $\Lambda\to0$ is found to satisfy the following asymptotic behavior~\cite{1994-Holzhey.etal-NuclearPhysicsB,2004-Calabrese.Cardy-JournalofStatisticalMechanicsTheoryandExperiment,2009-Calabrese.Cardy-JournalofPhysicsAMathematicalandTheoretical,2006-Ryu.Takayanagi-Phys.Rev.Lett.}:
\begin{equation}\label{scaling1d}
S\sim c_s\log{\left\{\frac{L}{a\pi}\sin{\left(\frac{\pi r}{L}\right)}\right\}}+d_s,
\end{equation}
where $c_s$ is related to central charge and $d_s$ captures the subleading corrections. 

In the case of DBC chain for $N=500$ oscillators, $n = 200$ and $\Lambda=0$, we numerically obtain 
$c_s\sim 0.1664$ and $d_s = 0.0868$. The value obtained is in close agreement ($c_s=1/6$) with that obtained in the conformal limit of free Boson theories~\cite{2009-Calabrese.Cardy-JournalofPhysicsAMathematicalandTheoretical}. 
Thus, the finite chain with DBC exactly captures the conformal limit scaling and provides the correct value of the entanglement entropy.

Third, from Eq. \eqref{eq:def-EntE}, the entanglement energy (for $\epsilon = 1$) turns out to be:
\begin{equation}
\tilde{E}=aE\sim c_e\log{\left\{\frac{L}{a\pi}\sin{\left(\frac{\pi r}{L}\right)}\right\}}+d_e,
\end{equation}
where we numerically obtain $c_e\sim0.0757$ and $d_e\sim0.0026$ for $\Lambda=0$. It is interesting to note that entanglement energy scales similar to entropy for $\Lambda = 0$. 

Fourth, in the case of NBC, to extract the non-divergent part we need to consider a small cut-off value for $\Lambda$. For $\Lambda\sim 10^{-15}$, numerical fitting gives $c_s\sim0.1719$, $d_s\sim5.771$, $c_e\sim-0.0809$ and $d_e\sim1.67\cross10^{4}$. Clearly, there are spurious effects coming from zero modes in the NBC chain that cause the entanglement energy to decrease with subsystem size, and also prevent the intercepts from being sub-dominant.

Fifth, the entanglement structure of both DBC and NBC chains converge far away from the critical limit ($\Lambda>0$), where it follows an area-law:
\begin{equation}
\label{eq:SE-1DScalar}
S=c_s;\quad E=\frac{c_e}{a}
\end{equation}
This confirms that entanglement entropy and energy have the same scaling relations in Minkowski space-time~\cite{1997-Mukohyama.etal-Phys.Rev.D}. It is important to note that this holds even in the zero-mode regime where the area law is violated in $(1+1)$-dimensions. 
Since, the entanglement energy scales similar to entanglement entropy in $\Lambda \to 0$ and $\Lambda > 0$, we obtain that the entanglement temperature \eqref{eq:def-EntT} to be a constant 
fixed by the UV cutoff $a$: 
\begin{equation}
\label{eq:Temp-1DScalar}
T=\frac{c_e}{ac_s}
\end{equation}

Lastly, like in CHO, we see that  $\{\Lambda=0\}$ correspond to two physically different limits: 
\begin{itemize}
	\item \textbf{Massless limit} $m_f\to0$: For the NBC chain, there is a zero-mode divergence of entanglement entropy and energy, whereas they are finite for the DBC chain (since there are no zero modes).
	\item \textbf{Continuum limit} $a\to0$: From the definition of IR cutoff $L$, it is clear that in the continuum limit, $N$ should diverge at least as fast as $a^{-1}$. This implies that the DBC chain also develops a zero mode in the continuum limit. As a result, entanglement entropy and energy diverge for both DBC and NBC chains.
\end{itemize}

This is the \emph{second key result} of this work and leads to the following conclusions:
\begin{enumerate}
\item As $N\to\infty$, the massless limit and the continuum limit despite being physically different, cannot be distinguished as far as entanglement entropy/energy divergence is concerned. 

\item  Any occurrence of divergence in DBC and NBC chains is always associated with the zero-modes. This further cements the inescapable connection between zero-modes and UV divergence in the discretized approach to field theory.
\end{enumerate}

\section{Massive Scalar Field in $(3+1)-$ dimensions}
\label{sec3d}

\begin{figure*}[!ht]
		\begin{center}
			\subfloat[\label{3ds}][$\Lambda=10^{-10}$]{%
				\includegraphics[width=0.4\textwidth]{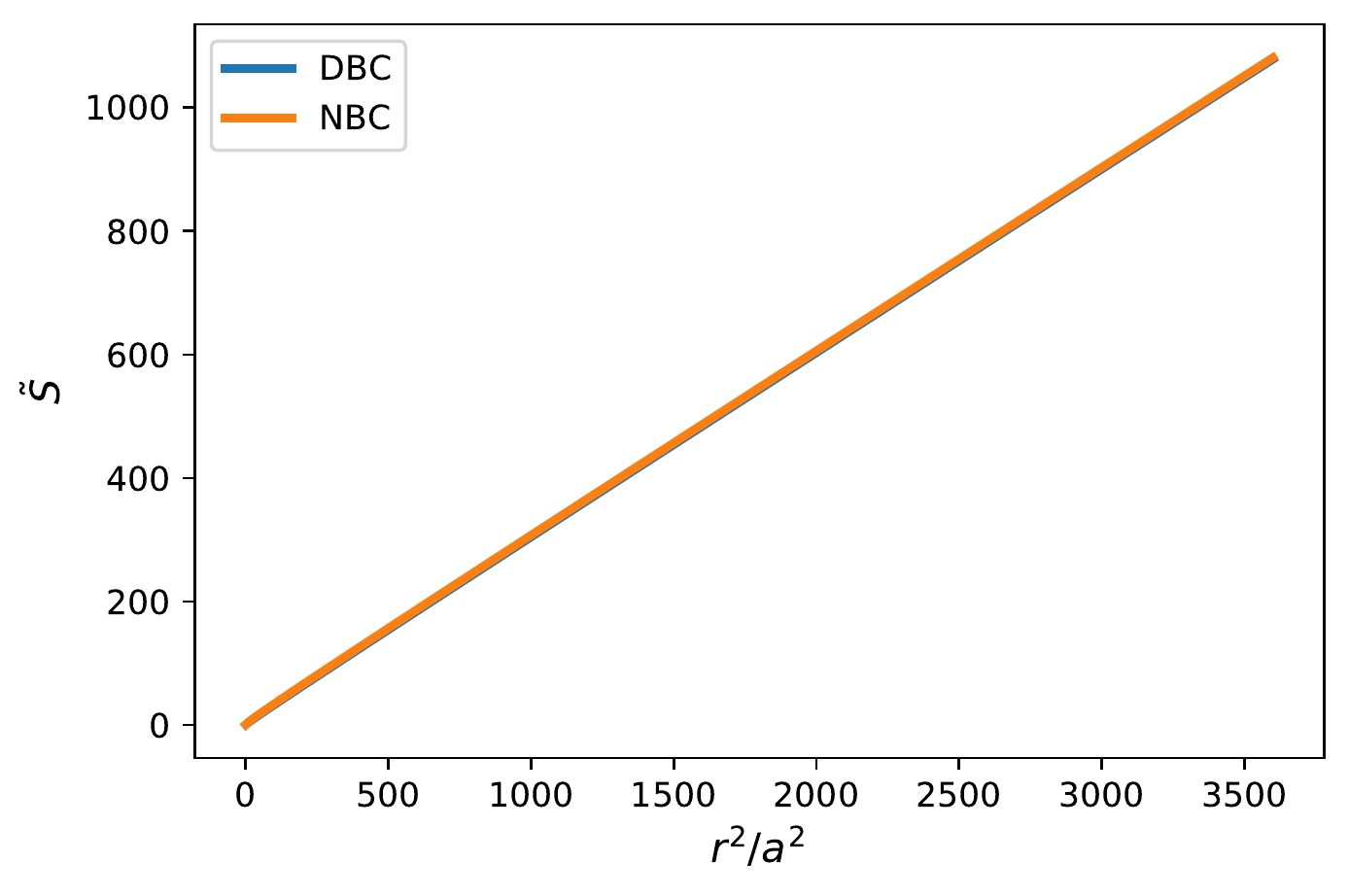}
			}
			\subfloat[\label{3de}][$\Lambda=10^{-10}$]{%
				\includegraphics[width=0.4\textwidth]{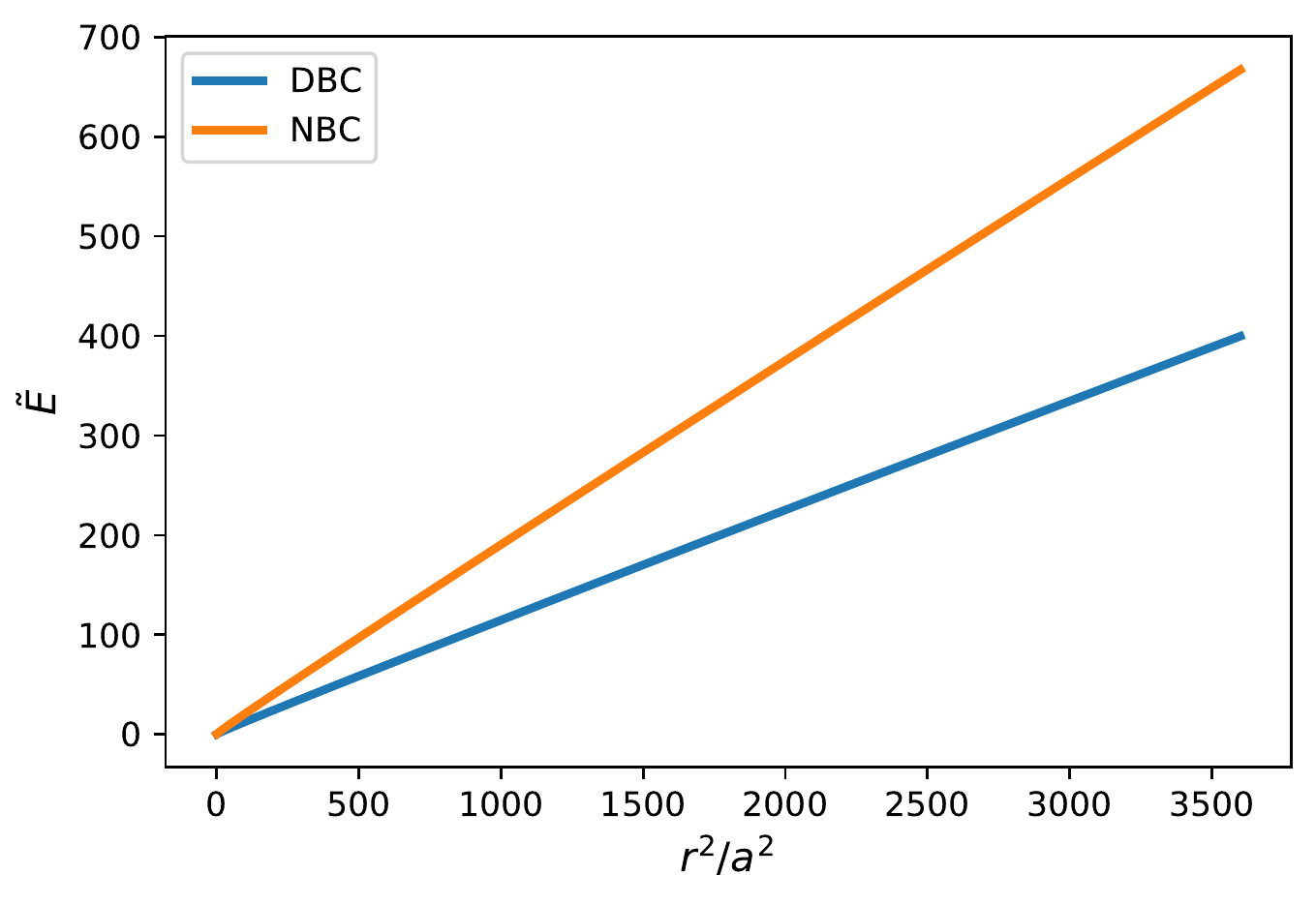}
			}\hfill
			\subfloat[\label{3ds}][$\Lambda=1$]{%
				\includegraphics[width=0.4\textwidth]{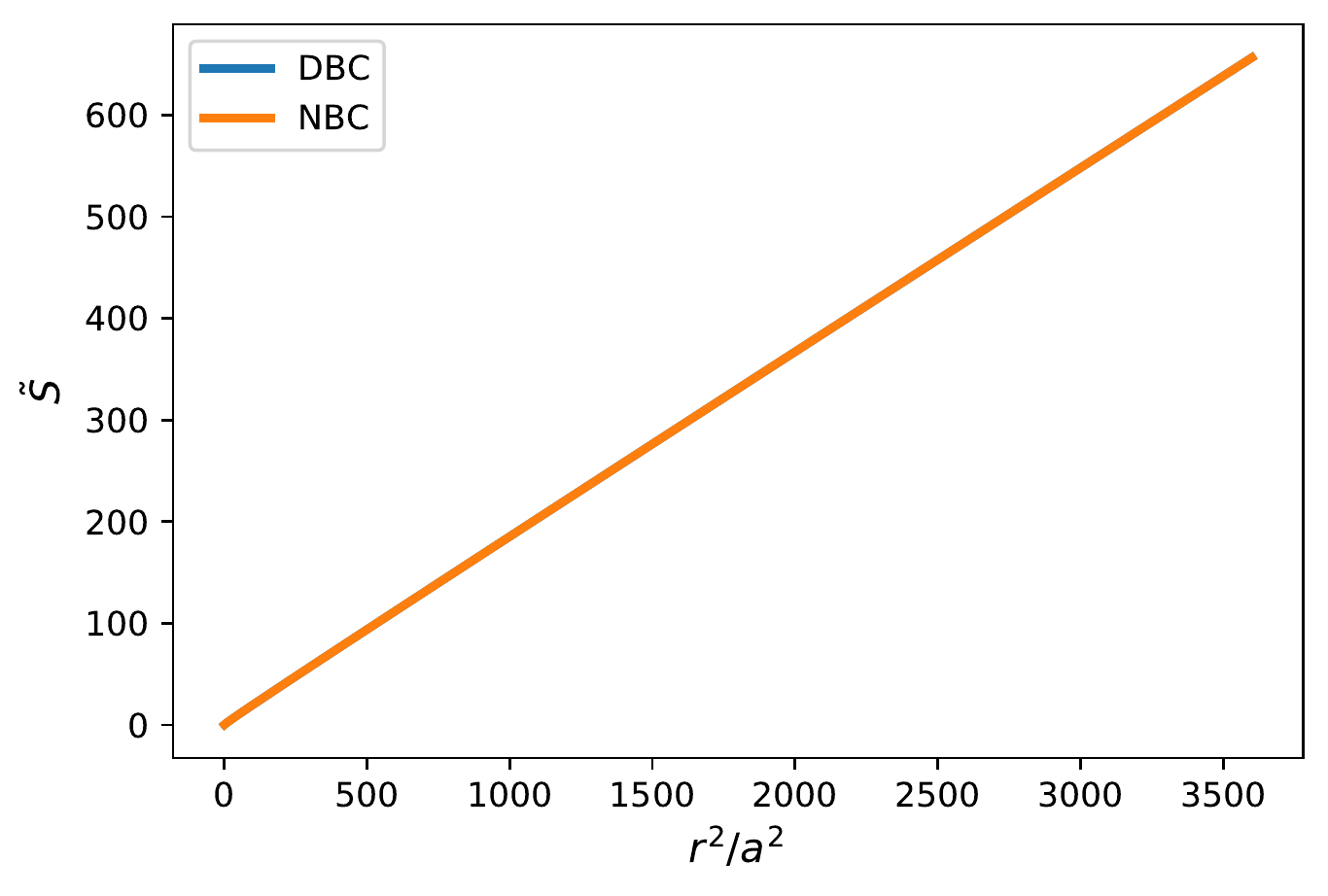}
			}
			\subfloat[\label{3de}][$\Lambda=1$]{%
				\includegraphics[width=0.4\textwidth]{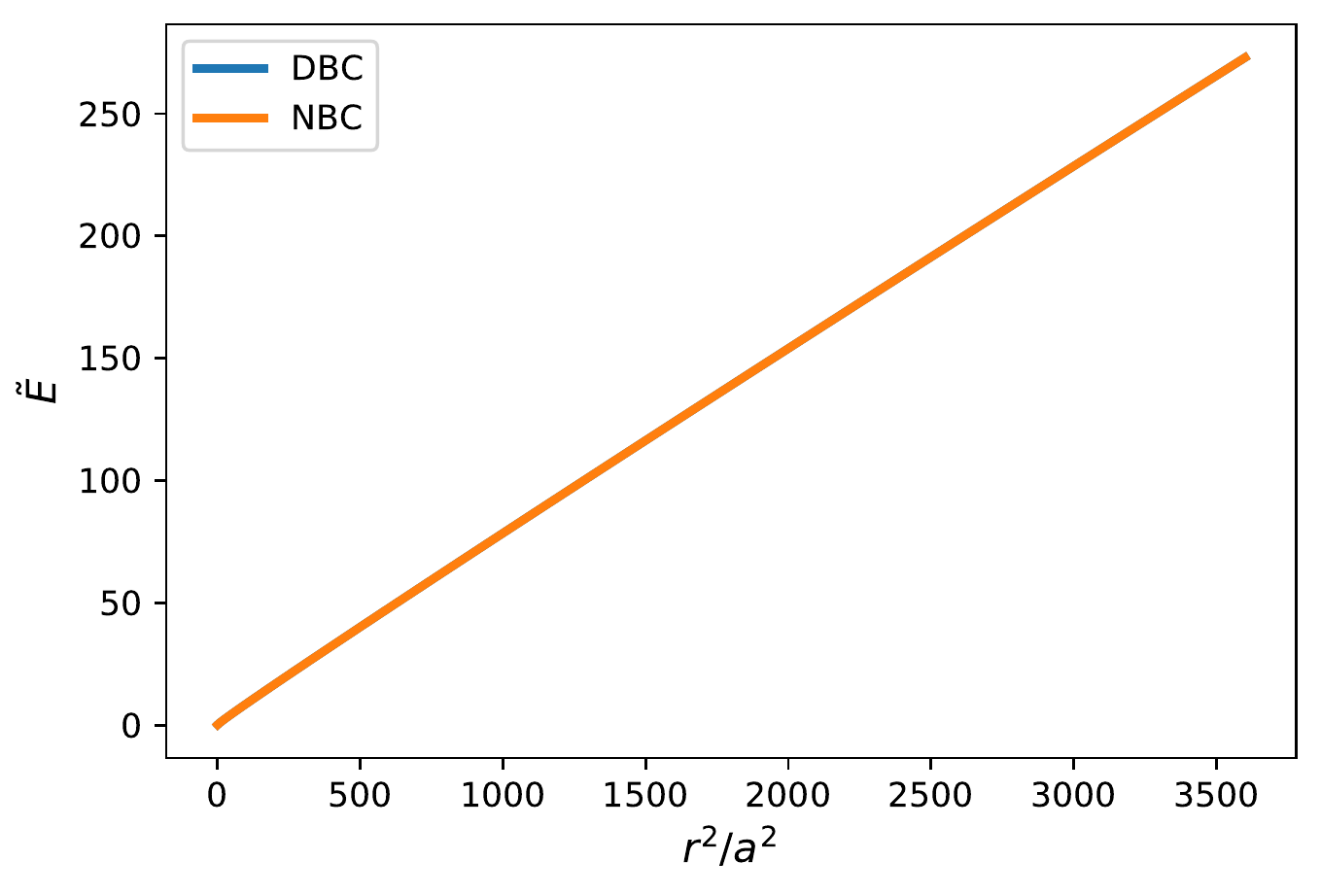}
			}
			
			\caption{Subsystem scaling of entanglement entropy and entanglement energy for $N=100$..}
			\label{3dse}
			\end{center}
		\end{figure*}

In this section, we extend the analysis to ($3+1$)-dimension flat space-time 
to understand the connection between scale-invariance of entanglement entropy and zero-mode divergence. The Hamiltonian of a massive scalar field is given below:
\begin{eqnarray}
H=\frac{1}{2}\int d^3x \left[\pi^2+(\nabla\varphi)^2+m_f^2\varphi^2\right]
\end{eqnarray}
In spherical coordinates, we make use of the partial wave expansion of scalar field:
\begin{align}
\pi(r,\theta,\phi)&=\frac{1}{r}\sum_{lm}\pi_{lm}(r)Z_{lm}(\theta,\phi)\\
\varphi(r,\theta,\phi)&=\frac{1}{r}\sum_{lm}\varphi_{lm}(r)Z_{lm}(\theta,\phi)\\
Z_{lm}(\theta,\phi)&=\begin{cases}Y_{lm}&m=0\\\sqrt{2}\Re{Y_{lm}}&m>0\\\sqrt{2}\Im{Y_{lm}}&m<0\end{cases}
\end{align}
The above expansion reduces the system into an effective ($1+1$)-dimensional Hamiltonian. On further discretizing the Hamiltonian, we get the following form:
\begin{multline}\label{eq:3dHami-Mink}
H=\frac{1}{2a}\sum_{lmj}\bigg[\pi_{lm,j}^2+\left\{\Lambda+\frac{l(l+1)}{j^2}\right\}\varphi_{lm,j}^2\\+\left(j+\frac{1}{2}\right)^2\left\{\frac{\varphi_{lm,j}}{j}-\frac{\varphi_{lm,j+1}}{j+1}\right\}^2\bigg],
\end{multline}
where $\Lambda=a^2m_f^2$ (as used in the previous section). Similar to the $(1+1)$-D case, the above Hamiltonian has factorized into a scale-dependent part $1/a$ and a scale-independent part $\tilde{H}$ which is invariant under the transformations $a\to\eta a$ and $m_f\to\eta^{-1}m_f$.

Since the angular momentum modes are uncoupled, the contributions to entanglement entropy and energy (for the scale-independent Hamiltonian $\tilde{H}$) can be collected as 
\[
\tilde{S}=\sum_l(2l+1)\tilde{S}_l \, \mbox{and} \, \tilde{E}=\sum_l(2l+1)\tilde{E}_l \, .
\]
Since both $\tilde{S}_l$ and $\tilde{E}_l$ converge as $l\to\infty$~\cite{1993-Srednicki-Phys.Rev.Lett.,1998-Mukohyama.etal-Phys.Rev.D}, we have employed an adaptive cut-off in the numerical calculations as and when the following condition is satisfied:
\begin{equation}
\max{\left\{\frac{\delta\tilde{S}}{\tilde{S}},\frac{\delta \tilde{E}}{\tilde{E}}\right\}}<10^{-5}
\end{equation}

As in the previous section, the coupling matrix is fixed depending on the boundary conditions. 
Like in Sec. \eqref{sec1d}, we consider the following two boundary conditions:
\begin{itemize}
	\item \textbf{Dirichlet condition} (DBC) : We impose this condition by setting 
	$\varphi_{lm,N+1}=0$. The non-zero elements of the coupling matrix are:
	\begin{align}
	K_{11}&=\Lambda+l(l+1)+\frac{9}{4}\nonumber\\
	K_{jj\neq1}&=\Lambda+\frac{l(l+1)+1/2}{j^2}+2\nonumber\\
	K_{j,j+1}&=K_{j+1,j}=-\frac{(j+1/2)^2}{j(j+1)}
	\end{align} 
	\item \textbf{Neumann condition} (NBC): In $(3+1)$-dimensions, after spherical expansion, $\partial_r\varphi=0$ corresponds to $\partial_r(\varphi_{lm}/r)=0$. This is equivalent to setting $N\varphi_{lm,N+1}=(N+1)\varphi_{lm,N}$ in \eqref{eq:3dHami-Mink} . The non-zero elements of the coupling matrix are therefore:
	\begin{align}
	K_{11}&=\Lambda+l(l+1)+\frac{9}{4}\nonumber\\
	K_{NN}&=\Lambda+\frac{l(l+1)}{N^2}+\left(1-\frac{1}{2N}\right)^2\nonumber\\
	K_{jj\neq1,N}&=\Lambda+\frac{l(l+1)+1/2}{j^2}+2\nonumber\\
	K_{j,j+1}&=K_{j+1,j}=-\frac{(j+1/2)^2}{j(j+1)}
	\end{align} 	
\end{itemize}

The coupling matrices for both the boundary conditions are not symmetric Toeplitz matrix, and hence, the eigenvalues can not be evaluated analytically.  However, for small values of $N$, it is easy to verify that DBC does not generate zero modes, while NBC does. In the thermodynamical limit ($N\to\infty$), DBC is expected to develop near-zero modes in the system. One key difference of the entropy scaling in higher dimensions from that in $(1+1)$-dimensions is that the area-law is obeyed for massive and massless cases. Therefore, $\Lambda\to0$ does not lead to any critical scaling behavior, unlike in $(1+1)$-dimensions.
	
\ref{3dse} contains the plot of entanglement entropy and energy for two different values of $\Lambda$.
we infer the following: Both entanglement entropy and energy obey area law in the case of a massive scalar field in ($3+1$)-dimensions, irrespective of the value of $\Lambda$:
\begin{equation}
\label{eq:SE-Scalar3dMink01}
\tilde{S}\sim c_s\frac{r^2}{a^2};\quad \tilde{E}\sim c_e\frac{r^2}{a^2}.
\end{equation}
For DBC chain with $\Lambda=0$, the numerically fixed pre-factors are  
$c_s\sim0.29$ and $c_e\sim0.11$. For the original Hamiltonian \eqref{eq:3dHami-Mink}, we have:
\begin{equation}
\label{eq:SE-Scalar3dMink02}
S\sim c_s\frac{r^2}{a^2};\quad E\sim c_e\frac{r^2}{a^3};\quad T=\frac{c_e}{ac_s}
\end{equation}
Like the $(1+1)$-D case, we notice that both entanglement entropy and energy scale similarly ~\cite{1997-Mukohyama.etal-Phys.Rev.D}, and therefore the temperature is a constant value fixed by the UV cut-off. 

As we approach the zero-mode limit, the entropy for NBC and DBC coincide at least down to $\Lambda=10^{-10}$. The value of $\Lambda$ at which the zero-mode effects on entropy become noticeable was found to be much lower ($<10^{-15}$), wherein the area-law pre-factor for NBC starts increasing drastically. For entanglement energy in NBC, this drastic increase is visible much earlier when $\Lambda\sim10^{-10}$. 

From the above results, we make the following conclusions: \\ 
(i) Unlike in $(1 + 1)-$dimensions, zero modes do not affect the area-law behavior of both entanglement entropy as well as energy. However, the zero-modes change the pre-factors drastically. As can be seen from Eq. \eqref{eq:SE-Scalar3dMink02}, entanglement energy is far more sensitive to zero modes than entropy in higher dimensions. \\
(ii) Like in $(1 + 1)-$dimensions, for both boundary conditions, any occurrence of entropy divergence is always associated with the development of zero modes in the system. To our knowledge, this is the first time a connection between the zero modes and divergence in entropy has been established in higher dimensions.
	
\section{Scalar Field in Spherically  Symmetric Space-times : Horizon Thermodynamics}\label{secgr}

In this section, we probe the entanglement properties of scalar fields in static, spherically symmetric space-times whose line element is of the form:
\begin{equation}
\label{eq:SpherLineele01}
ds^2=-f(r)dt^2+\frac{1}{f(r)}dr^2+r^2d\Omega^2.
\end{equation}
Depending on the form of $f(r)$, there is usually a handful of coordinate settings that help us understand the system better. Suppose the space-time in question has a horizon ($r_h$), such as in the Schwarzschild case, it is useful to rewrite the metric in terms of proper-length coordinates~\cite{1998-Mukohyama.etal-Phys.Rev.D}:
\begin{equation}
\label{eq:SpherLineele02}
ds^2=-f(r)dt^2+d\rho^2+r^2d\Omega^2,
\end{equation}
where,
\begin{equation}\label{prop}
\rho=\int_{r_h}^r \frac{dr}{\sqrt{f(r)}}.
\end{equation}
We see that the proper length only captures the region on that side of the horizon where $f(r)>0$. Therefore, to study entanglement properties, we bipartite this region only. In this coordinate, the physics in the region $f(r) < 0$ will remain inaccessible. 

The action for the massive scalar field in arbitrary space-time is~\cite{1982-Birrell.Davies-QuantumFieldsCurved,Jacobson2005}:
\begin{eqnarray}\label{ac1}
S=\frac{1}{2}\int d^4x\sqrt{-g}\left[g^{\mu\nu}\partial_{\mu}\varphi\partial_{\nu}\varphi-m_f^2\varphi^2\right].
\end{eqnarray}
For the proper-length coordinate \eqref{eq:SpherLineele02}, we use the following spherical decomposition of the scalar field with appropriate scaling:
\begin{align}
	\dot{\varphi}(\rho,\theta,\phi)&=\frac{f^{1/4}(r)}{r}\sum_{lm}\dot{\varphi}_{lm}(\rho)Z_{lm}(\theta,\phi)\\
	\varphi(\rho,\theta,\phi)&=\frac{f^{1/4}(r)}{r}\sum_{lm}\varphi_{lm}(\rho)Z_{lm}(\theta,\phi).
\end{align}
Substituting these in the action \eqref{ac1}, leads to the following effective $(1+1)$-D Lagrangian:
\begin{small}
	\begin{multline}\label{spherlag}
		L=\frac{1}{2}\sum_{lm}\int d\rho\Bigg[\dot{\varphi}_{lm}^2-r^2\sqrt{f(r)}\bigg\{\partial_{\rho}\bigg(f^{1/4}(r)\frac{\varphi_{lm}}{r}\bigg)\bigg\}^2\\-f(r)\left\{m_f^2+\frac{l(l+1)}{r^2}\right\}\varphi_{lm}^2\Bigg]
\end{multline}\end{small}
With the help of canonical conjugate momenta $\pi_{lm}=\dot{\varphi}_{lm}$, we can directly write down the Hamiltonian of the system. In order to regularize this Hamiltonian, we introduce lattice spacing $a$ in the proper length co-ordinate as $\rho=ja$. The IR cut-off here is on the proper length, which is fixed to be $\rho_L=(N+1)a$. For each lattice point $j$, we obtain the corresponding lattice point in rescaled radial co-ordinate $r'=r/a$, by inverting the following expression for $r_j$:
	\begin{equation}\label{disc}
	j=\int_{\Delta_h}^{r_j} \frac{dr'}{\sqrt{f(r')}}.
	\end{equation}
	where we have introduced the dimensionless parameters $\Delta_h=r_h/a$ and $r_j=r/a|_{\rho=ja}$. For convenience, we further define $f_j=f(r)|_{\rho=ja}$. It should be noted that the lattice points in radial co-ordinate $\{r_j\}$ are not equally spaced. We will eventually elaborate on the shorthand dimensionless expression \eqref{disc} for each space-time considered in this section. On employing the midpoint discretization scheme~\cite{Das2010}, we obtain a fully regularized Hamiltonian:
\begin{widetext}
	
		\begin{equation}\label{bh1}
		H=\frac{1}{2a}\sum_{lmj}\left[\pi_{lm,j}^2+r_{j+\frac{1}{2}}^2f_{j+\frac{1}{2}}^{1/2}\bigg\{f_j^{1/4}\frac{\varphi_{lm,j}}{r_j}-f_{j+1}^{1/4}\frac{\varphi_{lm,j+1}}{r_{j+1}}\bigg\}^2+f_j\left\{\Lambda^2+\frac{l(l+1)}{r_j^2}\right\}\varphi_{lm,j}^2\right],
		\end{equation}
\end{widetext}
where $\Lambda=a^2m_f^2$ (as used in the previous sections). Let us factorize the Hamiltonian as $H=\tilde{H}/a$. Contrary to earlier results, we now see that $\tilde{H}$ is not invariant under the scaling transformations $a\to\eta a$ and $m_f\to\eta^{-1}m_f$. This is because the Hamiltonian $\tilde{H}$ also features an additional parameter $\Delta_h$ (in the case of a space-times described by a single parameter), which also depends on $a$. Taking this into account, we consider the following scaling transformations:
\begin{equation}\label{bhscale}
a\to\eta a;\quad m_f\to\eta^{-1}m_f;\quad r_h\to \eta r_h
\end{equation}
Under these transformations, the parameters $\Lambda$ and $\Delta_h$ remain invariant. As a result, like in the previous cases, we can now factorize the Hamiltonian $H$ into a scale-dependent part $1/a$ and a scale-independent part $\tilde{H}$. 

This is the \emph{third key result} of this work and suggests that the techniques used in all the previous cases can be translated to the scalar field in spherically symmetric space-times. While $\Lambda$ is the rescaled scalar field mass, $\Delta_h$ corresponds to the rescaled horizon radius. Therefore, when $\Delta_h=0$, we see that $r_j=j$ and $f_j=1$, and the Hamiltonian (\ref{bh1}) reduces exactly to the flat space-time Hamiltonian \eqref{eq:3dHami-Mink}.  As we show later in this section, the analysis can be extended to spherically symmetric space-times with multiple horizons. The above formalism can be further simplified using near-horizon approximation (Appendix \ref{nearhorizon}) for those cases where an exact analytical expression for proper length cannot be obtained. %However, this is not needed for the models worked out in this paper.

In the rest of this section, we will focus on the massless case ($\Lambda=0$), and therefore impose the Dirichlet boundary condition $\varphi_{lm,N+1}=0$ to obtain a non-divergent scaling behavior. The coupling matrix $K$ for the scale-independent Hamiltonian $\tilde{H}$ has the following non-zero elements:
\begin{small}
	\begin{align}\label{grK}
		K_{11}=&f_1\left\{\Lambda^2+\frac{l(l+1)}{r_1^2}\right\}+\frac{r_{3/2}^2}{r_1^2}\sqrt{f_1f_{3/2}}\nonumber\\
		K_{jj\neq1}=&f_j\left\{\Lambda^2+\frac{l(l+1)}{r_j^2}\right\}\nonumber\\&+\frac{\sqrt{f_j}}{r_j^2}\left\{r_{j+\frac{1}{2}}^2\sqrt{f_{j+\frac{1}{2}}}+r_{j-\frac{1}{2}}^2\sqrt{f_{j-\frac{1}{2}}}\right\}\\
		K_{j,j+1}=&K_{j+1,j}=-\frac{r_{j+\frac{1}{2}}^2}{r_jr_{j+1}}\left\{f_{j+\frac{1}{2}}^2f_{j}f_{j+1}\right\}^{1/4}\nonumber
	\end{align}
\end{small}

Having fixed the elements of the coupling matrix, we now proceed to identify the properties of entanglement entropy, energy \eqref{eq:def-EntE} and temperature for different space-times with horizons. In evaluating the entanglement energy, we set 
$\epsilon = 1$. We will see that for all space-times, the numerically fixed pre-factors 
$c_e$ and $c_s$ consistently satisfy the following relation:
\begin{equation}
T \sim 1.26 \, T_{\rm H}.
\end{equation}
where $T_{\rm H}$ is the Hawking temperature of the horizon corresponding to the spherically symmetric space-time. Demanding that the entanglement temperature is indeed Hawking temperature will fix $\epsilon \sim 1.26$. In the rest of the section, we set $\epsilon = 1$.
	\begin{figure*}[!ht]
		\begin{center}
			\subfloat[\label{bh1a}]{%
				\includegraphics[width=0.4\textwidth]{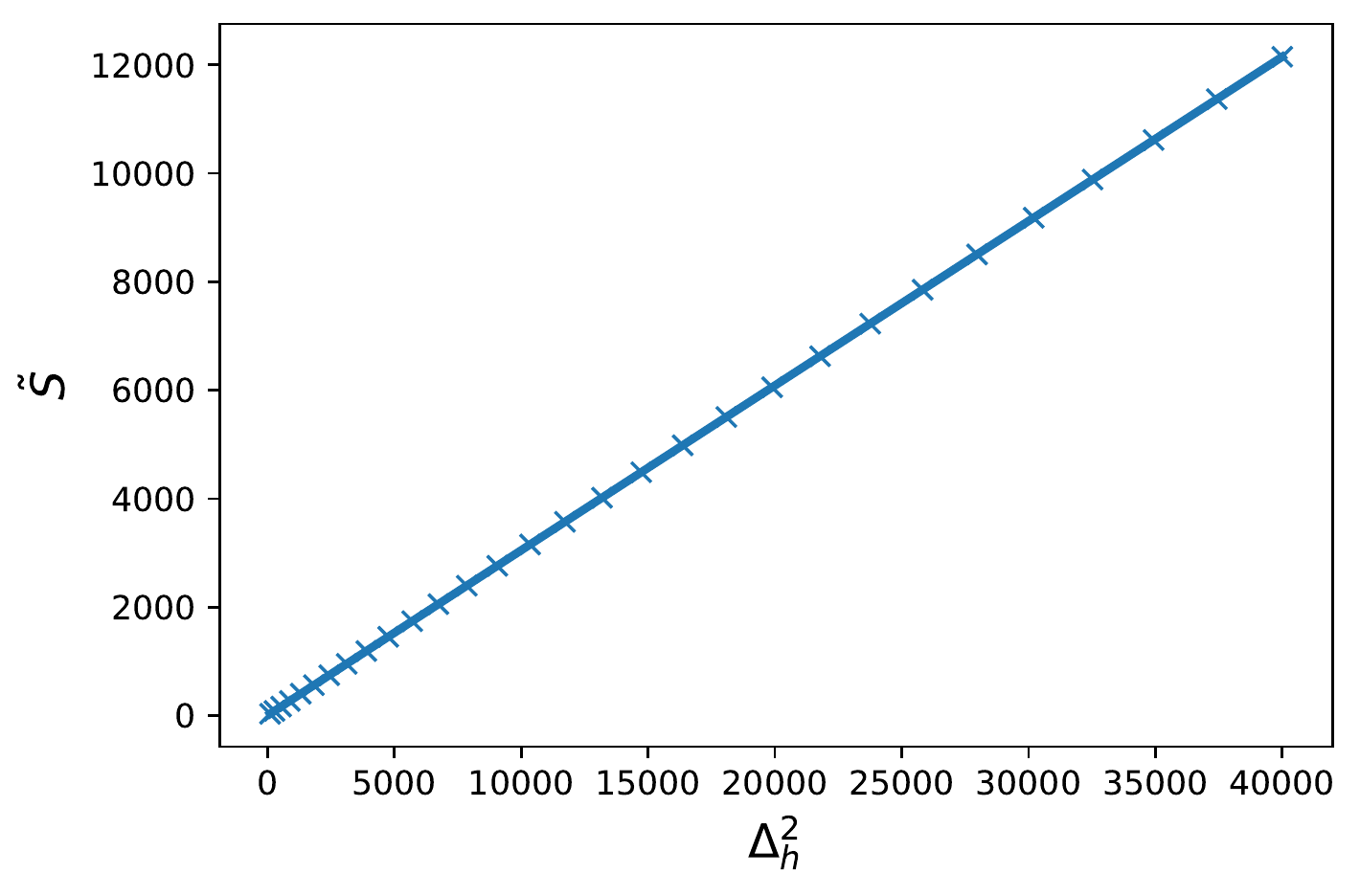}
			}
			\subfloat[\label{bh1b}]{%
				\includegraphics[width=0.4\textwidth]{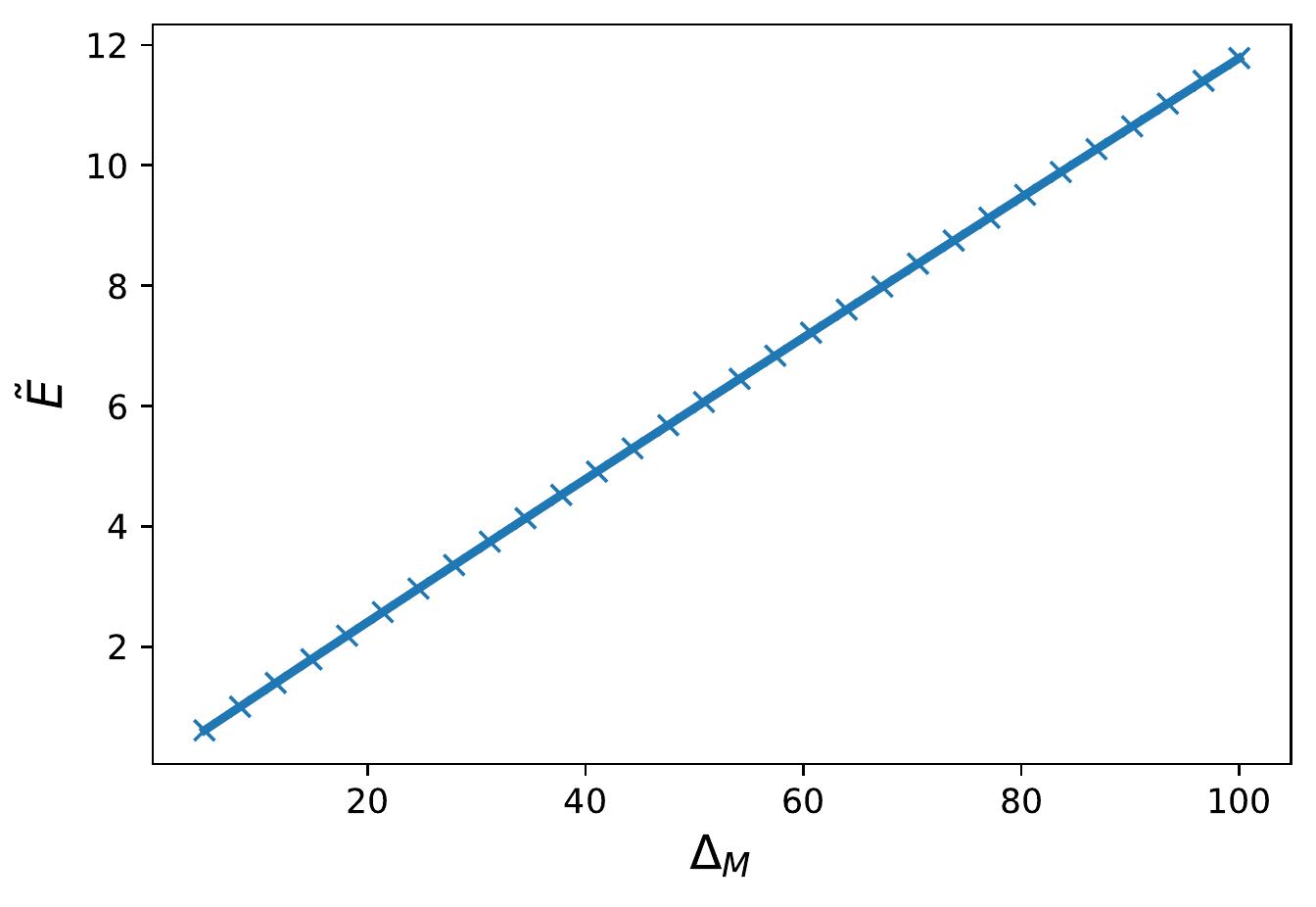}
			}\hfill
			
			\caption{Horizon scaling of entanglement entropy and entanglement energy for Schwarzschild black hole ($N=100$).}
			\label{BH1}
			\end{center}
		\end{figure*}

\subsection{Schwarzschild Black Hole}

In the Schwarzschild space-time, $f(r)=1-r_h/r$, and the proper length \eqref{prop}  
takes the form:
\begin{equation}
\rho=r\sqrt{1-\frac{r_h}{r}}+\frac{r_h}{2}\ln{\left[\frac{r}{r_h}\left\{1+\sqrt{1-\frac{r_h}{r}}\right\}^2\right]}.
\end{equation}
where the horizon radius is  $r_h=2M$. On discretizing $\rho=ja$, we get a scale-invariant expression that connects lattice-points in the proper length and radial co-ordinates as follows:
\begin{equation}
j=r_j\sqrt{1-\frac{\Delta_h}{r_j}}+\frac{\Delta}{2}\ln{\left[\frac{r_j}{\Delta_h}\left\{1+\sqrt{1-\frac{\Delta_h}{r_j}}\right\}^{2}\right]},
\end{equation}
where $\Delta_h=2M/a$ and $r_j=r/a$ are dimensionless. We also see that $f_j=1-\Delta_h/r_j$. This confirms that the Hamiltonian in (\ref{bh1}) is characterized by dimensionless parameters $\Lambda$ and $\Delta_h$, and is therefore invariant under the transformations:
\begin{equation}\label{scale1}
a\to\eta a;\quad m_f\to\eta^{-1}m_f;\quad M\to \eta M
\end{equation}
In order to further understand the implications of scale invariance, we first study the structure of entanglement mechanics of a scalar field in Schwarszchild background. Focusing on the scale-invariant Hamiltonian $\tilde{H}$, we vary the rescaled horizon $\Delta_h$, and assume that the entanglement energy/entropy of the horizon can be approximated by tracing out the closest oscillator near the horizon. This approximation is reasonable for large values of $\Delta_h$, wherein the radial distance of the closest oscillator from horizon is negligible ($r_1\sim \Delta_h$).

From \ref{BH1}, we observe the following scaling relations:
\begin{equation}
\label{eq:SE-Schw01}
\tilde{S}=c_s\Delta_h^2;\quad \tilde{E}=c_e\Delta_M,
\end{equation}
where a linear fit fixes the values $c_s\sim0.3$ and $c_e\sim0.06$. For the original Hamiltonian $H$, we then have:
\begin{equation}
\label{eq:SE-Schw02}
S=c_s\frac{r_h^2}{a^2};\quad E=c_e\frac{r_h}{a^2}
\end{equation}
The entanglement temperature for the system defined as $T=dE/dS$ is given by:
\begin{equation}
\label{eq:EntTemp-Sch}
T=\frac{c_e}{2c_s r_h}=\frac{\pi c_e}{c_s}T_H\sim 1.26 T_H \sim \frac{2 \pi}{5} T_H
\end{equation}
where $T_H$ is the Hawking temperature of the horizon in Schwarzschild space-time. From the above relations, we see that (i) entanglement energy scales linearly with horizon radius and therefore is fundamentally different from the area-law scaling observed in Minkowski space-time, (ii) 
entanglement temperature is independent of the UV cut-off $a$, and (iii) The entanglement mechanics follows the same laws of black-hole mechanics~\cite{2001-Wald-LivingReviewsinRelativity}. As we will show, this holds for all
known spherically symmetric space-times with horizon (including Cauchy horizon). In the next section, we will discuss the implications of 
Eq. \eqref{eq:EntTemp-Sch} in more detail.

Under the scaling transformations \eqref{scale1}, we  see that:
\begin{equation}
S\to S;\quad E\to\eta^{-1}E;\quad T\to\eta^{-1}T
\end{equation}
While entanglement entropy is scale-independent, entanglement energy and temperature are scale-dependent. As a result, when we scale down the UV-cutoff $a$, the entanglement energy and temperature increase drastically while the entropy remains invariant.

		\begin{figure*}[!ht]
		\begin{center}
			\subfloat[\label{dSa}]{%
				\includegraphics[width=0.4\textwidth]{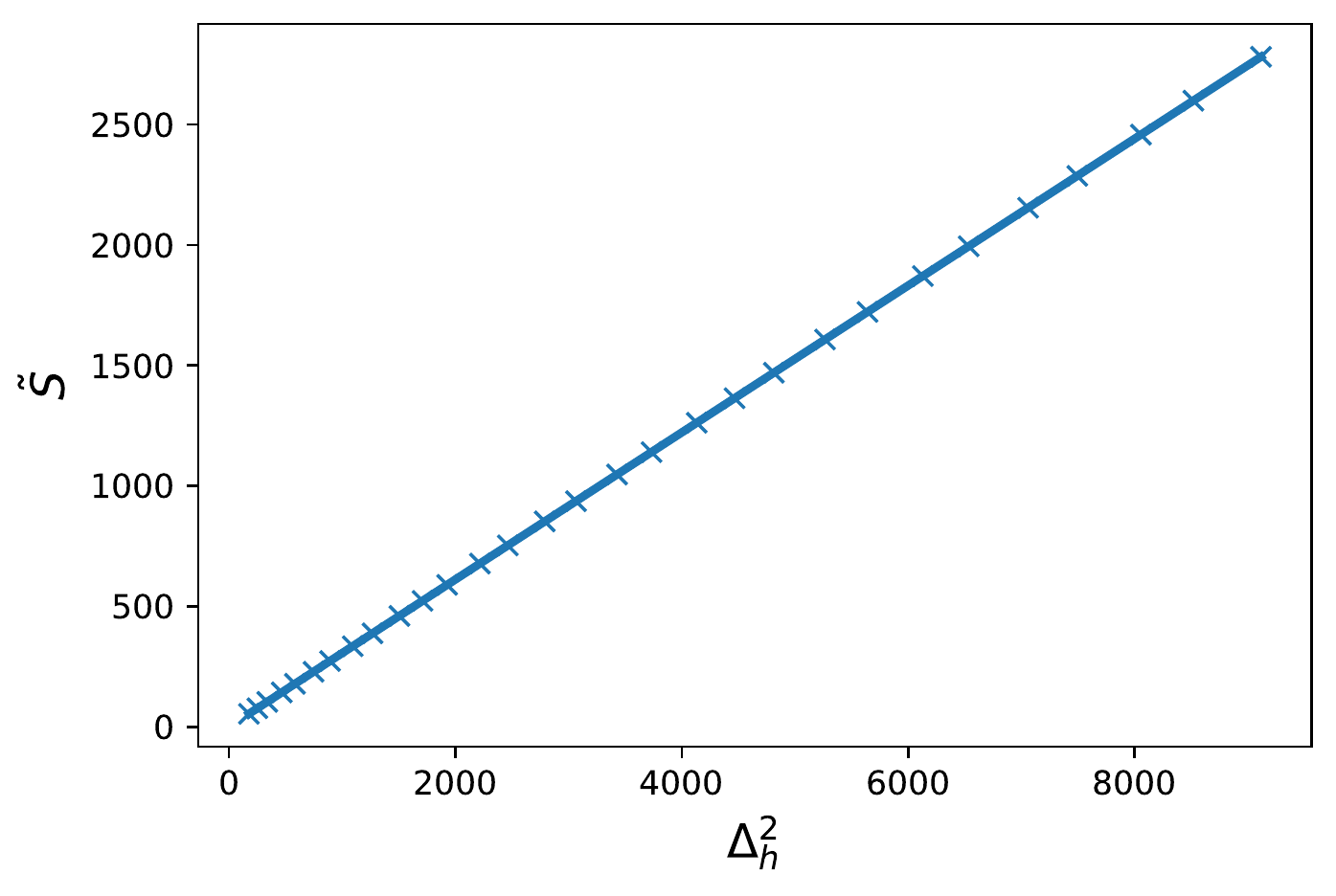}
			}
			\subfloat[\label{dSb}]{%
				\includegraphics[width=0.4\textwidth]{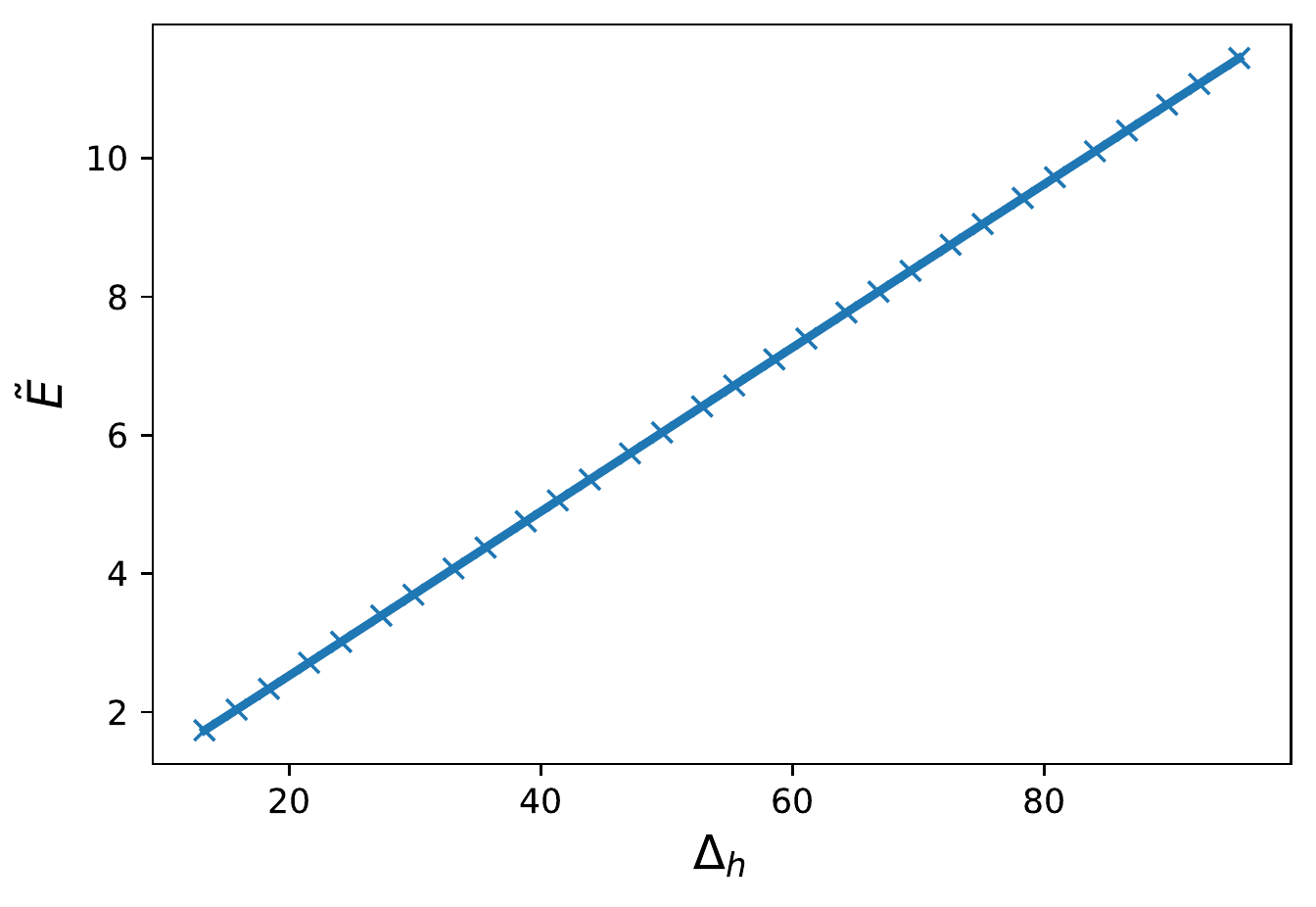}
			}\hfill
			
			\caption{Horizon Scaling of entanglement entropy and entanglement energy for de Sitter.}
			\label{dS}
				\end{center}
		\end{figure*}

\subsection{Static de-Sitter Patch}

In the static patch of de-Sitter space-time, $f(r)=1-r^2/l^2$ and the proper length \eqref{prop}
take the following form: 
\begin{equation}
\rho=l\cos^{-1}\left(\frac{r}{l}\right)
\end{equation}
where, $r_h=l$ is the cosmological horizon. On discretizing $\rho=ja$, we convert the 
above expression into a dimensionless form:
\begin{equation}
r_j=\Delta_h\cos{\frac{j}{\Delta_h}},
\end{equation}
where $\Delta_h=l/a$ and $r_j=r/a$. This implies that $f_j=\sin^2{\left(j/\Delta_h\right)}$. We also see that the IR cut-off on proper length given by $\rho_L=(N+1)a$, is automatically fixed as we restrict ourselves to the region $0\le r\le l$, that is, $r(\rho_L)=0$. Hence we obtain an additional constraint:
\begin{equation}\label{dsdisc}
N+1=\frac{\pi\Delta_h}{2};\quad l_N=\frac{2(N+1)a}{\pi}
\end{equation}
From this expression, we see that $\Delta_h$, and therefore radius $l$, can only take a discrete set of equally spaced values since $N$ is a positive integer. $A_N\propto N^2$ implies that the surface area of the cosmological horizon is also quantized. Like the Schwarzschild case, the rescaled Hamiltonian \eqref{bh1} is characterized by dimensionless parameters $\Lambda$ and $\Delta_h$, all of which remain invariant under the scaling transformations:
\begin{equation}\label{scale2}
a\to\eta a;\quad m_f\to\eta^{-1}m_f;\quad l\to \eta l.
\end{equation}
For the scale-invariant Hamiltonian $\tilde{H}$, we calculate the entropy and the energy for 
different values of the rescaled cosmological horizon $\Delta_h$ and is plotted in \ref{dS}.  
We obtain the following relations for entropy and energy:
\begin{equation}
\label{eq:SE-dS01}
\tilde{S}=c_s\Delta_h^2;\quad \tilde{E}=c_e\Delta_M,
\end{equation}
where a linear fit fixes the values $c_s\sim0.3$ and $c_e\sim0.12$.  

As mentioned earlier, Eq. \eqref{dsdisc} implies that $\Delta_h$ has a discrete spectrum. This means that both $\tilde{S}$ and $\tilde{E}$ are also discretized. For the original Hamiltonian $H$, we may therefore write:
\begin{equation}
\label{eq:SE-dS02}
S_N\sim c_s\frac{4(N+1)^2}{\pi^2};\quad E_N\sim c_e\frac{2(N+1)}{a\pi}.
\end{equation}
The discrete nature of de-Sitter space-time has been previously commented upon, and it is conjectured that the positive cosmological constant is associated to the bound on the degrees of freedom of the Universe~\cite{2001-BANKS-InternationalJournalofModernPhysicsA,2000-Bousso-JournalofHighEnergyPhysics,2004-Xiang.Shen-PhysicsLettersB}. Our results suggest that this is indeed the case not just for de-Sitter, but in any space-time where proper length is well-defined and bounded. Therefore, there is a natural scheme for quantizing the horizon radius, surface area, entanglement entropy, and entanglement energy. We will see this feature for the inner horizon of 
Reissner-Nordstr\"om black-hole, as well as for a Schwarzschild black hole in asymptotically de-Sitter space-time. In such cases, we will define entanglement temperature for the system as follows:
\begin{align}
T_N&=\frac{E_N-E_{N-1}}{S_N-S_{N-1}}=\frac{c_e}{2c_s\bar{l}}\nonumber\\
\bar{l}&=\frac{l_N+l_{N-1}}{2}
\end{align}
In the large $N$ limit, the entanglement temperature spectrum can be treated as a continuous function, and we obtain:
\begin{equation}
\label{eq:EntTemp-dS}
T=\frac{\pi \, c_e}{c_s}T_H\sim 1.26T_H \sim \frac{2 \pi}{5} T_H \, ,
\end{equation}
where $T_H$ is the Hawking temperature of cosmological horizon.  Note that the above
prefactor $1.26$ is identical to the one we obtained in Schwarzschild \eqref{eq:EntTemp-Sch}.
We see that the entanglement structure of de-Sitter horizon is analogous to the semi-classical black-hole mechanics. We will now show that the analysis can be extended to space-times with multiple horizons.

\subsection{Reissner-Nordstr\"om black hole}

The line-element for Reissner-Nordstr\"om black hole is given by \eqref{eq:SpherLineele02}, where
\[
f(r)=1- \frac{2M}{r} + \frac{Q^2}{r^2} \, .
\]
For $Q < M$, the roots are given by $r_\pm=M\pm\sqrt{M^2-Q^2}$ where $r_+$ corresponds to the event-horizon and $r_-$ refers to the internal Cauchy horizon. Thus, $f(r)$ is positive in two regions: 
1. $0 < r < r_-$ and 2. $r_+ < r < \infty$.  In the rest of this subsection, we will obtain entanglement properties of the horizons in these two regions.

\subsubsection{Cauchy horizon}

\begin{figure*}[!ht]
			\begin{center}
			\subfloat[\label{RN1a}]{%
				\includegraphics[width=0.4\textwidth]{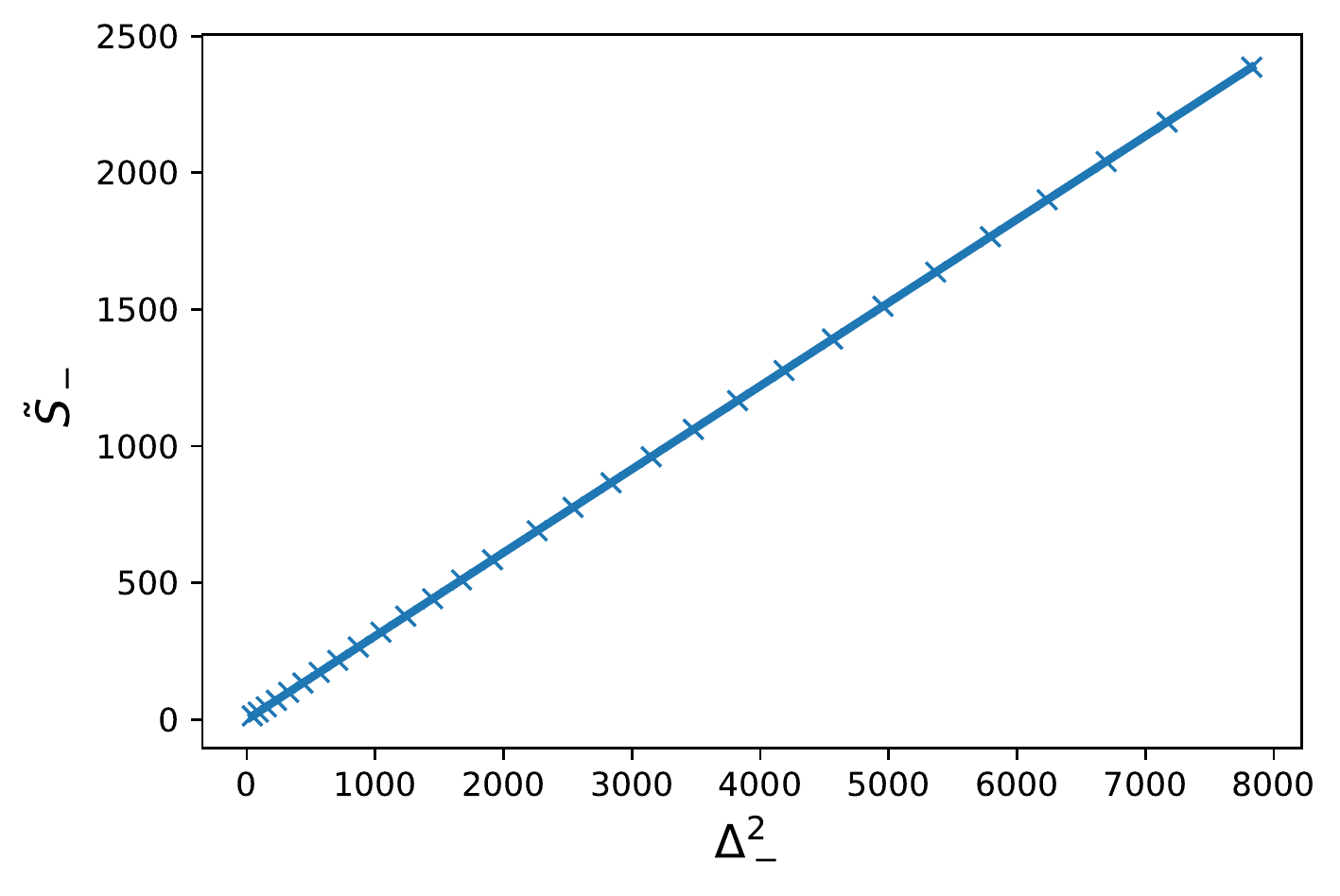}
			}
			\subfloat[\label{RN1b}]{%
				\includegraphics[width=0.4\textwidth]{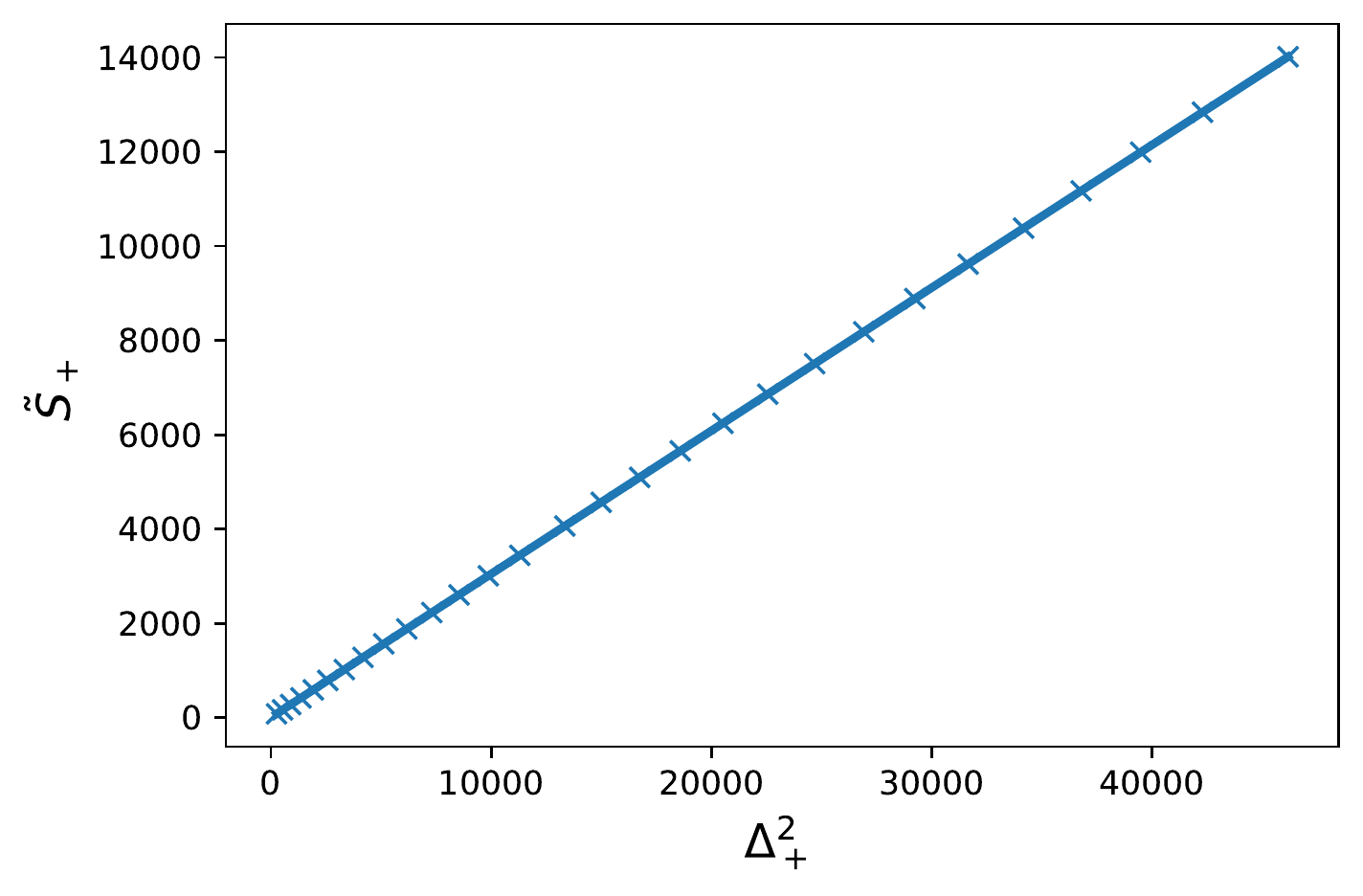}
			}\hfill
			\subfloat[\label{RN1c}]{%
				\includegraphics[width=0.4\textwidth]{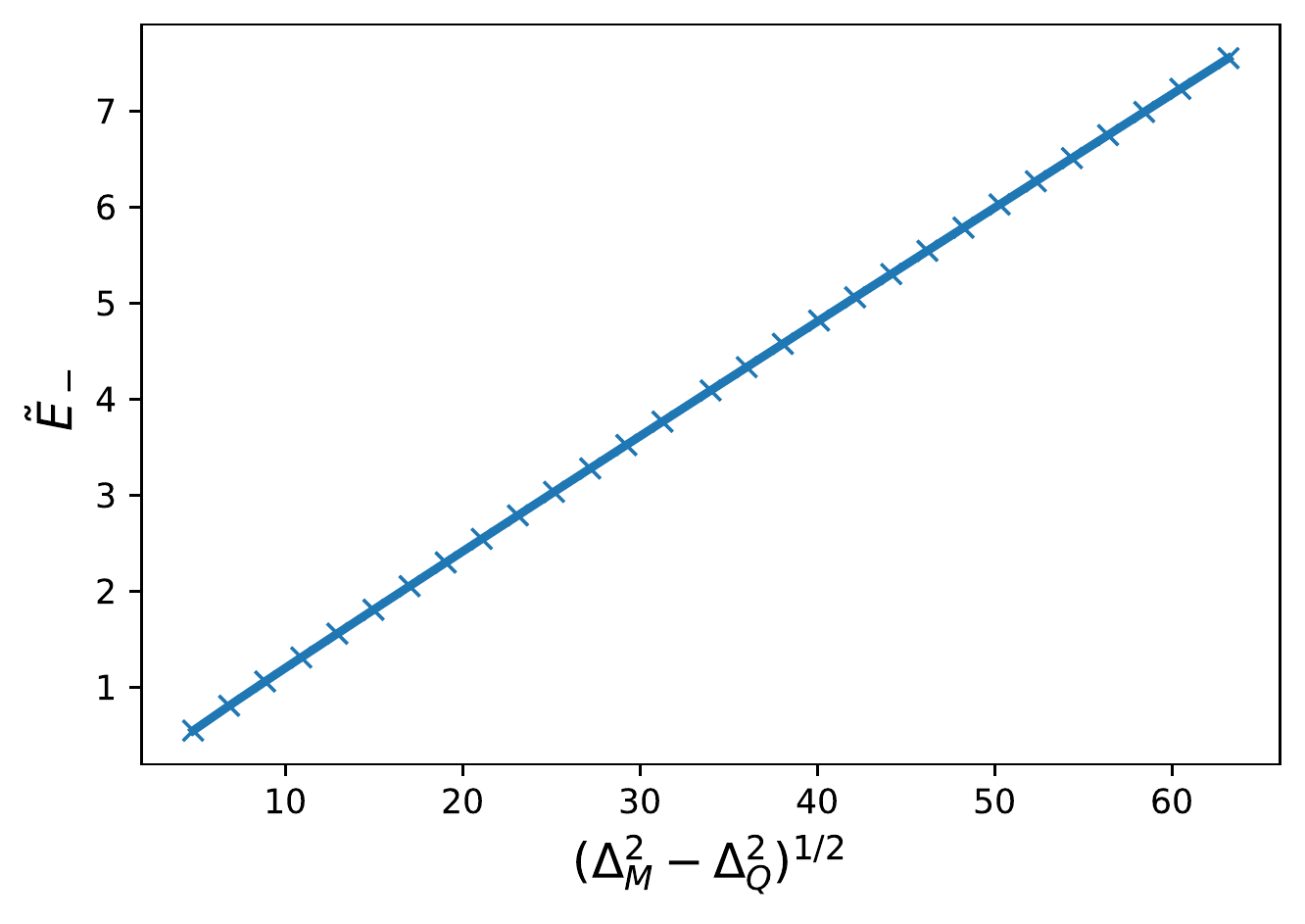}
			}
			\subfloat[\label{RN1d}]{%
				\includegraphics[width=0.4\textwidth]{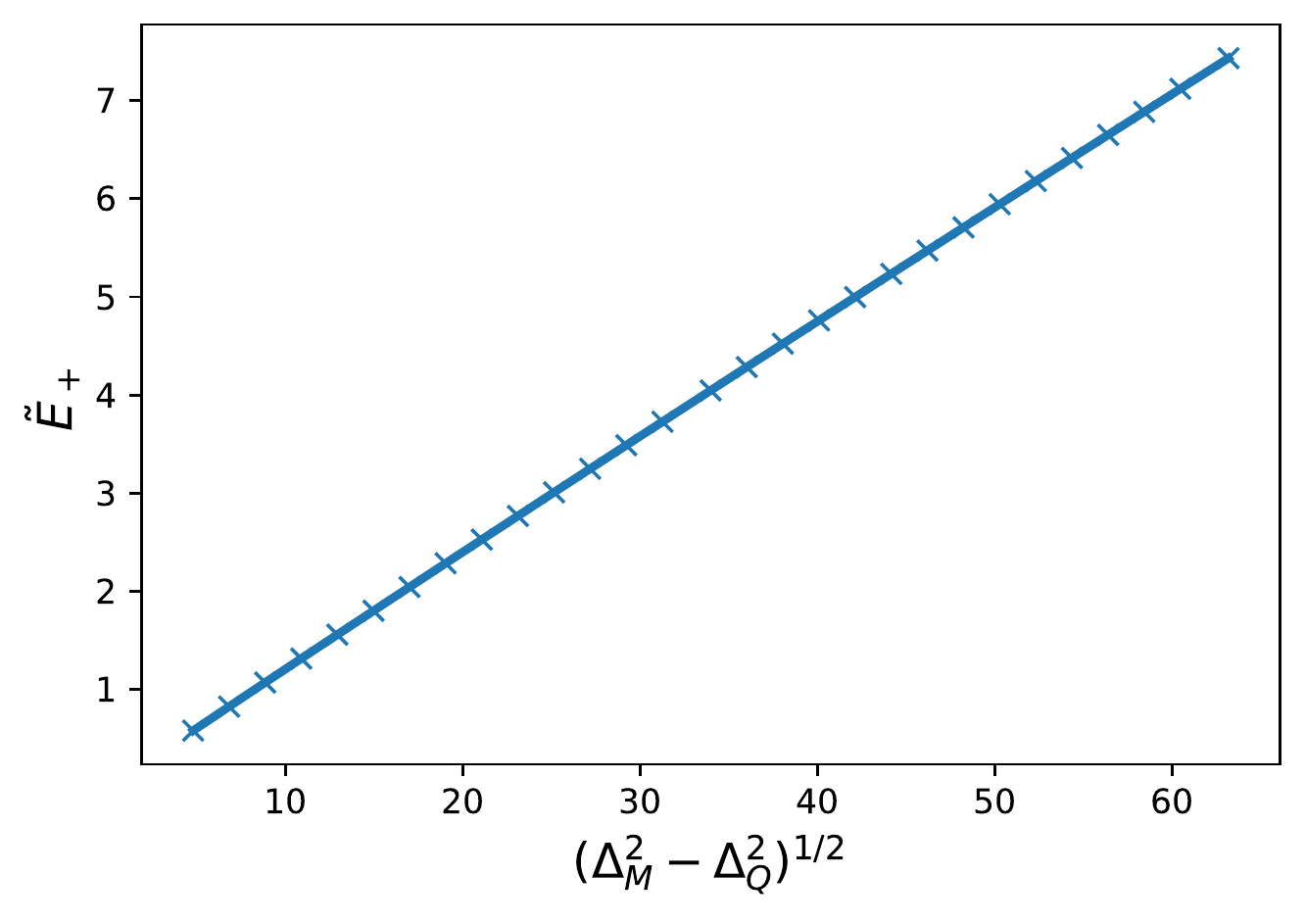}
			}
			\caption{Scaling of entanglement entropy and entanglement energy of both inner and outer horizons on varying $\Delta_Q$ ($\chi=1.1$) for Reissner-Nordstr\"om black hole.}
			\label{rn}
			\end{center}
		\end{figure*}

In terms of the dimensionless variable ($\chi$), the Cauchy horizon is 
\[
r_-=Q\{\chi-\sqrt{\chi^2-1}\}~~ \mbox{where}~~\chi=M/Q\in(1,\infty) \, .
\]
To ensure that the proper length is positive definite quantity, we reverse the limits of integration in Eq. \eqref{prop}, i. e.,
\begin{align}
\rho=&\int_{r}^{r_h} \frac{dr}{\sqrt{f(r)}}=-\sqrt{Q^2+r(r-2\chi Q)} \nonumber\\
&+\chi Q\ln{\left[\frac{Q\sqrt{\chi^2-1}}{\chi Q-2\left\{r+\sqrt{Q^2+r(r-2\chi Q)}\right\}}\right]}
\end{align}
On discretizing $\rho=ja$, we convert the above expression into a dimensionless form:
\begin{multline}
j=-\sqrt{\Delta_Q^2+r_j(r_j-2\chi \Delta_Q)}\\+\chi \Delta_Q\ln{\left[\frac{\Delta_Q\sqrt{\chi^2-1}}{\chi \Delta_Q-2\left\{r_j+\sqrt{\Delta_Q^2+r_j(r_j-2\chi \Delta_Q)}\right\}}\right]},
\end{multline}
where $\Delta_Q\equiv Q/a$ and $r_j\equiv r/a$ are both dimensionless. We also see that
\begin{equation}
f_j=1-\frac{2\chi\Delta_Q}{r_j}+\frac{\Delta_Q^2}{r_j^2}.
\end{equation}
Like in the two previous cases, the resulting Hamiltonian $H$ is factorized into a scale-dependent part $1/a$ and a scale-independent part $\tilde{H}$. The latter is completely characterized by dimensionless parameters $\Lambda$, $\Delta_Q$ and $\chi$, all of which are invariant under the scaling transformations: 
\begin{equation}\label{scale3}
a\to\eta a;\quad m_f\to\eta^{-1}m_f;\quad M\to \eta M;\quad Q\to\eta Q
\end{equation}
Therefore, we have obtained the set of transformations for the Cauchy horizon that leaves the entanglement entropy invariant. Like de-Sitter, we also see that the IR cut-off on proper length is fixed at $r=0$, leading to:
\begin{equation}
\Delta_Q\left[\chi\ln{\sqrt{\frac{\chi+1}{\chi-1}}}-1\right]= N+1
\end{equation}
Since $N$ is a natural number, the above expression is also a discretization relation. If we fix $\Delta_q$, then $\chi$ is discretized and vice versa. However, we will consider the scenario where the horizon changes on account of varying $\Delta_Q$ while keeping $\chi$ fixed. Physically, this corresponds to varying both mass and charge of the black hole proportionately to account for particles with a fixed mass-charge ratio ($\chi$) that are entering the event horizon. As a result, both mass and charge have equally spaced discrete spectra:
\begin{equation}\label{qn}
Q_N=\frac{(N+1)a}{\left(\chi\ln{\sqrt{\frac{\chi+1}{\chi-1}}}-1\right)};\quad M_N=\chi Q_N;
\end{equation}
The rest of the procedure follows from previous sections; the bipartited boundary is at the first oscillator closest to the horizon, which gives us a good approximation of the entanglement properties at the horizon itself. \ref{rn} contains the plot of entropy and energy of the Cauchy horizon. We will discuss the implications of the results along with the discussion on the event horizon.

\subsubsection{Event horizon}
In terms of the dimensionless variable ($\chi$), the  event horizon is 
\[
r_+=Q(\chi+\sqrt{\chi^2-1}) \, . 
\]
From Eq.~\eqref{prop}, we obtain:
\begin{multline}
	\rho=\sqrt{Q^2+r(r-2\chi Q)}\\+\chi Q\ln{\left[\frac{r-\chi Q+\sqrt{Q^2+r(r-2\chi Q)}}{Q\sqrt{\chi^2-1}}\right]} \, .
	\end{multline}
On discretizing $\rho=ja$, we convert the above expression into a dimensionless form:
\begin{multline}
	j=\sqrt{\Delta_Q^2+r_j(r_j-2\chi \Delta_Q)}\\+\chi \Delta_Q\ln{\left[\frac{r_j-\chi \Delta_Q+\sqrt{\Delta_Q^2+r_j(r_j-2\chi \Delta_Q)}}{\Delta_Q\sqrt{\chi^2-1}}\right]}
\end{multline}
where $\Delta_Q\equiv Q/a$ and $r_j\equiv r/a$ are both dimensionless. \ref{rn} contains the plot of entropy and energy of the event horizon.

From \ref{rn}, we obtain the following scaling relations for scale-invariant system $\tilde{H}$:
\begin{equation}
\label{eq:SE-RN01}
\tilde{S}_{\pm}=c_s\Delta_{\pm}^2;\quad \tilde{E}_{\pm}=c_e\sqrt{\Delta_M^2-\Delta_Q^2},
\end{equation}
where a linear fit fixes the values $c_s\sim 0.3$ and $c_e\sim 0.12$ for both horizons. 
It can also be seen from here that in the limit $\Delta_Q\to0$, we recover the values of $c_e$ and $c_s$ for Schwarzschild \eqref{eq:SE-Schw01}. As discussed above, $\tilde{S}_-$ and $\tilde{E}_-$ have discrete spectra. Since the entanglement energy is identical for both the horizons, we may therefore write for the total Hamiltonian ($H$)\footnote{Entanglement energy for RN in \cite{Mukohyama1998} was found to scale as $E_{ent}=c(\chi)r_B/a^2$ where $c(\chi)=\sqrt{\chi^2-1}/(\chi+\sqrt{\chi^2-1})$ and it was incorrectly concluded that $E_{ent}\propto r_B$ and $T\propto r_B^{-1}$. For an oscillator very close to horizon $r_B\sim r_+$, and from here it is straightforward to see that $E_{ent}\propto r_+-r_-$.}
\begin{equation}
\label{rnscale}
S_{-,N}=c_s\frac{r_{-,N}^2}{a^2};\quad  E_{-,N}=c_e\frac{\sqrt{M_N^2-Q_N^2}}{a^2},
\end{equation}
where $r_{\pm,N}$ are the respective horizons that are discretized according to \eqref{qn}. Although the discretization arises from the Cauchy horizon alone, since the energy is identical for both the horizons, the discrete spectra of $Q$ and $M$ must carry over to the event horizon as well. However, our results do not rely on the discretization of the entropy and energy of the outer horizon.

This is the \emph{fourth key result} of this work regarding which we would like to stress the following points: First, the above scaling relations for Reissner-Nordstr\"om black-hole is different from that of space-times with single horizons. In the case of Schwarzschild and de Sitter, 
the entanglement entropy scales as $E^2$. However, in the case of Reissner-Nordstr\"om black-hole,
entanglement entropy does not scale as $S\propto E^2$. Second, in all the three cases, the entanglement energy is proportional to the Komar energy~\cite{2010-Banerjee.Majhi-Phys.Rev.D,2016-Skakala.Shankaranarayanan-InternationalJournalofModernPhysicsD}. Komar energy is a classical expression and is related to the Hamiltonian of the Einstein-Hilbert action~\cite{1985-Chrusciel-AdHP}. 
The entanglement energy is obtained from the reduced density matrix of the quantum scalar field in a fixed background. Remarkably, entanglement energy is proportional to Komar energy. Our result
gives a completely new and independent insight into Komar energy. Third, the above insight 
leads us to the crucial link between the entanglement mechanics and black hole mechanics --- Smarr formula~\cite{1973-Bardeen.etal-CMP,2004-Padmanabhan-ClassicalandQuantumGravity,2005-Padmanabhan-PhysicsReports,2009-Kastor.etal-CQG}: 
\[
E_{\rm Komar} = 2 \, T_H \, S_{\rm BH}
\] 
Defining, entanglement temperature as:
\begin{equation}\label{newtemp}
T=\frac{E}{2S}.
\end{equation}
we get the following expression for the two horizons in Reissner-Nordstr\"om black-hole:
\begin{equation}
T_{N}^{(\pm)}=\frac{c_e}{2c_s}\left(\frac{r_{+,N}-r_{-,N}}{ r_{\pm,N}^2}\right).
\end{equation}
In the large $N$ limit, we obtain:
\begin{equation}
\label{eq:EntTemp-RN}
T^{(\pm)}=\frac{\pi c_e}{c_s}T_H^{(\pm)}\sim 1.26 T_H^{(\pm)} \sim \frac{2 \pi}{5} T_H^{(\pm)} \, ,
\end{equation}
where $T_H^{(-)}$ and $T_H^{(+)}$ are the Hawking temperatures of Cauchy and event horizon,
respectively.  Note that the factor $1.26$ is the same as we found in the case of Schwarzschild 
\eqref{eq:EntTemp-Sch} and de Sitter \eqref{eq:EntTemp-dS}. For Schwarzschild and de Sitter, the above expression of entanglement temperature also satisfies $T=dE/dS$, and hence the results are unchanged. Table \ref{tab:mapping} provides the key relations between the entanglement mechanics and black-hole thermodynamics:

\begin{table}[!htb]
\centering
\resizebox{0.46\textwidth}{!}{%
\begin{tabular}{@{}|l|l|@{}}
\toprule
Entanglement & Black holes  \\ 
\toprule
Energy $(E)$ & Komar Energy   $(E_{\rm Komar})$ \\
Entropy $(S)$& Black hole Entropy $(S_{\rm BH})$ \\
Temperature $(T)$ & Hawking Temperature  $(T_{\rm H})$ \\
\toprule
\end{tabular}
}
\caption{One-to-one mapping between the quantities in entanglement mechanics and black-hole thermodynamics}
\label{tab:mapping}
\end{table}

In the rest of this section, we show that the analogy between entanglement and black hole thermodynamics can be extended to asymptotic non-flat black-holes.

\subsection{Schwarzschild Anti-de Sitter}

A Schwarzschild black hole of mass $M$ in  Anti-de Sitter space-time (of radius $l$)
\[
f(r)=1- \frac{2M}{r} + \frac{r^2}{l^2} \, .
\]
has one horizon $r_h$ which is located at:
\begin{equation}
r_h=\frac{2l}{\sqrt{3}}\sinh{\left\{\frac{1}{3}\sinh^{-1}{\left(\frac{3\sqrt{3}M}{l}\right)}\right\}}.
\end{equation}
The proper length from the horizon is given by~\cite{1971-Byrd.Friedman-HandbookEllipticIntegrals}:
\begin{align}
\rho=& \, l \int_{r_h}^r\frac{\sqrt{r}dr}{\sqrt{(r-r_h)(r-z)(r-\bar{z})}}\nonumber\\
=&\frac{r_hl}{B+A}\sqrt{\frac{B}{A}}\Big[F(\vartheta,k)\nonumber\\&-\frac{1}{1+\alpha}\left\{\Pi\left(\vartheta,\frac{\alpha^2}{\alpha^2-1},k\right)-\alpha f_1\right\}\Big],
\end{align}
where\footnote{During the calculation, an error was noted in Eq 361.54 of Byrd \& Friedman\cite{1971-Byrd.Friedman-HandbookEllipticIntegrals}. This has been previously pointed out in \cite{2000-Kantowski.etal-}, and we have corrected the expression for $f_1$ accordingly.},
\begin{align}\label{sads1}
A^2&=l^2+3r_h^2;\qquad B^2=l^2+r_h^2;\qquad \alpha=\frac{B+A}{B-A}\nonumber\\
k^2&=\frac{(B+A)^2-r_h^2}{4AB};\qquad \cos{\vartheta}=\frac{(A-B)r+r_hB}{(A+B)r-r_hB}\nonumber\\
f_1&=\frac{\sqrt{AB}}{r_h}\ln{\left[\frac{2\sqrt{AB(1-k^2\sin^2\vartheta)}+r_h\sin\vartheta}{2\sqrt{AB(1-k^2\sin^2\vartheta)}-r_h\sin\vartheta}\right]}.
\end{align}

On discretizing the proper length $\rho=ja$, we obtain the above equations in dimensionless form:
\begin{align}
j=&\frac{\Delta_h \Delta_l}{B+A}\sqrt{\frac{B}{A}}\Big[F(\vartheta,k)\nonumber\\&-\frac{1}{1+\alpha}\left\{\Pi\left(\vartheta,\frac{\alpha^2}{\alpha^2-1},k\right)-\alpha f_1\right\}\Big],
\end{align}
where we have redefined:
\begin{align}\label{sads2}
A^2&=\Delta_l^2+3\Delta_h^2;\qquad B^2=\Delta_l^2+\Delta_h^2;\qquad \alpha=\frac{B+A}{B-A}\nonumber\\
k^2&=\frac{(B+A)^2-\Delta_h^2}{4AB};\qquad \cos{\vartheta}=\frac{(A-B)r_j+\Delta_hB}{(A+B)r_j-\Delta_hB}\nonumber\\
f_1&=\frac{\sqrt{AB}}{\Delta_h}\ln{\left[\frac{2\sqrt{AB(1-k^2\sin^2\vartheta)}+\Delta_h\sin\vartheta}{2\sqrt{AB(1-k^2\sin^2\vartheta)}-\Delta_h\sin\vartheta}\right]}.
\end{align}

It has been found that the spatial infinity for Schwarzschild-AdS corresponds to a finite value in the tortoise coordinates ~\cite{2018-Socolovsky-Adv.Appl.CliffordAlgebras}. However, in the proper length coordinate, it can be analytically shown that $f_1$ term (defined in \ref{sads2}) diverges in the limit 
$r \to \infty$. 

		\begin{figure*}[!ht]
	\begin{center}
			\subfloat[\label{sadsa}]{%
				\includegraphics[width=0.4\textwidth]{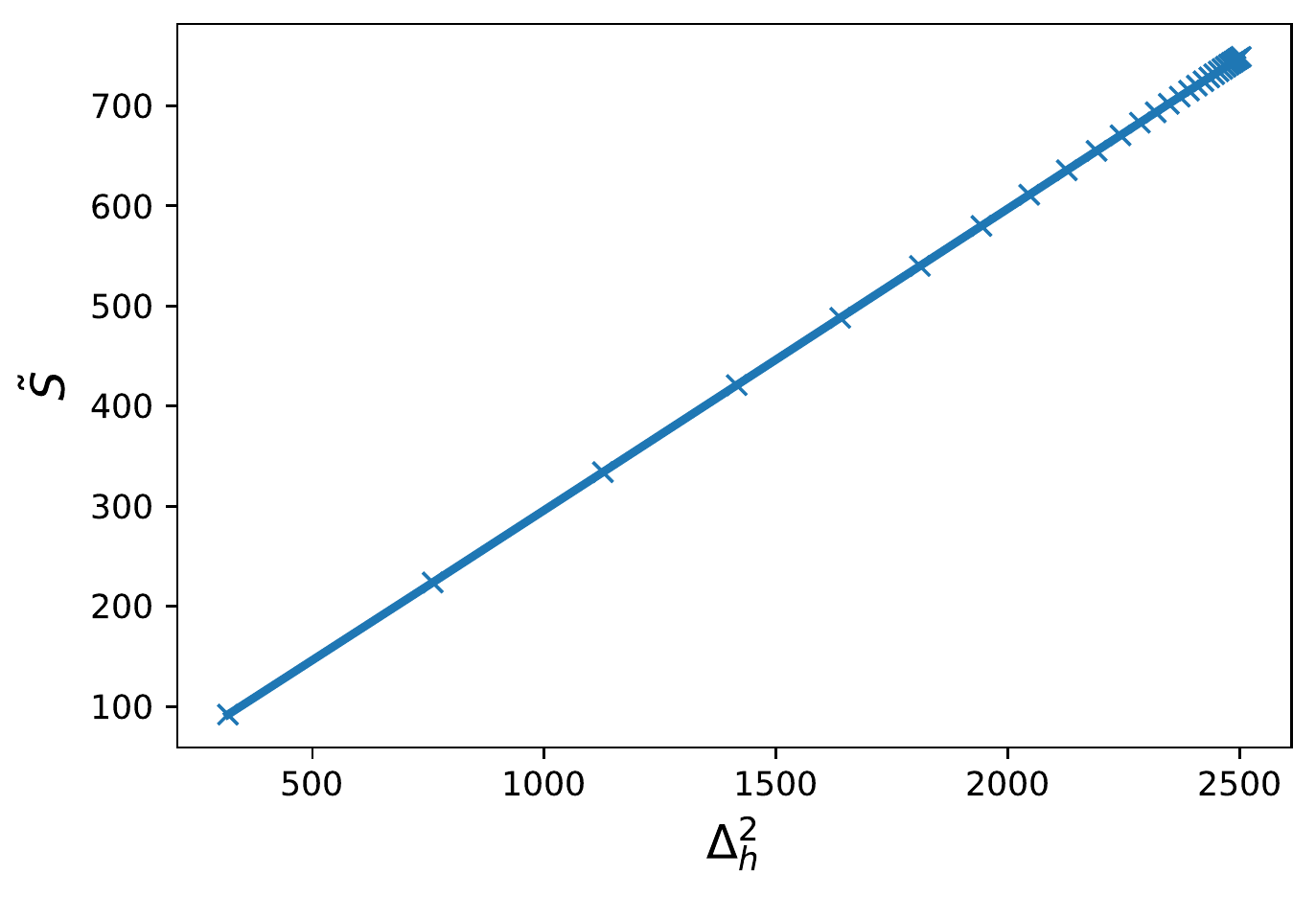}
			}
			\subfloat[\label{sadsb}]{%
				\includegraphics[width=0.4\textwidth]{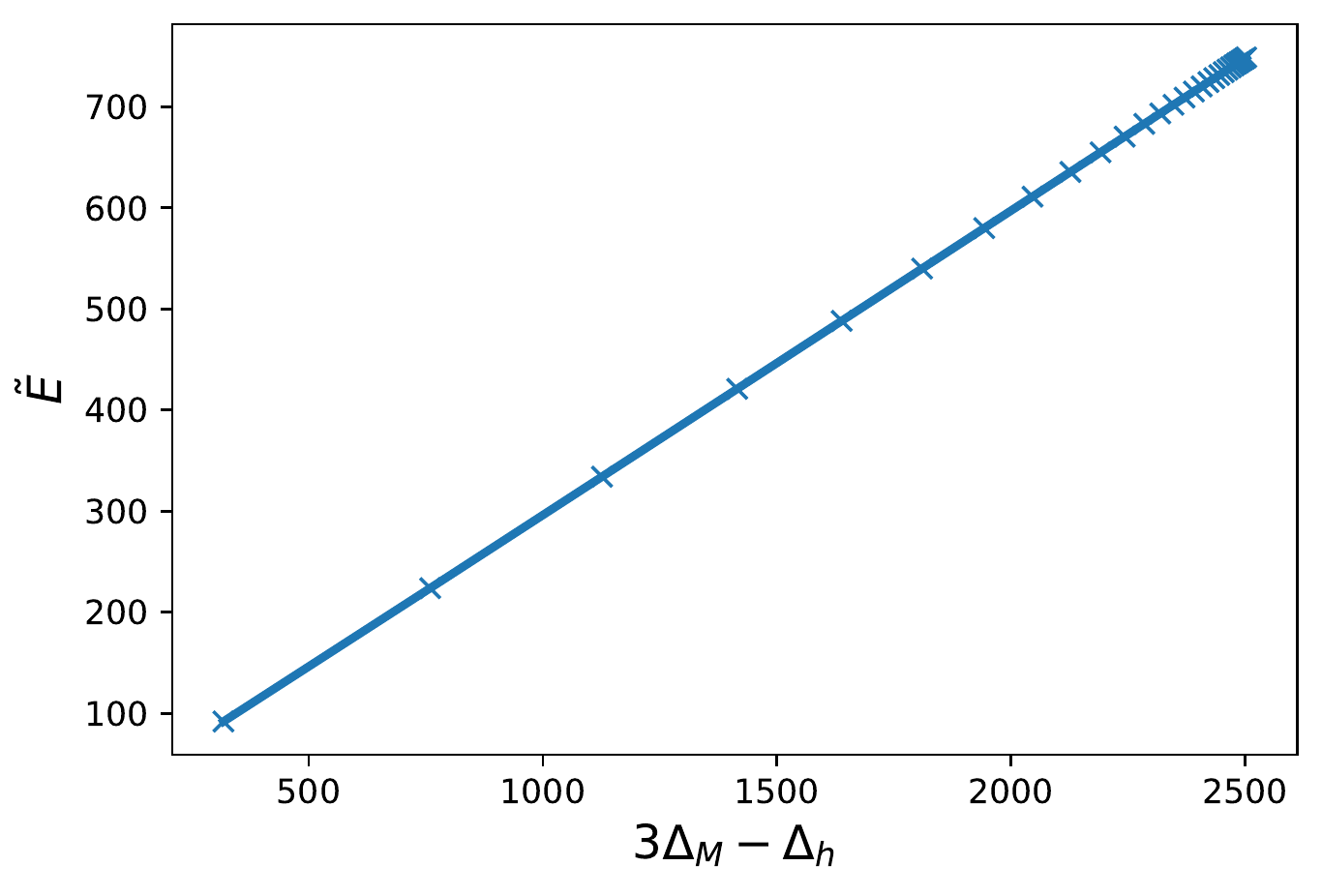}
			}\hfill
			
			\caption{Horizon scaling of entanglement entropy and entanglement energy on fixing $\Delta_M=25$ and $N=100$ for Schwarzschild-AdS black hole.}
			\label{sads}
	\end{center}
\end{figure*}

\ref{sads} contains the plot of entropy and energy for different values of the horizon radius.  From \ref{sads}, we obtain the following scaling relations:
\begin{equation}
\label{eq:SE-SAdS01}
\tilde{S} \sim c_s\Delta_h^2;\qquad \tilde{E}\sim c_e\left[3\Delta_M-\Delta_h\right]
\end{equation}
where $c_s\sim 0.3$ and $c_e\sim0.12$ are the best-fit numerical values.  
In the limit $\Delta_l\to\infty$, we recover the pre-factor values of Schwarzschild black hole \eqref{eq:SE-Schw01}. For the total Hamiltonian $H$, the scaling relations become:
\begin{equation}
\label{eq:SE-SAdS02}
S \sim c_s\frac{r_{h}^2}{a^2};\qquad E\sim c_e\frac{3M-r_h}{a^2}
\end{equation}
Using the definition of temperature in $\ref{newtemp}$, we see that:
\begin{equation}
\label{eq:EntTemp-SAdS}
T=\frac{c_e}{c_s}\left(\frac{3M-r_h}{r_{h}^2}\right)=\frac{\pi c_e}{c_s}T_H\sim 1.26 T_H 
\sim \frac{2 \pi}{5} T_H \, ,
\end{equation}
where $T_H$ is the Hawking temperature of the event horizon in Schwarzschild Anti-de Sitter. Note that the factor $1.26$ is the same as we found in the case of Schwarzschild 
\eqref{eq:EntTemp-Sch}, de Sitter \eqref{eq:EntTemp-dS} and Reissner-Norstr\"om \eqref{eq:EntTemp-RN}. The results show that the mapping in Table \ref{tab:mapping} is valid for asymptotically non-flat space-times. We discuss the importance of this mapping in Sec. \ref{sec:Ent-BHconnection}.

\subsection{Schwarzschild de-Sitter space-time}
A Schwarzschild black hole (of mass $M$) in a de-Sitter space-time (of radius $l$) 
\[
f(r)=1- \frac{2M}{r}- \frac{r^2}{l^2} \, ,
\]
also has two horizons --- $r_b$ (event horizon) and $r_c$ (cosmological horizon)\cite{2003-Shankaranarayanan-Phys.Rev.D}:
\begin{small}
	\begin{equation}
	r_-=-\frac{2l}{\sqrt{3}}\cos{\frac{\theta}{3}};\quad r_b=\frac{2l}{\sqrt{3}}\cos{\frac{\pi+\theta}{3}};\quad r_c=\frac{2l}{\sqrt{3}}\cos{\frac{\pi-\theta}{3}}
	\end{equation}
\end{small}
where $r_-$ is the third negative root,  $\theta=\cos^{-1}(3\sqrt{3}\chi)$ and $\chi=M/l\in\left[0,1/(3\sqrt{3})\right]$. $f(r)$ is positive in the region between the two horizons. Hence, 
in this region, we have definitions for proper length --- one w.r.t. the event horizon $r_b$ which 
we refer as $\rho_b$ and the second w.r.t. the cosmological horizon $r_c$ which we refer as
$\rho_c$. As we show, both the definitions of proper length coincide for the Dirichlet boundary 
condition since $\Lambda = 0$ (cf. Sec. \ref{sec3d}). 

		\begin{figure*}[!ht]
			\begin{center}
			\subfloat[\label{SdS1a}]{%
				\includegraphics[width=0.4\textwidth]{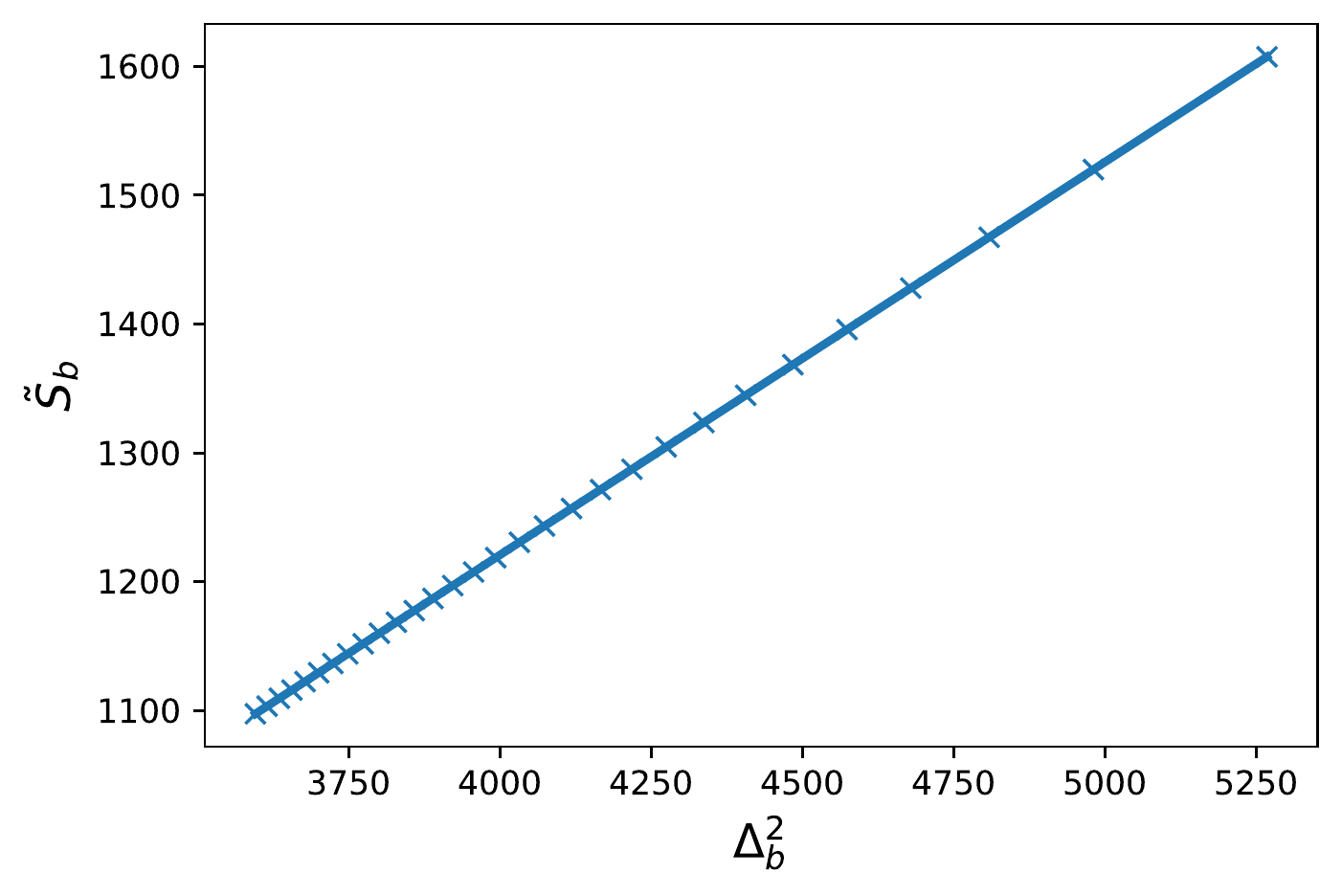}
			}
			\subfloat[\label{SdS1b}]{%
				\includegraphics[width=0.4\textwidth]{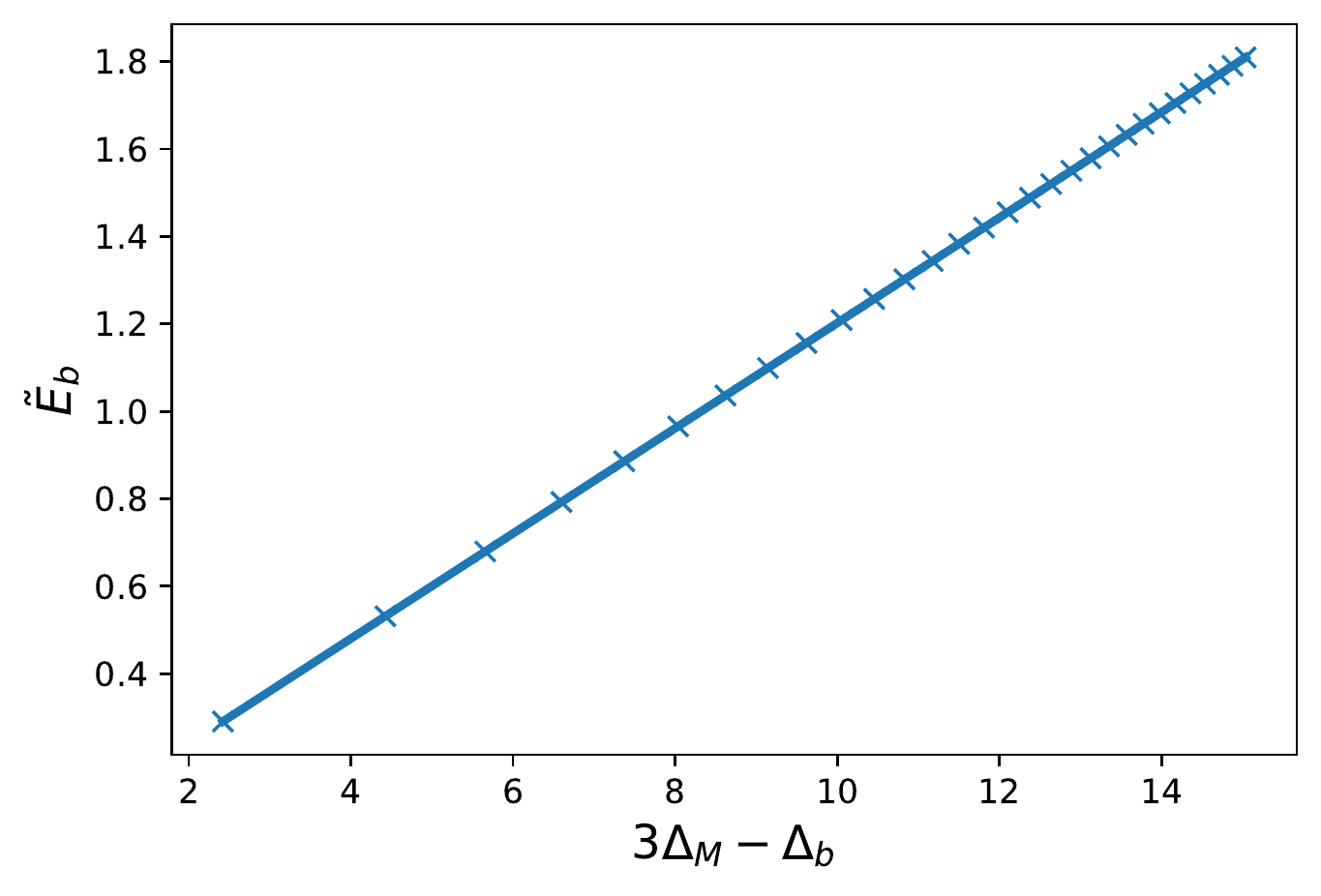}
			}\hfill
			\subfloat[\label{SdS1c}]{%
				\includegraphics[width=0.4\textwidth]{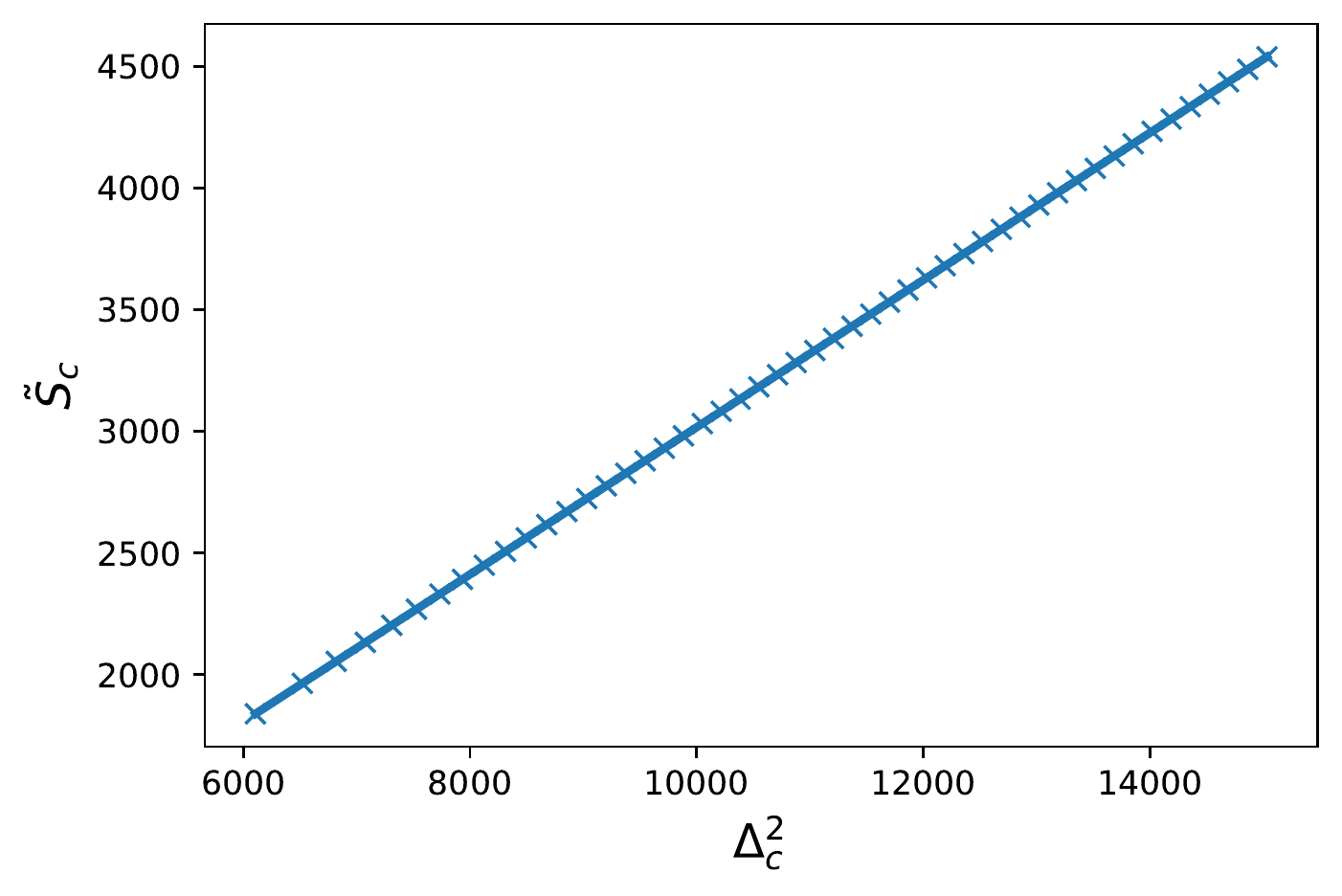}
			}
			\subfloat[\label{SdS1d}]{%
				\includegraphics[width=0.4\textwidth]{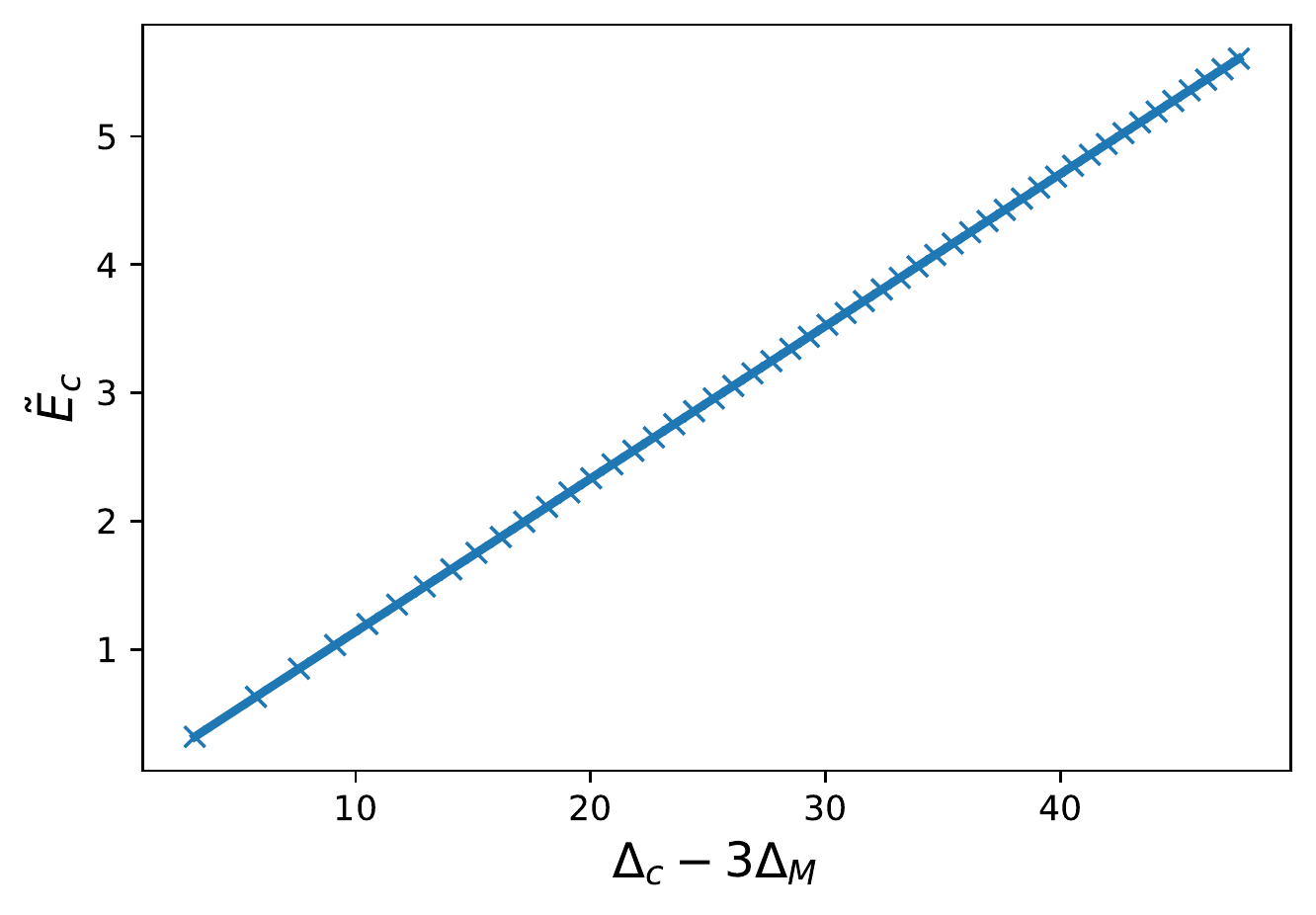}
			}
			\caption{Scaling of entanglement entropy and entanglement energy with respect to event and cosmological horizons on fixing $\Delta_M=25$ and $\Lambda=0$.}
			\label{SdS}
				\end{center}
		\end{figure*}

\subsubsection{Event Horizon}

The proper length with respect to the event horizon $r_b$ is obtained by solving the following 
integral~\cite{1971-Byrd.Friedman-HandbookEllipticIntegrals}:
\begin{align}
	\rho_b&=l\int_{r_b}^r\frac{\sqrt{r}dr}{\sqrt{(r-r_-)(r-r_b)(r_c-r)}}\nonumber\\
	&=\frac{2r_bl}{\sqrt{r_c(r_b-r_-)}}\Pi\left(\vartheta,\alpha^2,k\right),
\end{align}
where,
\begin{small}
	\begin{equation}\label{sdsvar}
	\vartheta=\sin^{-1}{\sqrt{\frac{r_c(r-r_b)}{r(r_c-r_b)}};\quad \alpha^2=1-\frac{r_b}{r_c}};\quad k^2=\frac{r_-(r_b-r_c)}{r_c(r_b-r_-)}.
	\end{equation}
\end{small}
On discretizing proper length $\rho_b=ja$, we convert the above expression into a dimensionless form:
\begin{equation}
j=\frac{2\Delta_b\Delta_l}{\sqrt{\Delta_c(\Delta_b-\Delta_-)}}\Pi\left(\vartheta,\alpha^2,k\right) \, .
\end{equation}
In terms of the dimensionless variables $\Delta_l=l/a$ and $r_j=r/a$, we have 
\begin{equation}
f_j=1-\frac{2\Delta_M}{r_j}-\frac{r_j^2}{\Delta_l^2} \, ,
\end{equation}
and,
\begin{small}
	\begin{align}\label{sdsvar2}
	\Delta_-&=-\frac{2\Delta_l}{\sqrt{3}}\cos{\frac{\theta}{3}};\quad \Delta_b=\frac{2\Delta_l}{\sqrt{3}}\cos{\frac{\pi+\theta}{3}}; \nonumber \\
 \Delta_c & =\frac{2\Delta_l}{\sqrt{3}}\cos{\frac{\pi-\theta}{3}}; \quad 
\vartheta =\sin^{-1}{\sqrt{\frac{\Delta_c(r_j-\Delta_b)}{r_j(\Delta_c-\Delta_b)}}};  \nonumber \\
\alpha^2&= 1-\frac{\Delta_b}{\Delta_c};\quad k^2=\frac{\Delta_-(\Delta_b-\Delta_c)}{\Delta_c(\Delta_b-\Delta_-)} \, .
	\end{align}
\end{small}
As a result, the Hamiltonian $\tilde{H}$ is fully characterized by dimensionless parameters $\Lambda$, $\Delta_l$ and $\Delta_M$, all of which are invariant under the scaling transformations:
\begin{equation}\label{scale4}
a\to\eta a;\quad m_f\to\eta^{-1}m_f;\quad M\to \eta M;\quad l\to\eta l
\end{equation}

In the case of SdS, the IR cut-off on proper length is again automatically fixed as we restrict ourselves to the region $\tilde{r}_b\leq r_j \leq \tilde{r}_c$:
\begin{equation}\label{sdsdis1}
N+1=\frac{2\tilde{r}_b\Delta_l}{\sqrt{\tilde{r}_c(\tilde{r}_b-\tilde{r}_-)}}\Pi\left(\frac{\pi}{2},\alpha^2,k\right)
\end{equation}
This is a discretization relation similar to what was obtained for dS and RNBH. We will consider the case where we fix $\Delta_M$ and vary $\Delta_l$ by varying $N$. This is to ensure that $\chi$ is always between $[0, 1/(3 \sqrt{3}]$.

The top panel of \ref{SdS} contains the plot of the entanglement entropy  and energy for the  event horizon. From \ref{SdS}, we obtain the following scaling relations for the scale-invariant Hamiltonian 
($\tilde{H}$):
\begin{equation}
\label{eq:SE-SdS01}
\tilde{S}_b \sim c_s\Delta_b^2;\qquad \tilde{E}_b\sim c_e(3\Delta_M-\Delta_b)
\end{equation}
where, $c_s\sim0.3$ and $c_e\sim0.12$ are the best-fit numerical values. In the limit 
$\Delta_l\to\infty$, we recover the prefactors of the Schwarzschild black hole~\eqref{eq:SE-Schw01}. For the total Hamiltonian $(H)$, the scaling relations become:
\begin{equation}
\label{eq:SE-SdS02}
S_{b,N} \sim c_s\frac{r_{b,N}^2}{a^2};\qquad E_{b,N}\sim c_e\frac{3M-r_{b,N}}{a^2}
\end{equation}
Using the definition of temperature from \eqref{newtemp}, we get:
\begin{equation}
T^{(b)}=\frac{c_e}{c_s}\left(\frac{3M-r_{b,N}}{r_{b,N}^2}\right).
\end{equation}
In the large $N$ limit, we see that:
\begin{equation}
\label{eq:EntTemp-SdS01}
T^{(b)}=\frac{\pi c_e}{c_s}T_H^{(b)}\sim 1.26 T_H^{(b)} \sim \frac{2 \pi}{5} T_H^{(b)} \, ,
\end{equation}
where $T_H^{(b)}$ is the Hawking temperature of the event horizon in SdS~\cite{1977-Gibbons.Hawking-PRD,2003-Shankaranarayanan-Phys.Rev.D}. Note that the factor 
$1.26$ is the same as we found in the case of Schwarzschild 
\eqref{eq:EntTemp-Sch}, de Sitter \eqref{eq:EntTemp-dS}, Reissner-Norstr\"om \eqref{eq:EntTemp-RN} and Schwarschild Anti-de Sitter \eqref{eq:EntTemp-SAdS}.

\begin{figure}[!hbt]
	\centering
	\includegraphics[scale=0.58]{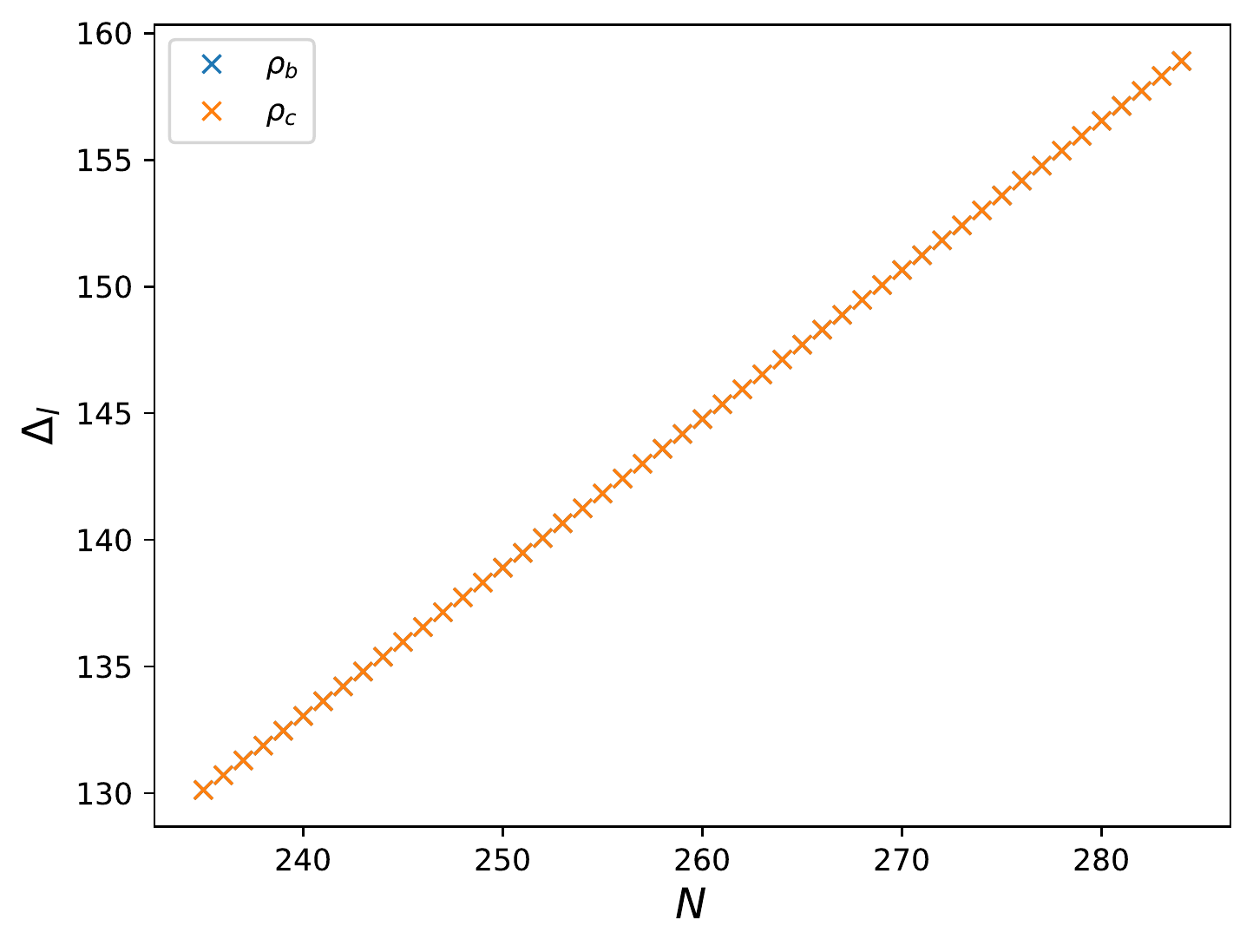}
	\caption{Discretization of $\Delta_l$ from the cutoffs on proper lengths $\rho_b$ and $\rho_c$, for $\Delta_M=25$.}
	\label{sdsrho}
\end{figure}

\subsubsection{Cosmological Horizon}

To explore the scaling properties of cosmological horizon, we define proper distance $r_c$ as follows~\cite{1971-Byrd.Friedman-HandbookEllipticIntegrals}:
\begin{align}
\rho_c&=l\int_{r_c}^r\frac{\sqrt{r}dr}{\sqrt{(r-r_-)(r-r_b)(r_c-r)}}\nonumber\\
&=\frac{2l}{\sqrt{r_c(r_b-r_-)}}\left[r_-F(\vartheta,k)-\left(r_c-r_-\right)\Pi\left(\vartheta,\alpha^2,k\right)\right],
\end{align}
where,
	\begin{equation}
	\sin{\vartheta}={\sqrt{\frac{(r_b-r_-)(r_c-r)}{(r_c-r_b)(r-r_-)}};\quad \alpha^2=\frac{r_b-r_c}{r_b-r_-}},
	\end{equation}
and the definition of $k$ is the same as Eq.~\eqref{sdsvar}. On discretizing proper length $\rho_c=ja$, we convert the above expression into a dimensionless form:
\begin{small}
\begin{equation}
j=\frac{2\Delta_l}{\sqrt{\Delta_c(\Delta_b-\Delta_-)}}\left[\Delta_-F(\vartheta,k)-\left(\Delta_c-\Delta_-\right)\Pi\left(\vartheta,\alpha^2,k\right)\right],
\end{equation}
\end{small}where in terms of dimensionless variables $\Delta_l=l/a$ and $r_j=r/a$, we can also rewrite:
\begin{small}
	\begin{equation}
	\sin{\vartheta}={\sqrt{\frac{(\Delta_b-\Delta_-)(\Delta_c-r_j)}{(\Delta_c-\Delta_b)(r_j-\Delta_-)}};\quad \alpha^2=\frac{\Delta_b-\Delta_c}{\Delta_b-\Delta_-}},
	\end{equation}
\end{small}Except for $\vartheta$ and $\alpha$ given above, the parameters used here follow the same definition as in \eqref{sdsvar2}. Now we impose an IR cut-off on proper length to restrict ourselves in the region $\tilde{r}_b\leq r_j \leq \tilde{r}_c$:
\begin{small}
\begin{equation}\label{sdsdis2}
N+1=\frac{2\Delta_l}{\sqrt{\Delta_c(\Delta_b-\Delta_-)}}\left[\Delta_-K(k)-\left(\Delta_c-\Delta_-\right)\Pi\left(\frac{\pi}{2},\alpha^2,k\right)\right]
\end{equation}
\end{small}
This expression relates the number of oscillators $N$, $\Delta_l$ and $\chi$, and we only need to fix two of these to fix the third. We will fix $\Delta_M$ here as we did for the event horizon case, which leaves $\Delta_l$ discretized. From \ref{sdsrho}, we see that the spectra for $\Delta_l$ obtained from $\rho_b$ and $\rho_c$ coincide exactly, and therefore the two discretization relations \eqref{sdsdis1} and \eqref{sdsdis2} are identical.

From \ref{SdS}, we obtain the following scaling relations for the scale-invariant system $\tilde{H}$:
\begin{equation}
\label{eq:SE-SdS03}
\tilde{S}_c \sim c_s\Delta_c^2;\qquad \tilde{E}_c\sim c_e\left[\Delta_c-3\Delta_M\right]
\end{equation}
where $c_s\sim0.3$ and $c_e\sim0.12$ are the best-fit numerical values. In the limit $\Delta_M\to0$, we recover the prefactor values of de-Sitter space \eqref{eq:SE-dS01}. 
For the total Hamiltonian $H$, the scaling relations become:
\begin{equation}
\label{eq:SE-SdS04}
S_{c,N} \sim c_s\frac{r_{c,N}^2}{a^2};\qquad E_{c,N}\sim c_e\frac{r_{c,N}-3M}{a^2}
\end{equation}
Using the definition of temperature in \eqref{newtemp}, we have:
\begin{equation}
T^{(c)}=\frac{c_e}{c_s}\left(\frac{r_{c,N}-3M}{r_{c,N}^2}\right).
\end{equation}
In the large $N$ limit, we get:
\begin{equation}
\label{eq:EntTemp-SdS02}
T^{(c)}=\frac{\pi c_e}{c_s}T_H^{(c)}\sim 1.26 T_H^{(c)} \sim \frac{2 \pi}{5} T_H^{(c)},
\end{equation}
where $T_H^{(c)}$ is the Hawking temperature of the cosmological horizon in SdS. The entanglement entropy, energy, and temperature of both the horizons satisfy the thermodynamic properties of the individual horizons~\cite{1977-Gibbons.Hawking-PRD}. The results show that the mapping in Table \ref{tab:mapping} is valid for asymptotically non-flat space-times with multiple horizons. 

In the next section, we discuss the importance of this mapping between entanglement mechanics and black-hole thermodynamics.

\section{Entanglement mechanics and black-hole thermodynamics}
\label{sec:Ent-BHconnection}

In the previous section, we obtained the scaling relations of entanglement entropy and entanglement energy of a quantum scalar field in a variety of spherically symmetric asymptotic flat (and non-flat) 
space-times with one or more horizons. In two ways, these relations were fundamentally different from the entanglement structure of Minkowski space-time: First, entanglement entropy and energy \emph{do not} have the same scaling. Second, entanglement temperature is independent of the UV-cutoff $a$. This further implies that unlike in flat-space where the boundary is merely artificial, entanglement calculation close to a space-time horizon captures relevant information about its thermodynamic structure. 

We have shown that the entanglement energy scales as Komar energy of the horizon, however, 
$dE/dS$ does not scale as horizon temperature. The entanglement temperature of 
the quantum scalar field in any space-time with horizon(s) is defined such that it satisfies the universal relation \eqref{newtemp}
\[
T=\frac{E}{2S} \, .
\]
We have observed that that the following relations are universal: 
\begin{equation}
\label{eq:EntTem-HawTemp}
c_e\sim0.12;\quad c_s\sim 0.3;\quad \frac{T_{ent}}{T_H}=\frac{\pi c_e}{c_s}\sim1.26
\end{equation}
As mentioned in Sec. \eqref{secmath}, in deriving the above expression, we have set $\epsilon = 1$ for entanglement energy \eqref{eq:def-EntE} and is not unique~\cite{1998-Mukohyama.etal-Phys.Rev.D}. 
Choosing $\epsilon= 2\pi/5\sim1.26$, leads to:
\begin{equation}
c_e\to\frac{c_e}{1.26}\sim0.0955;\quad \frac{\pi c_e}{c_s}\sim 1
\end{equation}
From this, we obtain the following structure for entanglement energy, entropy, and temperature:
\begin{equation}
\label{eq:SmarrFor01}
T=T_H;\quad E=2TS.
\end{equation}
The above relation is satisfied for all the spherically symmetric space-times considered and is consistent with the thermodynamic properties of the horizons. We have thus established one-to-one mapping between entanglement mechanics and black hole thermodynamics. Having established, we can promote entanglement mechanics to entanglement thermodynamics!

\begin{table*}[!htb]
	\centering
	\resizebox{1.0\textwidth}{!}{%
		\begin{tabular}{@{}|l|l|l|l|l|l|@{}}
			\toprule
			Space-time & Entanglement Structure& Thermodynamic Structure & Smarr formula & Pressure & Potential  \\ 
			\toprule
			Schwarzschild & $\,\,\,\,S=(c_s/a^2)r_h^2;\quad E=(c_e/a^2)M$ & $S_{BH}=\pi r_h^2;\quad E_{Komar}=M$ &$M=2TS_{BH}$ & ---  & --- \\[6pt] \hline
			Reissner-Nordstr\"om & $S_+=(c_s/a^2)r_+^2;\,\,E_+=(c_e/a^2)\sqrt{M^2-Q^2}$ & $S_{BH}=\pi r_+^2;\quad E_{Komar}=\sqrt{M^2-Q^2}$ &$M=2TS_{BH}+Q^2/r_+$ & --- &$Q/r_+$  \\[6pt] \hline
			Schwarzschild-AdS&$\,\,\,\,S=(c_s/a^2)r_h^2;\quad E=(c_e/a^2)[3M-r_h^2]$ &$S_{BH}=\pi r_h^2;\quad E_{Komar}=3M-r_h^2$&$M=2TS_{BH}-r_h^3/l^2$& $3/8\pi l^2$&  --- \\[6pt] \hline
			Schwarzschild-dS& $\,\,S_b=(c_s/a^2)r_b^2;\,\,\,\,E_b=(c_e/a^2)[3M-r_b^2]$&$S_{BH}=\pi r_b^2;\quad E_{Komar}=3M-r_b^2$&$M=2TS_{BH}+r_b^3/l^2$&$-3/8\pi l^2$& --- \\[6pt]
			\toprule
		\end{tabular}
	}
	\caption{Summary of entanglement mechanics and event-horizon thermodynamics. Since we have set $\epsilon\sim1.26$, they also satisfy $T=T_H$ and $E=2TS$ universally.}
	\label{tab:mapping02}
\end{table*}

We now proceed to derive the Smarr formula of black-hole thermodynamics from entanglement energy. 
For the event horizon of Reissner Nordstr\"om black hole, let us rewrite the entanglement energy \eqref{eq:SE-RN01} as
\begin{equation}
\label{eq:Ener-RN}
E_+=\frac{c_e}{a}\sqrt{\Delta_M^2-\Delta_Q^2}=\frac{c_e}{a^2}\left(M-\frac{Q^2}{r_+}\right).
\end{equation}
Using the relation $E=2TS$, we can rearrange the above equation to obtain the Smarr formula for Reissner Nordstr\"om~\cite{2010-Banerjee.Majhi-Phys.Rev.D}:
\begin{equation}
\label{eq:SmarrFor02}
M=\frac{2a^2}{c_e}TS+\frac{Q^2}{r_+}=2TS_{BH}+\frac{Q^2}{r_+},
\end{equation}
where we have used $\pi c_e/c_s=1$ and $S_{BH}$ is the Bekenstein-Hawking entropy.

Similarly, in the case of Schwarzschild-AdS, we can rearrange entanglement energy relation \eqref{eq:SE-SAdS02} as follows:
\begin{equation}
\label{eq:Ener-SAdS}
E=\frac{c_e}{a}[3\Delta_M-\Delta_h]=\frac{c_e}{a^2}\left(M+\frac{r_h^2}{l^2}\right).
\end{equation}
Using the relation $E=2TS$, we can now obtain the Smarr formula for SAdS~\cite{2009-Kastor.etal-CQG}:
\begin{equation}
\label{eq:SmarrFor03}
M=2TS_{BH}-\frac{r_h^3}{l^2},
\end{equation}
where we have used $\pi c_e/c_s=1$. A similar procedure can be followed for the event horizon of SdS:
\begin{equation}
E_b=\frac{c_e}{a}[3\Delta_M-\Delta_b]=\frac{c_e}{a^2}\left(M-\frac{r_b^2}{l^2}\right).
\end{equation}
The resultant Smarr formula for the event horizon in SdS will be:
\begin{equation}
\label{eq:SmarrFor04}
M=2TS_{BH}+\frac{r_b^3}{l^2},
\end{equation}
where have used $\pi c_e/c_s=1$. This is the \emph{fifth key result} of this work. We observe the following --- although entanglement energy and entropy are not equal to the Komar energy and Bekenstein-Hawking entropy, respectively, there is a one-to-one mapping that exactly generates the Smarr-formula from entanglement. We have consolidated the results in Table \eqref{tab:mapping02}.
Therefore, we have built a consistent structure of entanglement thermodynamics, which also happens to suggest a quantum origin to the thermodynamic structure of space-time horizons~\cite{2019-Padmanabhan-IJMPD}.

Another interesting result from proper-length treatment is that when we quantize a field in a region where $\rho$ is bounded and well-defined, the horizon radius becomes discretized. We have observed the discretization in de-Sitter, the Cauchy horizon of Reissner Norstr\"om, and Schwarzschild de Sitter. 
We see that the rescaled horizon radius $\Delta_h$ has a discrete spectrum of equally-spaced values due to which the area, and hence entropy, energy, and temperature scale as
\[
S\sim N^2,~~E\sim N, ~~ T\sim N^{-1} \, . 
\]
This also implies a bound on the degrees of freedom in these space-times, as has been conjectured for the de-Sitter~\cite{2001-BANKS-InternationalJournalofModernPhysicsA}.
While the entropy is discrete, it is not equally spaced. This is in contrast to Bekenstein's argument 
that the horizon area  being an adiabatic invariant, must have equally spaced discrete spectrum~\cite{Bekenstein1998,2003-Dreyer-Phys.Rev.Lett.,2007-Corichi.etal-Phys.Rev.Lett.,2016-Skakala.Shankaranarayanan-InternationalJournalofModernPhysicsD}. Note that in the semi-classical 
limit, there is little distinction between our result and Bekenstein's horizon area discretization.

\section{Conclusions and Discussions}\label{secconc}

In this work, we have extensively studied entanglement thermodynamics in flat and spherically symmetric space-times with at least one horizon. 
In the discretized approach towards field theory, where the Physics is highly sensitive to the UV cut-off ($a$), we have exploited an inherent scaling symmetry of entanglement entropy to obtain a scale-independent treatment of entanglement thermodynamics. In each of the models studied, we were able to identify the scaling transformations that generate an infinite number of systems with the same entanglement entropy, distinguished only by their respective energies and temperatures. Under these scaling transformations, we see that:
\begin{equation}
\label{eq:SET-Scaling}
S\to S;\quad E\to\eta^{-1}E;\quad T\to\eta^{-1}T.
\end{equation}
An immediate consequence of the above relation is that when we rescale $a$ down to Planck length $l_p$, both energy and temperature increase drastically, whereas the entropy remains invariant. If we were to define an entanglement heat capacity here of the form $C=dE/dT$, we see that this quantity will also be scale-invariant.

The scale-invariant treatment of entanglement also reveals some fundamental properties involving zero modes. In Section \ref{seccho}, we found that the entropy divergence in $\alpha\to\infty$ limit and $\omega\to 0$ limit is associated with the presence of a single zero-mode in the system. The zero-mode occurs when the scale-invariant parameter $\lambda$ vanishes. In Section \ref{sec1d}, we studied two different boundary conditions to isolate the effects of zero modes in a $(1+1)$-D scalar field. 
For a finite-size chain with Neumann boundary condition with  $\Lambda=0$, the entanglement entropy diverges due to the presence of a zero-mode. This corresponds to both $m_f\to0$ and $a\to0$ limits. For the DBC chain, there are no zero modes for finite $N$, and hence no entropy divergence even when $m_f\to0$. However, in the continuum limit ($N\to\infty$), the DBC chain eventually develops a zero mode, which causes the entropy to diverge. We thus conclude that any divergence of entanglement entropy in discretized field theory, including that of UV-divergence arising from the $a\to 0$ limit, can always be associated with the accumulation of zero-modes.

In Section \ref{sec3d}, we reproduced the above results for a $(3+1)$-D sphere. Here, despite using a cut-off as small as $\Lambda=10^{-10}$, we could see that the NBC entanglement entropy agrees exactly with that of DBC. This means that in higher dimensions, the zero-mode effects on entropy scaling start kicking in at a much lower cut-off value of $\Lambda$, and even then, the area-law scaling is preserved. On the other hand, for NBC as $\Lambda\to0$, entanglement energy is far more sensitive to zero modes, wherein the prefactor of area-law starts increasing drastically. 

In Section \ref{secgr}, we extended the formalism to space-times with horizons and numerically obtained scaling relations for entanglement energy and entropy. We show that the entanglement energy scales as the Komar energy of the horizon, and for all space-times considered the entanglement temperature satisfied $T\sim 1.26 T_H$. In Section \ref{sec:Ent-BHconnection}, we appropriately fix the pre-factor $\epsilon$ in the definition of entanglement energy (\ref{entenergy}) to establish the equivalence between entanglement thermodynamics and black-hole thermodynamics, wherein $E=2TS$ and $T=T_H$ are consistently satisfied. This equivalence is further proved by deriving the Smarr relations for RN, SdS, and SAdS from entanglement scaling relations. Similar properties have also been noted for holographic entanglement entropy\cite{2020-Saha.etal-}. Our results suggest that the horizon thermodynamics is of quantum origin.

The above results bring attention to the following interesting questions:
\begin{enumerate}
\item We have restricted our analysis to spherically symmetric space-times. Whether the results will also hold for rotating black hole space-times?
\item Can we extend the one-to-one correspondence between entanglement 
thermodynamics and black hole thermodynamics to higher dimensions? 
Does it hold for higher genus black holes?
\item What is the physical significance of the horizon radius discretization for 
asymptotic de Sitter black holes?
\item The scaling relation \eqref{eq:SET-Scaling} suggests that the entanglement thermodynamics is valid down to the Planck scale. However, it is expected that the gravity corrections will become relevant at these scales. Will these scaling relations  
\eqref{eq:SET-Scaling} hold for modified gravity theories and scalar fields non-minimally coupled to gravity~\cite{1998-Frolov.Fursaev-CQG}?  
\end{enumerate}

We hope to return to study some of these problems shortly.

\begin{acknowledgements}
The authors thank T. Padmanabhan for discussions. SMC is supported by DST-INSPIRE  Fellowship offered by the Dept. of Science and Technology, Govt. of India. The work is supported by the MATRICS SERB grant.

\end{acknowledgements}
\appendix

\section{Harmonic Chains}
\label{app:cho}

\subsection{Reduced density matrix for CHO}
\label{wkb}

For completeness, we provide the steps leading to entanglement entropy \eqref{eq:CHO-EntS}.
The ground state wave-function of the Hamiltonian \eqref{eq:CHO-Hamil02} is therefore:
\begin{equation}
\Psi_0(x_+,x_-)=\frac{(\beta_+\beta_-)^{1/4}}{\sqrt{\pi}}\exp{-\frac{\beta_+x_+^2}{2}-\frac{\beta_-x_-^2}{2}},
\end{equation}
where we have defined $\beta_{\pm}=m\omega_{\pm}/\hbar$. On shifting back to the original co-ordinates, we see that wave-function is entangled in $x_1$ and $x_2$. We can then perform a partial trace on the density matrix to obtain the reduced density matrix of a subsystem:
\begin{align}
	\rho_1(x_1,x_1')&=\int\limits_{-\infty}^{\infty}dx_2\Psi_0^*(x_1',x_2)\Psi_0(x_1,x_2) \\
	&=\sqrt{\frac{\gamma_1-\gamma_2}{\pi}}\exp{-\gamma_1\frac{(x_1^2+x_1'^2)}{2}+\gamma_2x_1x_1'} \, , \nonumber
\end{align}
where $\gamma_1$ and $\gamma_2$ are given by:
\begin{equation}
	\gamma_1=\frac{\beta_+^2+\beta_-^2+6\beta_+\beta_-}{4(\beta_++\beta_-)};\quad\gamma_2=\frac{(\beta_+-\beta_-)^2}{4(\beta_++\beta_-)}
\end{equation}
In order to find the eigenvalues of $\rho_1(x_1,x_1')$, we must solve the following integral equation for $p_n$:
\begin{equation}\label{inte}
\int\limits_{-\infty}^{\infty}dx_1'\rho_1(x_1,x_1')f_n(x_1')=p_nf_n(x_1)
\end{equation}
The solution for the above integral equation is well known~\cite{1993-Srednicki-Phys.Rev.Lett.}:
\begin{align}
	p_n&=(1-\xi)\xi^n,\\
	f_n(x)&=H_n(\sqrt{\varrho}x)\exp{-\varrho\frac{x^2}{2}},
\end{align}
where the new parameters $\varrho$ and $\xi$ are defined as follows:
\begin{equation}
	\varrho=\sqrt{\beta_+\beta_-};\quad \xi=\frac{\gamma_2}{\gamma_1+\varrho}.
\end{equation}
The parameter $\xi$ in the above expression, on further simplification, reduces to the following form:
\begin{equation}
\xi(\lambda) = \left\{\frac{\left(\lambda+2\right)^{1/4}-\lambda^{1/4}}{\left(\lambda+2\right)^{1/4}+\lambda^{1/4}}\right\}^2;\qquad\lambda=\frac{m\omega^2}{\alpha^2}.
\end{equation}

\subsection{Zero mode analysis for CHO}
\label{app:zeromode}

Let us focus on the dimensionless Hamiltonian $\tilde{H}$ in \eqref{cho1}, and look at the normal modes of the system. In the case where $\lambda=0$, we get:
\begin{equation}
\tilde{\omega}_-=\sqrt{2};\quad\tilde{\omega}_+=0.
\end{equation}
Due to the zero mode in the $\tilde{x}_+$ co-ordinate, we have a free particle in this limit. Since the free particle wave-function is non-normalizable, we will assume that this particle is in a box of size $\tilde{L}\gg 1$. The ground state wave-function takes the form:
\begin{equation}\label{fp1}
\Psi_0(\tilde{x}_+\tilde{x}_-)=\frac{1}{\sqrt{\tilde{L}}}\left(\frac{\tilde{\beta}_-}{\pi}\right)^{1/4}\exp{-i\tilde{k}_+\tilde{x}_+-\tilde{\beta}_-\frac{\tilde{x}_+^2}{2}},
\end{equation}
where $\tilde{\beta}_{\pm}=\tilde{\omega}_{\pm}/\hbar$. The reduced density matrix in this case will turn out to be:
\begin{equation}
\rho_1(\tilde{x}_1,\tilde{x}_1')=\frac{\sqrt{2}}{\tilde{L}}\exp{i\frac{\tilde{k}_+}{\sqrt{2}}(\tilde{x}_1-\tilde{x}_1')-\frac{\tilde{\beta}_-}{8}\left(\tilde{x}_1-\tilde{x}_1'\right)^2}
\end{equation}
To find the eigenvalues, we must solve the integral equation given in \eqref{inte}. Utilizing translational symmetry, we may guess the eigenfunctions to be $f(\tilde{x}_1)=\exp{-ik\tilde{x}_1}$. The eigenvalues are therefore given by the fourier transform of the reduced density matrix:
\begin{equation}
p_k=\frac{4}{\tilde{L}}\sqrt{\frac{\pi}{\beta_-}}\exp{-\frac{2}{\tilde{\beta}_-}\left(\frac{\tilde{k}_+}{\sqrt{2}}-k\right)^2}
\end{equation}
The entanglement entropy can therefore be calculated as follows:
\begin{align}
S&=-\int_{-\infty}^{\infty} \frac{dk}{(2\pi/\tilde{L})}p_k\ln{p_k}\nonumber\\
&=\frac{1}{\sqrt{2}}\ln{\left(\frac{e\tilde{\beta}_-\tilde{L}^2}{16\pi}\right)}
\end{align}

From the above expression, we see that when on removing the IR cutoff and taking $\tilde{L}\to\infty$, the entropy diverges. This is in fact the limit where the wave-function becomes non-normalizable. Instead, we may also fix $\tilde{L}$ by taking the WKB approximation of the ground state wave-function of $\tilde{x}_+$ oscillator:
\begin{align}
\psi_0(\tilde{x}_+)&=\frac{c}{\sqrt{\tilde{p}_+}}\exp{-\frac{i}{\hbar}\int d\tilde{x}_+\tilde{p}_+(\tilde{x}_+)}\\
\tilde{p}_+&=\sqrt{2\left(\tilde{E}_+^{(0)}-\frac{\lambda^2}{2}\tilde{x}_+^2\right)}
\end{align}
In the limit $\lambda\to0$, we see that $\tilde{p}_+\sim \sqrt{\hbar\lambda}$. We now normalize the wave-function using the turning points $u=\pm1/\tilde{k}_+$:
\begin{equation}
\int_{-u}^u \frac{c^2}{\tilde{p}_+}d\tilde{x}_+=1;\qquad c=\tilde{k}_+\sqrt{\hbar/2}
\end{equation}
The ground state wave-function therefore becomes:
\begin{equation}
\lim_{\lambda\to0}\psi_0(\tilde{x}_+)\sim \sqrt{\frac{\tilde{k}_+}{2}}\exp{-i\tilde{k}_+\tilde{x}_+}
\end{equation}
On matching this with \eqref{fp1}, we can fix the IR-cutoff from the relation $\tilde{L}\tilde{k}_+=2$. The expression for entanglement entropy now becomes:
\begin{equation}
\lim_{\lambda\to 0}S\sim \frac{1}{\sqrt{2}}\ln{\left(\frac{e}{2\sqrt{2}\pi\lambda}\right)}.
\end{equation}
For the entropy to be positive, we see that $\lambda<e/\sqrt{8}\pi\sim0.3$, and therefore gives a bound on $\lambda$ for the approximation to be physical. In the zero mode limit $\lambda\to0$, both the IR cutoff $\tilde{L}$ and entanglement entropy diverge. From this, we deduce that entropy divergence is a direct consequence of non-normalizability of free particles in the system.

\subsection{Entanglement entropy for harmonic chains}
\label{app:Ent-HC}

The algorithm to numerically calculate entanglement entropy from this coupling matrix is well known~
\cite{1986-Bombelli.etal-Phys.Rev.D,1993-Srednicki-Phys.Rev.Lett.,Das2010}. To summarize, we start with the ground state wave-function corresponding to the Hamiltonian \eqref{eq:HC-Hamil}:
\begin{equation}
\Psi_0(\vec{x})=\left[\frac{\det\Omega}{\pi^N}\right]^{1/4}\exp{-\frac{1}{2}\sum_{ij}x_i\Omega_{ij}x_j},
\end{equation}
where $\Omega=K^{1/2}$. To obtain the following reduced density matrix, we trace out the first $n$ oscillators:
\begin{multline}
\rho_{red}(x^{\alpha},x'^{\beta})\sim\exp\bigg\{-\frac{1}{2}(\Gamma_1)_{\alpha\beta}\left(x^{\alpha}x^\beta+x'^{\alpha}x'^{\beta}\right)\\+\left(\Gamma_2\right)_{\alpha\beta}x^{\alpha}x'^{\beta}\bigg\},
\end{multline}
where,
\begin{align}
K_{AB}&=\begin{bmatrix}
	(K_{in})_{ab}&(K_{int})_{a\beta}\\
	(K_{int}^T)_{\alpha b}&(K_{out})_{\alpha\beta}\end{bmatrix}\nonumber\\	
\Omega_{AB}&=\begin{bmatrix}A_{ab}&B_{a\beta}\\(B^T)_{\alpha b}&C_{\alpha\beta}\end{bmatrix}\nonumber\\ 
(\Omega^{-1})_{AB}&=\begin{bmatrix}
	\tilde{A}_{ab}&\tilde{B}_{a\beta}\\(\tilde{B}^T)_{\alpha b}&\tilde{C}_{\alpha\beta}\end{bmatrix}\nonumber\\
\Gamma_1&=C-\Gamma_2\nonumber\\
 \Gamma_2&=\frac{1}{2}B^TA^{-1}B
\end{align}
Here, all upper-case Latin indices take $1,\cdots,N$, all lower-case indices take $1,\cdots,n$, and all Greek indices take $(n+1), \cdots, N$. Let us rewrite $\Gamma_1=V^T\Gamma_DV$ where $\Gamma_D$ is diagonal, and define the following matrix:
\begin{equation}\label{beta}
\bar{\beta}=\Gamma_D^{-1/2}V\beta V^T\Gamma_D^{-1/2}.
\end{equation}
After performing a series of diagonalizations, the entanglement entropy \eqref{eq:Ent-HC} can be obtained.

\section{Near-Horizon Approximation}\label{nearhorizon}
For all the models studied in Section \eqref{secgr}, we were able to obtain an exact analytical expression for proper length $\rho$. However, for a general static, spherically symmetric space-time with multiple horizons, obtaining such an analytical expression may be difficult. Since we are interested in the entanglement properties of the field very close to the horizon, we can simplify our formalism by employing the near-horizon approximation of the metric:
\begin{equation}
f(r)\approx (r-r_h)f'(r_h)
\end{equation}
Proper length can therefore be approximated as:
\begin{equation}
\rho\approx 2\sqrt{\frac{r-r_h}{f'(r_h)}}.
\end{equation}
On discretizing proper length as $\rho=ja$, we get the following expression:
\begin{equation}
j=2\sqrt{\frac{r_j-\Delta_h}{\sigma_h}}
\end{equation}
where we have introduced dimensionless parameters $\Delta_h=r_h/a$, $r_j=r/a$ and $\sigma_h=af'(a\Delta_h)$. We obtain the same Hamiltonian as in \eqref{bh1}, but the near-horizon approximation gives us the following relations:
\begin{align}
r_j&=\Delta_h+\frac{j^2\sigma_h}{4}\\
f_j&=\frac{j^2\sigma_h^2}{4}
\end{align}

This can be substituted back into the Hamiltonian in \eqref{bh1} and the resultant coupling matrix in \eqref{grK}. On specifying the input parameter $\Delta_h$ (rescaled horizon radius), the parameter $\sigma_h$ is fixed with respect to surface gravity at the horizon. Given these two, we can simulate the entanglement thermodynamics of a scalar field near that horizon in a given spherically symmetric space-time. For example, in the case of Schwarzschild, we see that $\sigma_h=1/\Delta_h$ and for de-Sitter, $\sigma_h=-2/\Delta_h$. The accuracy of these results improves for larger and larger $\Delta_h$. The most significant difference from the exact treatment is that we cannot get a discretization relation as we did in space-times where proper length $\rho$ is bounded.


%\bibliography{ref}

\end{document}